\documentclass[12pt]{article}
\usepackage{bkqmb_mod}
\usepackage{psfig,epsfig,graphicx,citesort}

\newif\ifstandalone 
        \standalonetrue 
\newif\ifdraft
\ifstandalone {\def\titlehead#1{}} \fi

\makeatletter  
       \@addtoreset{equation}{section}
       \renewcommand\theequation{\thesection.\@arabic\c@equation}
\makeatother

  \ifdraft
   \newcommand{\note }[1]{{\bf [{ #1~}]}\marginpar{~~~~~~~\rule[0ex]{0.5mm}{2ex}} }
  \else
    \newcommand{  \note }[1]{}
  \fi
 \newcommand{\SKIP}[1]{}

\newcommand{\newfig}[5]{
  \begin{figure}[b!t]
    \begin{center}  \null\ 
      \psfig{figure=#2,#5=#3}
    \end{center}
 \vskip-5mm
    \caption[*]{
               #4} \label{#1}
  \end{figure}
}

\newcommand{\newfigw}[4]{\newfig{#1}{#2}{#3}{#4}{width}}  
\newcommand{\newfigh}[4]{\newfig{#1}{#2}{#3}{#4}{height}} 
  
\newlength{\lll}
\newcommand{\myfig}[1]{\setlength{\lll}{7ex}
                       \begin{minipage}{\lll}\hspace*{.01\lll}
                  \includegraphics[width=.6\lll]{Figures/#1.eps}
                       \end{minipage}}

\newcommand{\beq}[1]{\begin{equation}\label{#1}}

\newcommand{\eeq}{\end{equation}}
\newcommand{\1}{\\[1.5ex]} 
\newcommand{\2}{\\[3ex]} 
\newcommand{\beqa}[1]{\beq{#1}\begin{array}{lllllllll}}
\newcommand{\eeqa}{\end{array}\eeq}
\newcommand{\eq}[1]{eq.\ (\ref{#1})}       
\newcommand{\Eq}[1]{Eq.\ (\ref{#1})}       
\newcommand{\half}{{$1\over2$}}            
\newcommand{\Half}{{1\over2}}            
\newcommand{\ident}{\equiv}
\newcommand{\gsim}{\raisebox{-3pt}{$\stackrel{>}{\sim}$}}
\newcommand{\lsim}{\raisebox{-3pt}{$\stackrel{<}{\sim}$}}
\newcommand{\tr}{tr}
\newcommand{\hH}{\hat{H}}
\newcommand{\vS}{\vec{\hat{S}}}
\newcommand{\bra}[1]{\left\langle #1 \right|}
\newcommand{\ket}[1]{\left|       #1 \right\rangle}
\newcommand{\up}{\uparrow}
\newcommand{\down}{\downarrow}
\newcommand{\identity}{1\!\!1}
\newcommand{\en}{\epsilon}
\newcommand{\ul}[1]{\underline{#1}}
\newcommand{\ebh}{e^{-\beta \hH}}
\renewcommand{\H}{\hat{H}}
\newcommand{\h}{\hat{h}}

\newcommand{\EM}[1]{\smallskip\noindent\underline{\em #1}}
\newcommand{\EMM}[1]{\underline{\em #1}}

\newcommand{\bit}{\begin{itemize}}
\newcommand{\eit}{\end{itemize}}
\newcommand{\fig}[1]{figure \ref{#1}}       
\newcommand{\union}{\bigcup}
\newcommand{\element}{\in}
\newcommand{\TO}{\!\rightarrow\!}
\renewcommand{\SS}{{\cal S}^\prime}
\newcommand{\GG}{G^\prime}
\newcommand{\SSp}{S_p^\prime}

\newcommand{\scr}[1]{\mbox{\scriptsize #1}}
\renewcommand{\S}{{\cal S}} 
\newcommand{\hS}{\hat{S}}
\newcommand{\F}{{\cal F}} 
\renewcommand{\O}{{\cal O}} 
\newcommand{\tauexp}{\tau_{exp}}
\newcommand{\ij}{{\left< ij \right>}}
\newcommand{\tJ}{$t$-$J$ }

\newcommand{\dtau}{\Delta\tau}
\newcommand{\mscript}[1]{\mbox{\scriptsize #1}}
\newcommand{\goesto}{\rightarrow}

\newcommand{\wl}{worldline}
\newcommand{\WL}{worldline}
\newcommand{\psS} {p(\S       \TO \S^\prime)}
\newcommand{\pSs} {p(\S^\prime\TO \S)       }

\newcommand{\psgSg}  {p\left(\, (\S  ,G  ) \TO (\SS ,G   ) \,\right)} 
\newcommand{\psgSG}  {p\left(\, (\S  ,G  ) \TO (\SS ,\GG ) \,\right)} 
\newcommand{\pSGsg}  {p\left(\, (\SS ,\GG) \TO (\S , G   ) \,\right)}

\newcommand{\boldbar}[1]{\overline{#1}}

\title{The Loop Algorithm%
  \footnote{Third edition, July 2002. To appear in Adv. Phys.}
  }
\titlehead{The Loop Algorithm}
\author{ H.G. Evertz}
\address{\vspace*{3ex}
        Institut f\"ur Theoretische Physik\\ Technische Universit\"at Graz, 8010 Graz, Austria\\ 
        e-mail: evertz@tugraz.at\\
        Tel.\ +43-316-873-8178, Fax -8677}

\begin{document}

\maketitle

\begin{abstract}
A review of the Loop Algorithm, its generalizations,
and its relation to some other Monte Carlo techniques is given.
The loop algorithm is a Quantum Monte Carlo 
procedure which employs nonlocal changes of worldline configurations,
determined by  local stochastic decisions. 
It is based on a formulation of  quantum models of any dimension 
in an extended ensemble of worldlines and graphs, and is 
related to Swendsen-Wang algorithms.
It can be represented directly on an operator level,
both with a continuous imaginary time path integral and with the 
stochastic series expansion (SSE).
It overcomes many of the difficulties of traditional worldline simulations. 
Autocorrelations are reduced by orders of magnitude.
Grand-canonical ensembles, off-diagonal operators, and variance reduced estimators are accessible.
In some cases, infinite systems can be simulated.
For a restricted class of models, the fermion sign problem can be overcome.
Transverse magnetic fields are handled efficiently,
in contrast to strong diagonal ones.
The method has been applied successfully to a variety of models for spin
and charge degrees of freedom,
including Heisenberg and  XYZ spin models, hard-core bosons, Hubbard, and $t$-$J$-models.
Due to the improved efficiency, 
precise calculations of asymptotic behavior and
of quantum critical exponents have been possible.
\end{abstract}

\ifstandalone 
      \makeatletter
         \renewcommand{\baselinestretch}{0.5} 
         \renewcommand{\tableofcontents}{\section*{\contentsname
                                                 } \@starttoc{toc}\vfill}
      \makeatother
    \vfill\pagebreak\tableofcontents\vfill
         \renewcommand{\baselinestretch}{1} 
\fi


 \section{Introduction and Summary}\label{Introduction}
%
A pedagogical review of the loop algorithm, its generalizations,
and the range of present applications is given, including some new results.
The loop algorithm \cite{EvertzLM93,EvertzM92,EvertzM94,talk91}
is a Quantum Monte Carlo procedure. 
It is applicable to numerous models 
both in imaginary time worldline formulation \cite{Scalettar-WL99} 
and within the stochastic series expansion (SSE)
\newcommand{\SSElist}{SandvikK91,Sandvik92,Sandvik97}
\cite{\SSElist}.
It overcomes many of the difficulties of traditional worldline simulations
by performing {\em nonlocal} 
changes of worldline configurations,
which are determined by {\em local} stochastic decisions.
The loop algorithm is based on a formulation of the worldline system in 
an extended ensemble which consists of both the original variables 
(spins or occupation numbers) and of 
graphs (sets of loops),
either on the level of matrix elements \cite{EvertzLM93,EvertzM92,EvertzM94,KawashimaG95a},
or of loop-operators \cite{AizenmanN94,Nachtergaele93,BrowerCW98,Sandvik99b,HaradaK00}. 
It is related to Swendsen-Wang \cite{SwendsenW87} 
cluster algorithms for classical statistical systems.
It has been applied and generalized by a large number of authors.
Before we delve into technical details, let us summarize the main features.
\vspace*{-0.5ex} \begin{itemize} \itemsep=0pt
\item[(a)] Autocorrelations between successive Monte Carlo configurations are drastically 
           reduced, thereby reducing the number of Monte Carlo sweeps required for a given system, 
           often by orders of magnitude. 
\item[(b)] The grand-canonical ensemble (e.g.\ varying magnetization, occupation number, 
           winding numbers) is naturally simulated.
\item[(c)] The continuous time limit can be taken \cite{BeardW96}, completely eliminating 
           the Trotter-approximation. 
           In fact, the loop algorithm can be formulated directly in continuous time.
\item[(d)] Observables can be formulated in terms of loop-properties,
           as so called Improved Estimators,
           reducing the errors of measured quantities.
\item[(e)] Off-diagonal operators can be measured through improved estimators \cite{BrowerCW98}.
\item[(f)] Transverse fields can be simulated efficiently 
           \cite{RiegerK99,CoxGHSW99,ChandrasekharanSW99}.
\item[(g)] For a restricted class of models, including fermionic ones, it has been shown
           \cite{ChandrasekharanW99,ChandrasekharanCHW00,Chandrasekharan99,%
                 CoxGHSW99,%
                 ChandrasekharanSW99,Chandrasekharan00,ChandrasekharanO00a,ChandrasekharanO00b,%
                 CoxH00,Osborn00,ChandrasekharanO01,Chandrasekharan01a,ChandrasekharanSW01,%
                 ChandrasekharanCOW02,HeneliusS00}
           that by clever use of improved estimators the sign problem can be overcome.
\item[(h)] Bond disorder and depleted lattices can be trivially included.
           The algorithm  remains completely unchanged in any dimension.
           Generalizations to higher spin representations 
           \cite{KawashimaG95a,KawashimaG94,Kawashima95,HaradaTK98,TodoKT98,TodoK01,KimGWB98},
           biquadradic interactions \cite{HaradaK00,HaradaK01},
           and to fermionic models 
           \cite{KawashimaGE94,Kawashima96,AmmonEKTF98,BrunnerM98,BrunnerAM99,SenguptaSC02}
           exist.
\end{itemize}
Each of the points (a)-(g) can save orders of magnitude in computational effort over
the traditional local worldline method.
In addition, the algorithm is easier to program than traditional worldline updates.
The method has some limitations: 
\vspace*{-0.5ex}\begin{itemize} \itemsep=0pt
\item[(a)] Long range interactions make the algorithm more complicated
           and less effective.
\item[(b)] More seriously, strong asymmetries in the Hamiltonian
           will make the original algorithm exponentially slow.
           This includes large magnetic fields (or chemical potential) with $\beta h\,\gsim\, 3$
           and other non ``particle-hole-symmetric'' terms like 
           softcore bosons.
           The difficulty disappears when such a field
           can be put into transverse direction (section \ref{TransverseField}).
           Otherwise alternative methods are preferable (see section \ref{Related}).
\end{itemize}

Some of the usual limitations of worldline methods also remain in the loop algorithm.
The most serious limitation remains (so far) the sign problem,
which still occurs in most fermionic models as well as in frustrated spin systems.
Further generalizations of the meron idea (section \ref{Merons}) may help here in the future.

The loop algorithm has already been used for many physical models.
The original formulation \cite{EvertzLM93,EvertzM92,EvertzM94,talk91}
of the algorithm (in vertex language) applies directly to general
spin \half\ quantum spin systems in any dimensions, 
e.g.\ the 2D Heisenberg model \cite{WieseY94}, 
where improved estimators for this algorithm were first used.
At the root of the loop algorithm is an exact mapping of the physical model to 
an extended phase space which includes loops in addition to the original 
worldlines.
In ref.\ \cite{KawashimaG95a} it was shown in a general framework
that this mapping is a Fortuin-Kasteleyn-like representation.
A related mapping to a loop-model was independently used in a rigorous study of 
spin models \cite{AizenmanN94,Nachtergaele93}.
The algorithm was generalized to anisotropic XYZ-models \cite{EvertzM94},
with explicit update probabilities given
in ref. \cite{Kawashima95} and in section \ref{ProbXXZ}.
The method has been adapted and extended
to fermion systems like the Hubbard model \cite{KawashimaGE94,CoxGHSW99} and 
to the \tJ model \cite{Kawashima96,AmmonEKTF98,BrunnerM98,BrunnerAM99}.
The meron method \cite{ChandrasekharanW99} to overcome the fermion sign problem 
in a restricted class of models was developed
and also applied to non-standard Hubbard-like models 
\cite{Osborn00,ChandrasekharanO01,Chandrasekharan01a,ChandrasekharanCOW02},
to antiferromagnets in a transverse field \cite{CoxGHSW99,ChandrasekharanSW99},
and to a partially frustrated spin model \cite{HeneliusS00}.
The loop algorithm was extended to 
quantum spin systems with higher spin representation
\cite{KawashimaG95a,KawashimaG94,HaradaTK98,TodoKT98,TodoK01,KimGWB98}, 
also for the XYZ-case \cite{Kawashima95},
and to cases with transverse fields \cite{RiegerK99}.
The extension to more than (1+1) dimensions is immediate \cite{EvertzLM93,EvertzM92}:
the algorithm remains completely unchanged, only the geometry of the 
plaquette lattice changes.
In ref.\ \cite{BeardW96} it has been shown that the continuous time limit
can be taken, and in ref.\ \cite{BrowerCW98} that  any n-point function can be measured,
with diagonal and off-diagonal two-point functions being especially simple.

A related development along a somewhat different line 
are  the ``Worm'' algorithm in continuous time \cite{ProkofevST98b},
``operator-loop-updates'' in SSE \cite{Sandvik99b}
and the recent method of ``directed loops'' \cite{SyljuasenS02},
which are applicable to a larger class of models, especially with strong asymmetries.

There have been many  successful large scale applications
of the loop algorithm,
both in imaginary time, and, more recently,
in a variant called 
``deterministic operator loops'' \cite{Sandvik99b,HeneliusS00}
within the SSE formulation, 
to fermionic and  especially to numerous Heisenberg-like models,
for spin \half\ and higher spins, 
with and without anisotropy, disorder, or impurities,
from spin chains up to three dimensional systems,
including, e.g.,  a high statistics calculation of quantum critical exponents
on regularly depleted lattices 
of up to $20000$ spatial sites at temperatures down to $T=0.01$ \cite{TroyerKU96,TroyerIU97,TroyerI97}.

Section \ref{algorithm} describes the algorithm in its traditional form,
with a brief review of the worldline representation,
an intuitive  outline of the loop algorithm,
and a detailed step by step formal derivation
of the algorithm, followed by a brief summary.
We compute explicitly the update probabilities for the XXZ-model, and give a concise
recipe for the Heisenberg antiferromagnet.
Ergodicity is treated, 
it is shown that in some important cases a transformation
of the worldline model to a pure loop model can be done,
the original single cluster version is treated,
and arbitrary lattices are covered.
The state-of-the-art continuous time version is described (including a brief recipe)
in section \ref{ContTime}.
In section \ref{ImprEst} we introduce improved estimators,
 and in section \ref{sec:InfLattice}
simulations on infinite size systems.
In section \ref{Performance} we discuss the performance of the loop algorithm,
possibilities and limitations, and some implementation issues.

In section \ref{oper} the operator formulation of the method is introduced,
which provides an alternate derivation directly in continuous time,
and also within the stochastic series expansion,
which is discussed there,
including a description of the loop algorithm within SSE.

Section \ref{Generalizations} describes a number of generalizations,
some of them immediate.
Section \ref{Related} mentions related algorithms, and
section \ref{Applications} points to some of the physics problems to which 
the loop algorithm has been applied so far.
The appendix provides a prescription to ensure
the essential (yet often neglected) 
requirement for correct Monte Carlo simulations
that convergence and statistical errors are properly determined.

 \section{Algorithm}\label{algorithm}
%
We begin with the traditional formulation of  loop algorithm and loop representation
by way of a finite Trotter time worldline formulation.
An alternative derivation on an operator level is provided in section \ref{oper},
and is also applicable within the stochastic series expansion.

The loop algorithm, as usually presented,
acts in the \wl\ representation which is reviewed e.g.\ in ref.\  \cite{Scalettar-WL99}. 
We will develop the formal procedure for the general anisotropic (XYZ-like) case.
As an example we shall use the particularly simple but important case of
the one-dimensional quantum XXZ model \cite{Scalettar-WL99}.
It includes the {\em Heisenberg model} and {\em hard core bosons} as special cases.
We will see that the same calculation 
is valid for the loop algorithm in any spatial dimension
and already covers most of the important applications.
The simplest and most important case is  the loop algorithm for the
isotropic Heisenberg antiferromagnet, which will be summarized in section \ref{Recipe}.

 \subsection{Setup: Worldline representation and equivalent vertex models}\label{Setup}
%
Let us first recall the \WL\  representation
for the example of the XXZ model on a one-dimensional chain of $N$ sites \cite{Scalettar-WL99}.
The Hamiltonian is 
\beqa{HXXZ}
\hat{H} & = \sum_{\langle{\bf ij}\rangle} \hat{H}_{i,j}
        & = \sum_{\langle{\bf ij}\rangle}\left(
         J_{x}(\hat{S}^{x}_{{\bf i}}\hat{S}^{x}_{{\bf j}}+
         \hat{S}^{y}_{{\bf i}}\hat{S}^{y}_{{\bf j}})+
         J_{z}\hat{S}^{z}_{{\bf i}}\hat{S}^{z}_{{\bf j}} \right)
         -h \sum_{{\bf i}} \hat{S}^{z}_{{\bf i}}
                                            \1
  &   & = \sum_{\langle{\bf ij}\rangle}\left(
        {J_{x} \over 2} (\hat{S}^{+}_{{\bf i}}\hat{S}^{-}_{{\bf j}}+
        \hat{S}^{-}_{{\bf i}}\hat{S}^{+}_{{\bf j}})+
        J_{z}\hat{S}^{z}_{{\bf i}}\hat{S}^{z}_{{\bf j}}\right)
        -h \sum_{{\bf i}} \hat{S}^{z}_{{\bf i}}                       \;,
                                            \cr
\eeqa
where $\vec{\hat{S}}_i$
are quantum spin \half\ operators at each site $i$,
$\hat{S}_i^+$, $\hat{S}_i^-$ are the associated raising and lowering operators,
$h$ is a magnetic field,
and $\langle{ \bf ij}\rangle$ are pairs of neighbouring sites.
We use periodic boundary conditions (arbitrary ones are possible).

\pagebreak[3]\noindent
After splitting the Hamiltonian into commuting pieces
\beqa{split}
 \hat{H} &=& \hat{H}_{even} + \hat{H}_{odd} \\
 \hat{H}_{even, odd} & = & \sum_{i:\, \mscript{even}, \mscript{odd}} \hat{H}_{i,i+1} \,,
\eeqa
performing a Trotter-Suzuki breakup \cite{Trotter59,Suzuki76}
\beq{Zxxz}
  Z^{XXZ}   = tr \, e^{-\beta \hat{H}} 
    \,=\,    \lim_{L_t\goesto\infty} Z^{XXZ}_{tr} 
    \,=\,    \lim_{L_t\goesto\infty} tr \left( e^{-\frac{\beta}{L_t} \hat{H}_{even}} \,
                                              e^{-\frac{\beta}{L_t} \hat{H}_{odd}} 
                                      \right)^{\mbox{\normalsize $L_t$}} \,,
\eeq
and inserting complete sets of $\hat{S}^z$ eigenstates,
we arrive at the worldline representation
\beq{WL}
Z^{XXZ}_{tr} = \sum_{\{S_{il}^z\}} W(\{S_{il}^z\}) 
  = \sum_{\{S_{il}^z\}} \prod_{p}  W_p (\{S_p\}) \,,
\eeq
where the summation $\sum_{\{S^z_{il}\}}$ extends over all ``configurations'' 
$\S=\{S^z_{il}\}$ 
of ``spins'' $S_{il}^z = \pm \Half$, 
which live on the sites $(i,l)$,\, $i=1..N$,\, $l=1 .. 2L_t$,
of a (1+1)-dimensional checkerboard lattice.
The index $l=1,..,2L_t$\ corresponds to imaginary time.
The product $\prod_p$ extends over all shaded plaquettes of that lattice
(see \fig{figcheckerboard}),
and $S_p$ stands for the 4-tuple of spins at the corners of 
a plaquette $p=(\,(i,l),(i+1,l),(i,l+1),(i+1,l+1)\,)$.
%
%
\newfigh{figcheckerboard}{Figures/worldlines.eps}{5.5cm}{ 
Example of a worldline configuration on a checkerboard lattice of
shaded plaquettes. 
(a) Worldline picture. (b) The {\em same} configuration as a vertex picture.
Space direction (index $i$) is horizontal, 
imaginary time direction (index $l$) vertical.
The variables $S^z_{i,l}$ are defined on each lattice site.
Worldlines  (arrows upwards in time)   denote $S^z_{il}=+\Half$, 
empty sites (arrows downwards in time) denote $S^z_{il}=-\Half$. 
The Hamiltonian $\hat{H}_{i,i+1}$ acts on the shaded plaquettes.
}
%
\pagebreak[3]
The weight $W_p$ at each plaquette 
\footnote{We keep a plaquette index $p$ with $W_p$ to cover spatially varying Hamiltonians.}
\beq{Wp}
 W_p(S_p) = 
  \langle S^{z}_{i,l  } S^{z}_{i+1,l  } | e^{-(\Delta\tau) \hat{H}_{i,i+1}} |
          S^{z}_{i,l+1} S^{z}_{i+1,l+1} \rangle \;,
\eeq
where $\Delta\tau\ident \beta/L_t$, 
is given by the matrix elements%
\footnote{The notation $a,b,c$ is standard for vertex models \cite{Baxter-book89},
          the notation $1,3,2$ (in different order) is that used in
          refs.\ \cite{KawashimaG95b,KawashimaG95a,Kawashima95}.}
\footnote{\label{bipartite}
The matrix element $b$ is originally proportional to $\sinh{(-\frac{\dtau}{2} J_x)}$.
It is positive for ferromagnetic XY couplings $J_x<0$.
For antiferromagnetic XY couplings  $J_x>0$, the minus signs cancel on bipartite lattices.
Equivalently, $b$ can be made positive on a bipartite
lattice by rotating $\hat{S}^{x,y}\TO -\hat{S}^{x,y}$ on one of the two sublattices.
We have already assumed such a rotation in \eq{matrixelements}.
}
of $\hat{\cal H}=(\dtau){\hat{H}_{i,i+1}}$:%
%
\newcommand{\ME}[2]{\langle {#1} | e^{-\hat{\cal H}} | {#2} \rangle}
\newcommand{\id}{\!\!\ident\!\!}
\newcommand{\gl}{\!\!=     \!\!}
\beq{matrixelements}
 \begin{array}{lccccccccccc}
W(1^+)  &\id& a_{+}&\id&\ME{++}{++}&\gl&\          & & e^{-\frac{\dtau}{4} J_z} & e^{+\frac{\dtau}{2} h}\1
W(1^-)  &\id& a_{-}&\id&\ME{--}{--}&\gl&\          & & e^{-\frac{\dtau}{4} J_z} & e^{-\frac{\dtau}{2} h}\1
W(2^\pm)&\id& c    &\id&\ME{+-}{+-}&\gl&\ME{-+}{-+}&\gl& e^{+\frac{\dtau}{4} J_z} & 
                                                           \cosh{(\frac{\dtau}{2} |J_x|)} \1
W(3^\pm)&\id& b    &\id&\ME{+-}{-+}&\gl&\ME{-+}{+-}&\gl& e^{+\frac{\dtau}{4} J_z} & 
                                                           \sinh{(\frac{\dtau}{2} |J_x|)} . 
 \end{array}
\eeq

\begin{figure}[!t]
%
\begin{center} \null\                   
      \psfig{figure=Figures/plaquettes.eps,width=11cm} 
 \end{center}
 \vskip-5mm
    \caption[*]{%
       Allowed plaquette configurations $S_p$.
       Top line: worldline picture. Bottom line: the same plaquettes in vertex picture.
       In the XXZ case (= six-vertex case),
       only the plaquettes with continuous worldlines, $S_p = 1^\pm, 2^\pm, 3^\pm$,
       have nonzero weight, \eq{matrixelements}.
       In the anisotropic XYZ case (= eight-vertex case) the plaquettes $4^\pm$ also have nonzero weight.
} \label{figplaquettes}
%
    \begin{center} \null\ 
      \psfig{figure=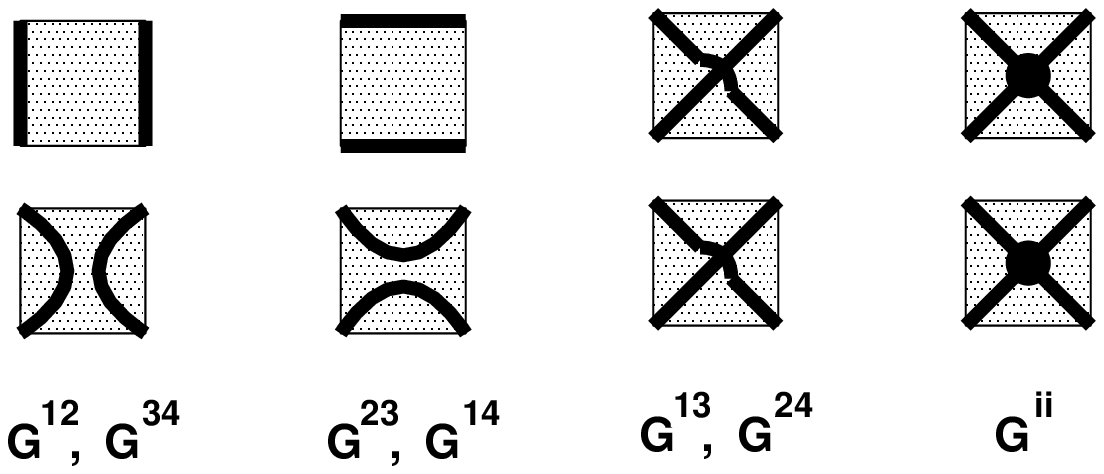,height=4cm}
    \end{center}
 \vskip-5mm
    \caption[*]{%
Graphical representations of plaquette breakups. 
Top row: worldline picture. Middle row: the same graphs in the
vertex picture.
A ``breakup'' specifies the direction in which the two loop segments that enter each plaquette will continue.
This direction can be vertical, horizontal, or diagonal.
Each graphical representation applies in general to two breakups ``$G^{ij}$'', 
as denoted below the pictures.
In the six-vertex-case, breakups $G^{i4}$ do not occur, thus
the three non-freezing breakups $G^{ij}$, $i\ne j$ 
are one-to-one equivalent to the three graphical representations.
Breakup ``$G^{ij}$'' is allowed 
(i.e.\ compatible with continuity of the arrow directions of the worldlines )
in plaquette configurations $S_p=i^\pm$ or $j^\pm$ (see \fig{figplaquettes}).
Namely, flipping the two spins on either one of the two lines in the graphical representation of
$G^{ij}$, $i\ne j$,
maps between configurations $i$ and $j$.
Breakup $G^{ii}$, called ``freezing'', forces all four spins to flip together,
thus mapping between $i^+$ and $i^-$.
For example, the diagonal breakup $G^{13}$ is compatible with plaquette configurations $1^\pm$
and $3^\pm$, which is most obvious in the vertex picture.
Starting from, e.g., plaquette $1^-$ and flipping two diagonally connected spins 
results in plaquettes $3^+$ or $3^-$.
} \label{figBreakups}
  \end{figure}

Since $[\hat{H}_{i,i+1},\hat{S}^z_{tot}]=0$, 
there are only the {\em six} nonvanishing matrix elements given in \eq{matrixelements},
namely those that conserve 
$\hat{S}^z_i + \hat{S}^z_{i+1}$, 
as shown pictorially in \fig{figplaquettes}.
Therefore, the locations of $S^z_{il} = +\Half$ in \fig{figcheckerboard}(a)
can be connected by continuous worldlines.
The worldlines close  in imaginary time-direction because of the trace in \eq{Zxxz}.
%

For models with fermions or hard core bosons one inserts occupation number eigenstates
instead of $S^z_{ij}$. Nearest neighbor hopping then again 
leads to the six-vertex case \cite{Scalettar-WL99} 
of \fig{figplaquettes}.
The term ``worldline'' derives from this case, since here they connect
sites occupied by particles.

We will find it useful to also {visualize} worldline configurations in 
a slightly different way, namely as configurations 
of a {\em vertex model} \cite{Baxter-book89}.
To do this, we perform a one-to-one mapping of each worldline configuration
to a vertex configuration.
We stay on the same lattice of shaded plaquettes.
We represent each spin $S^z_{il}$ by an arrow between the centers 
of the two shaded plaquettes
to which the site $(i,l)$ belongs. The arrow points upwards (downwards) 
in time for $S^z_{il}=+\Half$$(-\Half)$. 
The worldline-configuration in \fig{figcheckerboard}(a) is thus mapped to the vertex 
configuration of \fig{figcheckerboard}(b).
The one-to-one mapping of the worldline-plaquettes 
is shown in \fig{figplaquettes}.
The conservation of $\hat{S}^z_{tot}$ on each shaded plaquette
means in vertex language
that for each vertex (center of shaded plaquette)
two arrows point towards the vertex
and two arrows point away from it.
If one regards the arrows as a vector field, then this means a 
\beq{divzero}
\mbox{condition \  ``{\bf divergence = zero}'' \  for the arrows.}
\eeq
Note again that vertex language and worldline language refer to the same
configurations; they differ only in the pictures drawn.

We have now mapped the XXZ quantum spin chain 
to the  {\em six-vertex model} of statistical
mechanics \cite{Baxter-book89}, though with unusual boundary conditions,
since the vertex lattice here is tilted by 45 degrees with respect to that
of the standard six-vertex model.
Let us look more closely at the case of vanishing magnetic field, $h=0$.
This model has been exactly solved in (1+1) dimensions \cite{Baxter-book89}.
The exact phase diagram is shown in \fig{figphases}, 
in terms of the plaquette weights
given in \eq{matrixelements} and  in \fig{figplaquettes}.
%
\newfigh{figphases}{Figures/phasediagram.eps}{5cm}{
Exact phase diagram of the classical six-vertex model \cite{Baxter-book89} at $h=0$.
The weights $a,b,c$ are defined in \eq{matrixelements}.
Phase III is massless (infinite correlation length).
At $\frac{a}{c}+\frac{b}{c}=1$ there is a Kosterlitz-Thouless phase transition \cite{KosterlitzT73}
to the massive (finite correlation length)  phase IV,
and at $\frac{a}{c}-\frac{b}{c}=1$ there is a first order 
KDP phase transition \cite{Lieb67,LiebW72}
to the ferroelectric phase I.
The parameter regions of the Heisenberg model are denoted in brackets:
$J_z \ge |J_x|$ (AF), $J_z \le -|J_x|$ (FM), and $|J_z|<|J_x|$ (XY-like). 
The weights corresponding to the XY-model (or free fermions), i.e.\ $J_z=0$, are
located on the circle $a^2+b^2=c^2$.
When $\Delta\tau\rightarrow 0$, the point $(a/c=1,b/c=0)$ is approached in all cases.
}

It is interesting to note where the 
couplings of the Trotter-discretized XXZ-model at $h=0$
are located in this phase diagram (see \eq{matrixelements}):
For the Heisenberg antiferromagnet $J_z=|J_x| >0$  we have
$a+b=c$, i.e.\ we are {\em on} the Kosterlitz-Thouless line.
As $\dtau\TO 0$, we approach the point $a/c=1$, $b/c=0$.
For the Heisenberg ferromagnet $J_z=-|J_x| <0$  we have
$a-b=c$, i.e.\ we are {on} the KDP transition line,
approaching the same point $a/c=1$, $b/c=0$ as $\dtau\TO 0$.
When $|J_z|<|J_x|$, the same point is approached from inside the massless (XY-like) region.
When $|J_z|>|J_x|$, it is approached from below the respective transition line,
i.e.\ from the massive (Ising-like) phase IV when $J_x>0$ (AF) and from phase I when $J_x<0$ (FM).
Note that the local couplings $a,b,c$ do not change in higher dimensions (see section \ref{Dimensions}).

%
\underline{\em Anisotropic XYZ case:}~
For generality later on, let us briefly describe the anisotropic case
without magnetic field,
in which $J_x\ne J_y$ in the Hamiltonian
\beqa{Hxyz}
\hat{H}  & =  \sum_{\langle{\bf ij}\rangle}
         J_{x} \hat{S}^{x}_{{\bf i}}\hat{S}^{x}_{{\bf j}}  \,+\,
         J_{y} \hat{S}^{y}_{{\bf i}}\hat{S}^{y}_{{\bf j}}  \,+\,
         J_{z} \hat{S}^{z}_{{\bf i}}\hat{S}^{z}_{{\bf j}}  \;.
\eeqa
We also get this case if we quantize the XXZ-model along an axis different from
the $z$-axis.
The treatment is the same as for the XXZ-model.
Again we use $\hat{S}^z$ eigenstates to insert complete sets, and arrive at the
following nonvanishing matrix elements 
on the (1+1)-dimensional checkerboard lattice,
\newcommand{\sh}{\mbox{sh}}
\newcommand{\ch}{\mbox{ch}}
\beq{XYZmatrixelements}
 \begin{array}{lccccccccccc}
W(1^\pm)&\id&a&:=&\ME{++}{++}&\gl&\ME{--}{--}&\gl& e^{-\frac{\dtau}{4} J_z} & 
                                                           \ch{( \frac{\dtau}{4} (J_x-J_y))}\;, \1
W(2^\pm)&\id&c&:=&\ME{+-}{+-}&\gl&\ME{-+}{-+}&\gl& e^{+\frac{\dtau}{4} J_z} & 
                                                           \ch{( \frac{\dtau}{4} (J_x+J_y))}\;, \1
W(3^\pm)&\id&b&:=&\ME{+-}{-+}&\gl&\ME{-+}{+-}&\gl& e^{+\frac{\dtau}{4} J_z} & 
                                                           \sh{(-\frac{\dtau}{4} (J_x+J_y))}\;, \1
W(4^\pm)&\id&d&:=&\ME{++}{--}&\gl&\ME{--}{++}&\gl& e^{-\frac{\dtau}{4} J_z} & 
                                                           \sh{(-\frac{\dtau}{4} (J_x-J_y))}\;, 
 \end{array}
\eeq
which reduce 
\footnote{On bipartite lattices. See footnote \ref{bipartite}.}
to \eq{matrixelements} when $J_x=J_y$. 
We see that now there is an additional type of vertex with weight $d$,
shown as type $4^\pm$ in \fig{figplaquettes}, 
in which all four arrows point either towards or away from the center.
This vertex type may be thought of as a source (resp.\ sink) of arrows.
\Eq{divzero} becomes the 
\beq{divzero4}
 \mbox{condition \  ``{\bf divergence = zero mod 4}'' \  for the arrows.}
\eeq
The vertices and their weights now correspond to the eight-vertex model \cite{Baxter-book89}.
We will see that very little changes 
for the loop algorithm in this case \cite{EvertzM94}.

 \subsection{Outline of the Loop Algorithm}\label{outline}
%
The traditional way to perform Monte Carlo updates on a worldline configuration 
\cite{Scalettar-WL99,Suzuki-book94} 
consists of proposing {\em local deformations} of worldlines
and accepting/rejecting them with suitable probability.
In contrast, the updates for the loop algorithm  are very nonlocal.
We will first describe the basic idea 
for the example of the XXZ case
and outline the resulting procedure. 
We postpone the formal discussion and the calculation
of Monte Carlo probabilities to the next section.%
\footnote{{\em Notation: } 
From now on we will synonymously use ``plaquette'' or ``vertex''
to refer to the shaded plaquettes of the checkerboard lattice.
We also use interchangeably the terms ``spin direction'', ``arrow direction'', 
and ``occupation number''
to refer to the 2 possible states $S^z_{il}$ at each site $(il)$ of the checkerboard lattice.
We denote both probabilites and plaquettes by the letter $p$.
$S_p$ and $W_p$ are the spin configuration at plaquette $p$ and its weight,
and $G_p$ will be a breakup at $p$.
``Six-vertex-case'' (=``XXZ-case'') and ``eight-vertex-case'' 
will refer to the local plaquette constraints (i.e.\ nonzero weights), 
{\em not} necessarily to the respective models of statistical mechanics 
themselves.
}
An alternative derivation on an operator level is provided in section \ref{oper}.

Two observations lead to the loop algorithm:

(1) The Hamiltonian acts locally on individual plaquettes.
Thus the detailed balance condition on Monte Carlo probabilities can be 
satisfied locally on those plaquettes.

(2) The allowed configurations of arrows in the six-vertex-case have zero divergence, \eq{divzero}.
Therefore any two allowed configurations 
can be mapped into each other by changing the 
arrow-direction on a set of closed {\bf loops}, where 
{\em along each of these loops, the arrows have to point in constant direction}.
{These} are the loops that are constructed in our algorithm.
The path of the loops will be  determined locally on each plaquette (see below).

An  example is given in \fig{figLoopexample}.
Note that the loops are not worldlines; 
instead they consist of the locations of (proposed) {\em changes} 
in the worldline occupation number (=arrow directions = spin directions). 
Also, the loops are not prescribed, instead they will be determined stochastically,
with probabilities that depend on the current worldline configuration.
Since both the zero divergence condition and detailed balance
can be satisfied locally at the plaquettes,
we will be able to construct the loops by {\em local} stochastic decisions on the plaquettes,
yet achieve potentially very nonlocal worldline updates.

Let us construct a loop (see \fig{figLoopexample}).
This is most easily done using the vertex picture,
where the loop follows the arrows of the spin-configuration.
We start at some arbitrary site $(il)$ of the plaquette lattice and follow the 
arrow of the current spin configuration into the next plaquette.
There we have to specify the direction in which we will continue.
For any allowed spin configuration  (see \fig{figplaquettes})
there are two possibilities to continue along an arrow.
We choose one of these directions and follow the arrow into the next plaquette.
Then we continue to choose a direction at each of the plaquettes which we traverse.
If we reach a plaquette a second time, there is only one direction left to go,
since the loop should not overlap itself to avoid undoing previous changes%
\footnote{Removing this constraint and allowing the loop to move in any direction
          eventually leads to the ``directed loops''
          discussed in section \ref{DirectedLoops} !
          }.
Eventually we will (on a finite lattice) reach the starting point again,
thus closing the loop.
If we flip all the arrows (=spins) along the loop, we maintain the continuity of
arrows (worldlines) at each plaquette, and thus reach a new valid spin configuration.

We can now start to construct another loop
(not to overlap the first one), starting at some other arbitrary site
which has not been traversed by the first loop.
Note that by deciding the directions in which the first loop travelled,
we have at those plaquettes also already determined the direction
in which a second loop entering the same plaquettes will move,
namely along the two remaining arrows.
Thus, what we actually decide at each plaquette through which a loop travels,
is a {\bf ``breakup''} of the current arrow configuration into two disconnected parts,
denoting the paths that the two loop segments on this plaquette take.
The possible breakups of this kind are shown and described in \fig{figBreakups}. 
For each of the six possible arrow configurations $i^\pm$, \fig{figplaquettes},
there are two breakups which are compatible with the 
constraint that the arrows along the loop have to point in a constant direction;
namely those breakups labelled $G^{ij}, (i\ne j)$ in \fig{figBreakups}.

There is another kind of ``breakup'' that maintains  condition (\ref{divzero}).
Here all 4 spins on a plaquette are forced to be flipped together. 
We call this choice ``freezing'' (labeled $G^{ii}$ in \fig{figBreakups}),
since for a flip-symmetric model like the six-vertex model at $h=0$
it preserves the current weight $W(i^\pm)$. 
Freezing $G^{ii}$ can also be viewed as consisting of either one  of the two breakups $G^{ij}$ $(j\ne i)$,
with the two loop segments on this plaquette being ``glued'' together.
In this view freezing causes sets of loops to be glued together.
We shall call such a set (often a single loop) a ``{\bf cluster}''.
All loops in a cluster have to be flipped together.
(For an alternative, see section \ref{operfreeze}.)

Overall, we see that {\em by specifying a breakup for every shaded plaquette,
the whole vertex configuration is subdivided into a set of clusters which 
consist of closed loops.}
Each site of the checkerboard lattice is part of one such loop.
We shall call such a division of the vertex configuration into loops  a ``{\bf graph}'' $G$. 
Flipping the direction of arrows (= spin or occupancy)
on all sites of one or more clusters of a graph (a ``cluster flip'' which  
consists of ``loop flips''),
leads to a new allowed configuration.
In the loop algorithm the loops are constructed by specifying breakups
with suitable probabilities that depend on the current configuration
(see below).
In the vertex picture, the graph $G$ resides on the same arrows as the spins.
In the worldline picture, the elements of $G$ look slightly differently,
as seen in figures \ref{figLoopexample} and \ref{figBreakups} . 
Note that by introducing loops, we have effectively extended the space of variables, 
from spins, to spins and breakups. This point will be formalized in the next section.

%
\newfigh{figLoopexample}{Figures/loop.eps}{6.5cm}
{ Example of a loop update. 
(a) Worldline picture. The thick solid line denotes a single worldline,
the dot-dashed line depicts a possible loop.
By ``flipping the loop'', i.e.\ flipping the spin direction (arrow direction)
on all sites along the loop, the worldline will be deformed into the dashed shape.
Since the loop can potentially be very large, this deformation can be very nonlocal.
(b) The same situation as a vertex picture.
The loop is represented by the thick arrows.
By construction, in the vertex picture each loop follows arrows of the spin-configuration.
This means that the loop runs upwards in time-direction 
along sites with spin $S^z_{il}=+\Half$, 
i.e.\ along worldlines, 
and downwards in time-direction where there is no worldline.
}
In summary, the {\em basic procedure} for one Monte Carlo update then consists of 
two {\em stochastic} mappings:
First from spins to spins plus loops, and then to new spins.
I.e., starting with the current configuration of worldlines:
\begin{itemize}
\item[(1)]{\bf Select a breakup} (i.e.\ specify in which direction the loops will travel)
          for each shaded plaquette
          with a probability that depends on the current spin configuration at that 
          plaquette.
          These probabilities are discussed below.
          Identify the clusters which are implicitely constructed by these breakups.
          This involves a search through the lattice.
\item[(2)] {\bf Flip each cluster with suitable probability}, where
          ``flipping a cluster'' means to change the direction of 
          all arrows along the loops in this cluster
          (or, equivalently, changing spin direction or occupation number, respectively).
          The combined cluster flips  result in a new spin configuration.
          The flip probabilities depend in general on the Hamiltonian and 
          on the current spin configuration. 
          In the ideal case, for example the isotropic Heisenberg model in any dimension,
          each individual loop can be flipped independently with probability $1/2$.
\end{itemize}
An example is given in \fig{figLoopProcedure}. 
Notice that in this example the flip of a loop which happened
to wind around the lattice in spatial direction
led to a change in spatial winding number of the worldline configuration,
i.e.\ an update that {\em cannot} be done by local deformations of worldlines.
%
%
\newfigw{figLoopProcedure}{Figures/loopupdate.eps}{\textwidth}{
Example of a worldline update with the loop algorithm.
For clarity, we show a situation with only two worldlines.
We start with the worldline configuration $\S$ in the left picture.
The stochastic breakup decision on each plaquette, 
with probabilities depending on this worldline configuration, 
defines loops (in general clusters of loops), only two of which are drawn.
In this example we then flip both loops, i.e.\ flip the spin direction
(= worldline occupancy) along the loops.
This results in the new worldline configuration $\SS$ in the right picture,
which, as in this example, can be very different from the original one.
Since one of the loops happened to wind around the lattice in spatial direction,
its flip produced a worldline configuration with nonzero winding number.
Note that for the next worldline update, the current loop configuration is discarded,
and a completely new set of breakups will be determined with probabilities
depending on the new worldline configuration.
}

Little changes in the general XYZ-like (eight-vertex-like) case \cite{EvertzM94}.
The loops now have to change direction \cite{EvertzM94} at every breakup 
of type $(i,4)$. Alternatively, one can also omit assigning a direction to loops.

Let us now cast the general ideas into a valid procedure. 
Sections \ref{KDframework} to \ref{Loopsummary} are formal and comprehensive, with detailed explanations.
A summary is given in section \ref{Loopsummary},
and explicit weights for XXZ and XYZ cases in section \ref{ProbXXZ}.
A recipe for the most important (yet particularly simple) case, the Heisenberg antiferromagnet, 
is provided in section \ref{Recipe}.
See also sections \ref{ContTime} and \ref{LoopSSE}.
Previous formal expositions can be found in the original loop algorithm papers \cite{EvertzM92,EvertzM94}
(the best formal description there is that for the eight-vertex case in ref.\ \cite{EvertzM94}),
as well as, in a general setting and
in a more suitable language closer to the Fortuin-Kasteleyn mapping
of statistical mechanics, in the papers by Kawashima and Gubernatis \cite{KawashimaG95a,Kawashima95}.
We shall use both the worldline picture 
and the vertex picture of ref.\ \cite{EvertzM92,EvertzM94},
in order to provide a bridge between the existing formulations and to
make the simple geometry of the problem as obvious as possible.
In section \ref{PureLoop} we point out that for many models it is possible 
to sum over all spin variables to obtain a pure loop model.
We then cover the continuous time limit.
Finally, we introduce improved estimators, the single cluster version,
 and describe the performance of the loop algorithm.

 \subsection{Kandel-Domany framework}\label{KDframework}
A brief overview of the basics of Monte Carlo algorithms is given in 
Appendix \ref{MonteCarlo}.
The derivation \cite{EvertzM92,EvertzM94} of the loop algorithm 
is similar to that for the Swendsen Wang 
cluster algorithm in statistical mechanics \cite{SwendsenW87}
which uses the Fortuin-Kasteleyn  mapping of the Ising model to an extended phase space.
(For an excellent review see e.g.\ ref.\ \cite{Sokal92}).
A general formalism for such a mapping  was given 
by Kandel and Domany \cite{KandelD91}.
Here we use a more suitable language similar  to that of the general framework of 
Kawashima and Gubernatis \cite{KawashimaG95a},
who made the Fortuin-Kasteleyn-like nature of the mapping obvious.

For future use, we first write down the general scheme, without
yet making reference to individual spins, loops, or plaquettes.
We start with a set $\{\S\}$ of configurations $\S$ and a set $\{G\}$ of graphs $G$,
which together constitute the extended phase space. 
The partition function
\beq{Z}
  Z= \sum_\S W(\S)
\eeq
depends only on $\S$.
In addition we now {\em choose} 
a new weight function $W(\S,G)$ which must satisfy
\beqa{WSG}
  \sum_G   W(\S,G)  &=  &  W(\S) \,,\\
           W(\S,G)  &\ge&  0     \,.
\eeqa
Thus we have a Fortuin-Kasteleyn-like \cite{KasteleynF69,FortuinK72} representation:
\beq{FortKast}
Z = \sum_\S \sum_G \, W(\S,G)
\eeq

A Monte Carlo update now consists of 2 steps: 
\bit
\item[i)] 
          Given a configuration $\S$ (which implies $W(\S)\ne0$), 
          choose a graph $G$ with probability
          {\beq{PGS}
            p(\, \S \,\TO\, (\S,G)\,)  \;=\; \frac{W(\S,G)}{W(\S)} \;.
          \eeq} \  
\item[ii)]
          Given $\S$ and $G$ (this implies $W(\S,G)\ne0$), 
          choose a new configuration $(\SS,\GG)$ with a probability 
          $\psgSG$ that satisfies detailed balance with 
          respect to $W(\S,G)$:
          \beq{WSGdetbal}
             W(\S,G) \times \psgSG  \;=\; W(\SS,\GG) \times \pSGsg  \;,
          \eeq
          for example the heat-bath-like probability
          \beq{KDHeat}
            \psgSG \;=\; \frac{W(\SS,\GG)}{W(\S,G) + W(\SS,\GG) + \mbox{\em const}} \;.
          \eeq
\eit
Then the mapping $\S\TO\SS$ also satisfies detailed balance
with respect to the original weight $W(\S)$. Proof:
\newcommand{\sumGG}{\sum_{G,\GG}}
\beqa{ProofKD}
W(\S) \,\psS 
 &\gl&  W(\S)  &\;\sumGG\, p(\, \S \TO( \S  ,G)\,)       \;\; \psgSG                          \\
 &\gl&  W(\S)  &\;\sumGG\,    \frac{W( \S  ,G)} {W( \S) }\;\; \pSGsg\;\frac{W(\SS,\GG)}{W(\S,G)}\\
 &\gl&  W(\SS)&\;\sumGG\,    \frac{W(\SS,\GG)}{W(\SS)}\;\; \pSGsg                          \\
 &\gl&  W(\SS)&\;\sumGG\, p(\,\SS\TO(\SS,\GG)\,)      \;\; \pSGsg                          \\
 &\gl&  W(\SS)&\;\pSs \;.
\eeqa
(Within a Monte Carlo simulation the denominators in \eq{ProofKD} cannot vanish.)

 \subsection{Exact mapping of plaquette models}\label{applic}
%
We apply the Kandel-Domany  formalism to a model defined on plaquettes,
with
\beqa{WS}
W(\S)        &=&  A_{global}(\S) \,\times\, W^{plaq}(\S) \;, \\
             &=&  A_{global}(\S) \,\times\, \prod_p W_p(S_p)  \;.
\eeqa
To cover the general case, we have split off a global weight factor $A_{global}$.
This split is not unique.
\footnote{ 
For most of the subsequent discussion, we will implicitely assume
that the global weight can be factorized into independent contributions
from different clusters.
}
We devise an algorithm for $W^{plaq}(\S)\ident \prod_p W_p(S_p)$.
Because of its product structure, we can perform the decomposition into graphs
separately on every plaquette.
Thus in analogy with \eq{WSG} we look for a set of ``breakups'' $G_p$ and 
new weights $W_p(S_p,G_p)$ on every plaquette $p$
which satisfy
\beqa{WSGp}
  \sum_{G_p} W_p(S_p,G_p)  &=  &  W_p(S_p) \,,\\
             W_p(S_p,G_p)  &\ge&  0      \,.
\eeqa
This implies \eq{WSG} again, both for the plaquette part
\beq{WSG2}
 W^{plaq}(\S) = \prod_p \,\sum_{G_p}\; W_p(S_p,G_p) 
             = \sum_{\union_p G_p} \, \prod_p \,W_p(S_p,G_p) 
             \ident \sum_G \,W^{plaq}(\S,G) \;
\eeq
(where $G\ident\union_p G_p$, 
 $ W^{plaq}(\S,G) \ident \prod_p \, W_p(S_p,G_p)$)
and for the total weight $W(\S) = \sum_G   W(\S,G)$ with
\beqa{FKplaq}
 W(\S,G) &=& A_{global}(\S) \times W^{plaq}(\S,G) \\
        &=& A_{global}(\S)  \times \prod_p W_p(S_p,G_p) \;.
\eeqa
Thus we can apply the Kandel-Domany procedure.
Restricting ourselves to $\GG=G$, the two steps i), ii) 
in the previous section now become the procedure for the loop algorithm.
Starting with a configuration $\S$, a Monte Carlo update consists of:
\begin{itemize}
\item[(1)]  {Breakup:} For each plaquette, satisfy \eq{PGS} by choosing $G_p$ with probability 
            \beq{breakup}
              p( \, S_p \,\TO\, (S_p,G_p) \,) \;=\; \frac{W_p(S_p,G_p)}{W_p(S_p)} \;.
            \eeq
\item[(2)]  {Flip:} 
            Choose a new configuration $(\SS,G)$ with a probability 
            $\psgSg$ 
            that satisfies detailed balance with  respect to $W(\S,G)$.
\end{itemize}

In the next section we shall explicitely find a suitable 
set of breakups $G_p$ and plaquette weights $W_p(S_p,G_p)$.

\subsection{Structure of plaquette weight functions}\label{weights}
%
Given a graph $G$, we demand that $W^{plaq}(\S,G)$ does not change 
\beq{WSGinvariant}
  W^{plaq}(\S,G) = W^{plaq}(\SS,G)
\eeq
upon any spin update allowed by $W^{plaq}(\SS,G)\ne 0$.
Then it cancels in \eq{KDHeat}, which can now be written as
\beq{pflip}
  p_{flip}(\S\TO\SS) = \frac{A_{global}(\SS)}
                            {A_{global}(\S) + \sum_{\SS\ne\S,W(\SS,G)\ne 0}  A_{global}(\SS)} \;.
\eeq
The configurations $\SS$ for which $W(\SS,G)\ne 0$
will be those reached by cluster flips.
By enforcing \eq{WSGinvariant},
all cluster flips will  become independent of each other,
up to acceptance with $A_{global}$.
\Eq{WSGinvariant} is equivalent to
\beq{WSGdelta}
  W^{plaq}(\S,G) = \Delta(\S,G) \;V(G) \;,\;\; 
    \Delta(\S,G) := \left\{ \begin{array}{lll}
                              1,\; W^{plaq}(\S,G)\ne 0 ,\\
                              0,\;\mbox{otherwise}
                           \end{array} \right.  \;,
\eeq
which is the form introduced in ref.\ \cite{KawashimaG95a}.
Thus  
\beq{WSGdelta2}
 W(S,G) = \Delta(\S,G) \;V(G) \;A_{global}(\S).
\eeq
We shall achieve \eq{WSGinvariant} by enforcing it on every plaquette:
\beq{WSGpinvariant}
  W_p(S_p,G_p) = W_p(\SSp,G_p) \,.
\eeq
Then \eq{WSGdelta2} also holds on the plaquette level:
 $ W_p(S_p,G_p) = \Delta(\S_p,G_p) V(G_p) $.
The nontrivial part in this point of
view is that all allowed plaquette updates $S_p\TO\SSp$ 
match for different plaquettes, 
to give an overall allowed update $\S\TO\SS$.  As
we have seen in section \ref{outline}, it is the six- (or eight-) vertex
constraint, stemming from local conservation of $S^z$ (or $S^z \mbox{\ \em mod}\ 2)$ 
in the Hamiltonian,
that makes these plaquette updates match in the form of loops.
In other words, by enforcing \eq{WSGpinvariant}, we will achieve that
all clusters (sets of loops that are glued together at frozen plaquettes)
constructed during the breakup-step
can be flipped independently, up to acceptance with $p_{flip}$, \eq{pflip}.

Let us now find weights satisfying \eq{WSGpinvariant}.
Independent cluster flips require that \eq{WSGpinvariant} at least
include the case $\SSp=\boldbar{S}_p$, where all four spins at a plaquette are flipped,
$
 W_p(S_p,G_p) = W_p(\boldbar{S}_p,G_p)
$, 
which implies the requirement
\beq{Wsymm}
 W_p(S_p) \ident \sum_{G_p} W_p(S_p,G_p) = \sum_{G_p} W_p(\boldbar{S}_p,G_p) = W_p(\boldbar{S}_p)
\eeq
on the plaquette weights $W_p(S_p)$.
The first step in our construction is therefore to 
\beq{Aglobal}
 \mbox{\em Choose \ } A_{global} \mbox{\em \ \  such that \ \ }  W_p(S_p) = W_p(\boldbar{S}_p) \;.
\eeq
Such an $A_{global}$ always exists. It is not unique. 
The ideal case is $A_{global}=\mbox{\em const}$, since then for each cluster, $p_{flip}\ident\frac{1}{2}$ can be chosen.
See also section \ref{asymmetries}.

For worldline models, there are a total of eight allowed spin configurations 
$S_p = i^\pm = 1^\pm, 2^\pm, 3^\pm, 4^\pm$,
as shown in \fig{figplaquettes}.
With \eq{Wsymm}, the plaquette weight $W_p(S_p)$ depends only on  $i$.
Following ref.\ \cite{EvertzM94}, let us 
\beq{defbreakup}
\mbox{Define a different ``{\bf breakup}'' \ } 
     G_p := G^{ij} \ident G^{ji}
\mbox{\ for every transition \ } 
     i\leftrightarrow j  \;,
\eeq
such that the breakup $G^{ij}$ allows exactly the transitions $i\leftrightarrow j$.
Thus we define 
\beq{wij}
  W_p(S_p,G^{ij}) := \left\{ \begin{array}{llll}
                      w^{ij},\;& \mbox{if}\; S_p=i^\pm \;\mbox{or}\; S_p=j^\pm \;,\\
                      0        & \mbox{otherwise}
                           \end{array}\right. \;
\eeq
with suitable constants $w^{ij} \ident w^{ji}$.
We have satisfied \eq{WSGpinvariant} by construction.
By inspection of \fig{figplaquettes} 
we see that every transition $i\leftrightarrow j$, $i\ne j$,
corresponds to the flip of 2 spins on a plaquette
(all four spins for $i^+ \leftrightarrow i^-$).

We also see by inspection of \fig{figBreakups}
that, {\em given} the current worldline configuration $S_p=i^\pm$,
we can identify each of the 4 breakups $G^{ij}$, $j=1,2,3,4$,
with one of the {\em graphs} in \fig{figBreakups}.
Namely, flipping 2 of the spins connected in the graph for $G^{ij}$, $i\ne j$,
leads to one of the two plaquette configurations $j^\pm$, 
flipping the other two spins leads to the other  configuration,
flipping all four spins maps from $i^\pm$ to $i^\mp$.
For an example, see the figure caption.
Therefore, {\em given} a worldline configuration, the combined breakup $G=\union_p G_p$
can be represented%
\footnote{In general we should distinguish between the breakups $G^{ij}$,
of which there are 6 (10) in the six (eight) -vertex case,
and the fewer (4) graphical representation in \fig{figBreakups}.
Kawashima has shown that one can also give a common graphical representation 
of $G^{ij}$ for all $(ij)$ \cite{Kawashima95}.
This representation requires more than one loop-element per site.
}
as a graph consisting of the plaquette-elements in \fig{figBreakups}.
In many cases we can also transform the worldline model entirely into a loop graph model; see section \ref{PureLoop}.

Since $G_p$ connects pairs of sites, the breakups of all plaquettes
will combine to give a set of clusters consisting of loops,
as already described in section \ref{outline}.
When there is no freezing, i.e.\ no breakups $G^{ii}$ occur, then all clusters
consist of single loops.

We still need to find constants $w^{ij}\ident w^{ji}$, $i,j=1,2,3,4$,
such that the constraint \eq{WSGp} is satisfied,
which now reads 
\footnote{Eqs.\ (\ref{wij}),(\ref{wijconstraint}) 
          are eqs.\ (15),(16) in ref.\ \cite{EvertzM94}.}
\beqa{wijconstraint}
  \sum_{j} w^{ij}  &=  &  W(i)   \,,\1
           w^{ij}  &\ge&  0      \,,
\eeqa
(with $W(4)=0$ in the six-vertex-case).
This constraint underdetermines the $w^{ij}$.
There are 3 equations for 6 unknowns in the six-vertex case, 
and 4 equations for 10 unknowns in the eight-vertex case.
It can always be solved. One explicit solution is the following:
Let $W(k)$ be the smallest 
of the $n$ weights $W(j)$, $j=1,..,n$ ($n$ is $3$ or $4$).
\Eq{wijconstraint} is satisfied by
\beqa{wijsolution}
  w^{ij} = W(k)/n                      &\;\mbox{for} \; i\ne j \;, \\
  w^{ii} = W(i) - \sum_{j\ne i} w^{ij} &\;\mbox{for} \; i=1,..,n \;.
\eeqa

Experience tells us that for an efficient algorithm,
one should keep the loops as independent as possible.
Thus we should minimize the weights $w^{ii}$ which cause
loops to be glued together. 
Let $W(l)$ be the largest of the $n$ weights $W(j)$.
Given a solution $w^{ij}$ we can always find another one
in which no diagonal element $w^{ii}$ except at most $w^{ll}$ 
is nonzero \cite{Kawashima95}.
For example, to remove $w^{jj}, j>1$, define
\beqa{removefreezing}
  w^{\prime\,j,j}     &                    &=  0 \\
  w^{\prime\,j,j-1}   &= w^{\prime\,j-1,j} &=  w^{j-1,j}   &+& w^{jj} \\
  w^{\prime\,j-1,j-1} &                    &=  w^{j-1,j-1} &-& w^{jj} \;.
\eeqa
Iterating this transformation leads to the one surviving diagonal
element
\beq{survive}
  w^{\prime\, ll} = w^{ll} - \sum_{i\ne k} w^{ii} \;.
\eeq
More explicit solutions are given in section \ref{ProbXXZ}.

 \subsection{Summary of the loop algorithm}\label{Loopsummary}

Since the detailed derivation of the general formalism was a bit tedious,
we summarize the actual procedure here.
Start with a model in worldline representation with
  $Z= \sum_\S W(\S)$, \eq{Z}.
\begin{itemize}
\item[(1)] Choose a split 
           $W(\S) =  A_{global}(\S) \,\times\, \prod_p W_p(S_p)$, \eq{WS},
           such that 
           $W_p(S_p) = W_p(\boldbar{S}_p)$, \eq{Aglobal}.
\item[(2)] Find new weights $w^{ij}=w^{ji}\ge 0$ such that
           $\sum_{j} w^{ij} =  W(j)$, \eq{wijconstraint},
           while preferrably minimizing the ``freezing'' weights $w^{ii}$,
           see \eq{removefreezing}. (See also section \ref{operfreeze}).
\end{itemize}
Each Monte Carlo update from a worldline configuration  to a new one 
then involves the following steps:
\begin{itemize}
\item[(i)] (Breakup) For each shaded plaquettes, with current spin configuration $i^\pm$,
          choose a breakup $G^{ij}$ with probability 
          $ p = w^{ij} / W(i) $, proportional to the graph weight $w^{ij}$.
          (See eqs.\ (\ref{breakup},\ref{wij}) and section \ref{ProbXXZ}).
\item[(ii)] (Cluster identification) All plaquette breakups together 
          subdivide the vertex lattice into a set of clusters,
          which consist of closed loops. Loops which have a 
          frozen vertex (``$G^{ii}$'') in common
          belong  to the same cluster.    Identify which sites belong to which clusters. 
          (This is in general the most time consuming task).
\item[(iii)] (Flip) 
          Flip each cluster separately%
       \footnote{
           Alternatively one can perform a combined flip of a randomly chosen subset of clusters.
           However, when $A_{global}$ is not unity, this will in general produce bigger
           variations of $A_{global}$ and therefore lower acceptance rates.
                }, 
          one after the other, with (e.g.) heat-bath probability for $A_{global}$.
          In case $A_{global} \equiv 1$, this means that one can flip each cluster independently with probability \half.
          ``Flipping'' means to change the sign of $S^z_{il}$ on all sites in the cluster.
          If desired, one can artificially restrict the simulation to some sector
          of phase space, e.g.\ to the {\bf ``canonical ensemble''} of constant magnetization, by prohibiting
          updates that leave this sector,
          or one can select such sectors a posterior \cite{RoosM99,NishinoORYM99}. (See also section \ref{asymmetries}).
\end{itemize}

 \subsection{Graph weights for the XXZ, XYZ, and Heisenberg model}\label{ProbXXZ}

We now come back to our example
and compute \cite{EvertzM92,EvertzM94} one solution for
the weights $w^{ij}\ident w^{ji}$, and thus the  breakup and flip probabilities,
for the spin-flip symmetric six-vertex case, 
with weights $a$,$b$,$c$, \eq{matrixelements}.
This includes the Heisenberg model and the XXZ-model at $h=0$ (\eq{matrixelements})
in any dimension (see section \ref{Dimensions}).
Some solutions for the general XYZ model are also given.
We need to find a solution to \eq{wijconstraint}.
Here it reads%
\footnote{We have multiplied the weights in \eq{matrixelements} by $\exp{(-\Delta\tau J_z/4)}$
          and also provided the  expansion to order $\Delta\tau$ for later use 
          in the continuous time version.}
\beqa{xxzbreak}
w^{11} + w^{12} + w^{13}  & = W(1) &\ident& a &\simeq &1 - \frac{\Delta\tau}{2} \, J_z   \\
w^{22} + w^{12} + w^{23}  & = W(2) &\ident& c &\simeq &1                                 \\
w^{33} + w^{13} + w^{23}  & = W(3) &\ident& b &\simeq &    \frac{\Delta\tau}{2} \, |J_x| \;.
\eeqa
From \fig{figBreakups}
we see that $w^{12}$, $w^{23}$, and $w^{13}$ correspond to 
vertical, horizontal, and diagonal breakups, respectively. 
The weights $w^{ii}$ correspond to transitions
$i^\pm \TO i^\pm$, i.e.\ to flipping zero or four spins on a plaquette.
They freeze the value of the weight $W(i)$.
Experience tells us that we should minimize
freezing in order to get an efficient algorithm,
in which then loops are as independent as possible.
We will construct solutions with minimal freezing; others exist.
See also sections \ref{Performance} and  \ref{operfreeze}.

\Eq{xxzbreak} has different types of solutions in different regions
of the parameter space $(a,b,c)$.
Remarkably, these regions are exactly the same \cite{EvertzM92,EvertzM94}
as the phases of the two-dimensional classical six-vertex model \cite{Lieb67,LiebW72,Baxter-book89},
shown in \fig{figphases}.
The regions of \fig{figphases} have been spelled out in terms of the coupling constants $J_x$,$J_z$
at the end of section \ref{Setup}.

\pagebreak[3]
\EM{Region~IV (AF)}, has antiferromagnetic couplings $J_z \ge |J_x|$, thus $c \ge a+b$.
To minimize the freezing of weight $c$, we have to minimize $w^{22}$.
From \eq{xxzbreak}, $w^{22} = c-a-b + w^{11} + w^{33} + 2w^{13}$.
With $w^{ij} \geq 0$ this implies $w^{22,\scr{min}} = c-a-b$.
This minimal value of $w^{22}$ is achieved for
$w^{11}=w^{33}=0$, i.e.\ when we minimize all freezing.
The optimized nonzero parameters for region IV are then: 
\beqa{qFOUR}
w^{12} &=& a        &\simeq &1 - \frac{\Delta\tau}{2} \, J_z          & \mbox{(vertical breakup)},   \\
w^{23} &=& b        &\simeq &    \frac{\Delta\tau}{2} \, |J_x|        & \mbox{(horizontal breakup)}, \\
w^{22} &=& c-a-b    &\simeq &    \frac{\Delta\tau}{2} \, (J_z-|J_x|)  & \mbox{(freezing of opposite spins)},  \;
\eeqa
without any diagonal breakups.
This has to be modified for non-bipartite lattices; see section \ref{ergodicity}.
For an alternative to freezing, see section \ref{operfreeze}.

\EM{In region I (FM)} with ferromagnetic couplings $J_z \le -|J_x|$ and $a>b+c$ 
we get 
\beqa{qONE}
w^{12} &=& c        &\simeq &1                                        & \mbox{(vertical breakup)},\\
w^{13} &=& b        &\simeq &    \frac{\Delta\tau}{2} \,  |J_x|       & \mbox{(diagonal breakup)},\\
w^{11} &=& a-c-b    &\simeq &    \frac{\Delta\tau}{2} \, (|J_z|-|J_x|)& \mbox{(freezing of equal spins)}, \;
\eeqa
without any horizontal breakups.
( This is similar for region II, $b>a+c$, which does not correspond to a quantum model. 
There we obtain minimal freezing
from \eq{qFOUR} with indices $2$ and $3$ interchanged, and no vertical breakups.)
 
\EM{Region III (XY-like)} has $|J_z|<|J_x|$, and $a,b,c$ $\le \frac{1}{2} (a+b+c)$.
Here we  can set all freezing probabilities to zero, obtaining
\beqa{qTHREE}
2w^{12} &=& a+c-b    &\simeq & 2 - \frac{\Delta\tau}{2} \, (|J_x| +J_z)      & \mbox{(vertical   breakup)}, \\
2w^{23} &=& c+b-a    &\simeq &     \frac{\Delta\tau}{2} \, (|J_x| +J_z)      & \mbox{(horizontal breakup)}, \\
2w^{13} &=& b+a-c    &\simeq &     \frac{\Delta\tau}{2} \, (|J_x| -J_z)      & \mbox{(diagonal   breakup)}. \;
\eeqa

\EM{The isotropic Heisenberg model} is located on the boundaries of region III.
The antiferromagnet $J_z=|J_x|$ has   $c-a-b=0$, thus    only vertical ($w^{12}$) and horizontal ($w^{23}$) breakups.
The     ferromagnet $J_z=-|J_x|$ obeys $c-a+b=0$ and has only vertical ($w^{12}$) and diagonal ($w^{23}$) breakups.
There is no freezing for the isotropic model.

\EM{The classical Ising model} is reached in the limit $J_x/J_z = 0$,
since then $b= 0$, so that there is no more
hopping and all worldlines are straight.
Remarkably, in this limit the loop algorithm becomes \cite{KawashimaG95b} 
the Swendsen-Wang cluster algorithm \cite{SwendsenW87} !
Frozen plaquettes connect the sites of clusters in the Ising model,
i.e.\ they correspond to the ``freezing'' operation \cite{KandelD91}
of the efficient Swendsen-Wang method.

\EM{The classical BCSOS model} is simulated for $a=b$ \cite{EvertzLM93}. 
When $a=b=\frac{1}{2}c$, the loop algorithm constructs \cite{EvertzLM93} 
the {\em boundaries} of the clusters which the VMR-cluster algorithm \cite{EvertzHMPS91,EvertzHMPS92,HasenbuschLMP92,HasenbuschMP92,HasenbuschMP94}
for the (1+1)-dimensional BCSOS model produces,
i.e.\ it constructs these clusters more efficiently. 
The loop representation was used in ref.\ \cite{KondevH96} to obtain exact analytical results 
for this model,
and in ref.\ \cite{HasenbuschP97,HasenbuschMP96} to study the roughening transition of
the BCSOS model.%

\EM{General XYZ case:}
(See also ref.\ \cite{Kawashima95} for explicit solutions.)
The loop algorithm remains unchanged in the XYZ case (see section \ref{outline}), 
except that at breakups $G^{i4}$, the arrows flip direction.
In each of the four ordered regions of the XYZ model we have
 $ W(m) \ge \sum_{i\ne m} W(i) $ for one $m\in \{1,2,3,4\}$.
The nonzero breakup weights with minimum freezing are then
\beqa{xyza}
  w^{mm} &=& W(m) - \sum_{i\ne m} W(i) \\
  w^{im} &=& W(i) \;\;(i\ne m).
\eeqa
This also summarizes the solutions for regions I,II,IV above.
In the disordered region $2 W(m) < \sum_i W(i)$
we can set all freezing $w^{ii}$ to zero, 
and in general still have 6 free parameters $w^{ij}$  with only 4 constraints \eq{wijconstraint}.

 \subsection{Recipe for the spin $\Half$ Heisenberg antiferromagnet} \label{Recipe}
%
In order to make the loop algorithm as clear as possible,
we restate the procedure for the important yet simple case of the isotropic
spin~$\Half$ Heisenberg antiferromagnet.
See also section \ref{ContTime}, for the continuous time version,
and section \ref{LoopSSE} for the SSE variant. 

A Monte Carlo update leads from a worldline configuration $\S$ of spin variables $S^z_{il}=\pm 1$
to a new configuration $\S^\prime$.
On each shaded plaquette $p$, the local spin configuration $S_p$ 
takes one of the six possibilities shown in the left part of \fig{figplaquettes},
with weights $W_p(S_p)$ given in \eq{matrixelements}, 
satisfying $a+b=c$ in the isotropic antiferromagnetic case.
The weights $w^{ij}$ in \eq{qFOUR} are  all zero except for
$  w^{12}=a \,, w^{23}=b $, so that we get only vertical ($G^{12}$) and horizontal ($G^{23}$) breakups.
The update consists of the following steps: 
\begin{itemize}
\item[(i)] For each shaded plaquette, choose the horizontal breakup
         with probability 
                  \beqa{GAF}
           p(S_p,G^{23}) &=& (\delta_{S_p,2^\pm} + \delta_{S_p,3^\pm}) 
                \;\;{\frac{w^{23}}{W_p(S_p)}} &=& \left\{
             \begin{array}{lllll}
                0,                           &\; S_p = 1^\pm \;,\\
                \tanh(\frac{\Delta\tau}{2}J) &\; S_p = 2^\pm \;,\\
                1,                           &\; S_p = 3^\pm \;,
             \end{array}
                \right.
         \eeqa
         (see eqs.\ (\ref{matrixelements},\ref{breakup},\ref{wij},\ref{qFOUR})), 
         otherwise choose the vertical breakup.
\item[(ii)] Identify the clusters constructed in step (i). Since there is no freezing here,
          all clusters consist of single loops.
\item[(iii)] Flip each loop with probability \half, where flipping means to change the sign
           of $S^z_{il}$ on all sites along the loop.
           This gives the new configuration $\S^\prime$.
\end{itemize}

This procedure is even  simpler than local worldline updates.
Moreover, it remains completely unchanged in arbitrary dimensions (see section \ref{Dimensions}).
The single cluster version is described in section \ref{SingleLoop}.
Note that one can and should avoid the Trotter approximation altogether, by working directly 
in continuous time, for which a modification of this recipe will be given in section \ref{ContTime}.
or by using the stochastic series expansion, described in section \ref{LoopSSE}.

 \subsection{Ergodicity}\label{ergodicity} 
To establish correctness of the loop algorithm, we still have to show ergodicity for the overall algorithm,
including the existence of global configuration changes,
Ergodicity is obvious when all $w^{ij}>0$ for $i\ne j$,
and when $p_{flip}$ is always nonzero (which is normally the case when we use \eq{pflip} for $p_{flip}$).
Any two allowed configurations (i.e.\ $W(\S)\ne 0$) are, as always, mapped into
each other by a unique set of spin-flips (loop-flips), which are compatible with a 
set of breakups $G^{ij}$, $i\ne j$.
With $w^{ij}>0$,  this set of breakups has a finite
probability to occur, and with $p_{flip}>0$, 
the two configurations will be 
mapped into each other in a single Monte Carlo step with finite probability.
Note that the trivially ergodic case $w^{ij}>0$ can always be constructed, 
as seen in \eq{wijsolution};
this may not be an efficient algorithm, however.
On the other hand,  
one can always construct weights $w^{ij}$ such that (for $J_{xy}\ne 0$)
ergodicity is not achieved, for example by choosing  
$w^{ij}=\delta_{ij} W(i)$, i.e.\ only freezing.

When some of the $w^{ij}$ vanish, ergodicity has to be shown case by case.
With the weight choices of section \ref{ProbXXZ},
region III is trivially ergodic.
We  have to show ergodicity explicitely 
in each of the regions I, II, IV, because some $w^{ij}$ vanish there.

Region I (including the Heisenberg FM): $w^{23}=0$, i.e.\ there are only vertical and
diagonal breakups (see \fig{figBreakups}). 
These breakups permit a loop configuration which is identical to any 
given worldline configuration. That loop configuration will occur with finite probability.
Flipping all loops in this configuration leads to the empty worldline configuration.
Conversely,  any  worldline configuration can be generated from the empty one
in a single (!) update by such a choice of loops. Therefore the algorithm is ergodic,
mapping any two worldline configurations into each other in only two steps.

Region IV (including the Heisenberg AF): $w^{13}=0$, i.e.\ there are only vertical and
horizontal breakups.
On a {\em bipartite} lattice with open or periodic spatial boundary conditions, 
ergodicity can be shown easily. %
Start with any worldline configuration $\S=\{S_{xl}\}$.
Our {\em reference configuration} this time is not the ``empty'' configuration 
$S^\prime_{xl}=-\Half$,
but instead the staggered configuration 
$S^\prime_{xl} = (-1)^x (-\Half)$, i.e.\ the configuration with straight worldlines on one of
the two sublattices.
As always, there is a unique set of loops whose flips will map $\S$ into $\SS$.
By inspection we see that these loops contain only vertical and horizontal breakups
(horizontal where $\SS$ has diagonal worldline parts, vertical elsewhere).
Since these breakups have finite probability to occur, the whole set of loops will
be constructed with finite probability.
Thus, again, any worldline configuration will be mapped to the reference configuration
with finite probability, and vice versa, so that on a bipartite lattice the algorithm is ergodic.
Furthermore, on any lattice, the loop algorithm is at least as ergodic as the 
algorithm with conventional local updates. The latter consist of spin-changes
around non-shaded plaquettes, equivalent to the flip of a small loop with
two vertical and two horizontal breakups, which will occur with finite probability
in the loop algorithm.

Note that for a frustrated antiferromagnet, i.e.\  on a non-bipartite lattice,
the algorithm \eq{qFOUR} with only horizontal breakups is not ergodic \cite{AmmonEKTF98}:
Loops switch time-direction at every breakup, thus a loop with an odd number of spatial
hops is not possible. To ensure ergodicity, one has to include diagonal breakups with some 
weight $0<w^{13}<b$. Then ergodicity is trivial, since all $w^{ij}>0$ for $i\ne j$.
\Eq{xxzbreak} is now solved with $w^{23}=b-w^{13}$ and demands freezing $w^{22}=2 w^{13}$ of equal spins. 
The size of $w^{13}$ has to be chosen for optimal performance of the algorithm,
which, however, is subject to a severe sign problem.

For completeness, we mention region II (which does not occur in worldline models).
Here there is no vertical breakup. In case of periodic spatial boundary conditions,
interchange of ``space'' and ``time'' leads us to the situation of region I, for
which we have already shown ergodicity.

 \subsection{Transformation to a pure loop model}\label{PureLoop} 
%
Remarkably, by using the exact mapping eqs.\ (\ref{WSG},\ref{WSGp}) on which the loop algorithm is based,
we can transform quantum spin and particle models  into pure loop representations,
i.e.\ into a completely different setting than the original worldlines.
This is analoguous to the Fortuin-Kasteleyn representation \cite{KasteleynF69,FortuinK72} of the Potts model.
It was first achieved, independently, by Nachtergaele and Aizenman \cite{AizenmanN94,Nachtergaele93}
for the one-dimensional Heisenberg model, and was used to 
prove rigorous correlation inequalities \cite{AizenmanN94}.
Kondev and Henley used it to compute the exact stiffness and critical exponents
of a twodimensional vertex model \cite{KondevH96,Kondev97}.
See also section \ref{oper}.

We get to a loop model 
by explicitely summing over the spin degrees of freedom in \eq{FortKast}.
Using eqs.\ (\ref{FortKast},\ref{FKplaq},\ref{WSGdelta}) we see that
\beq{Z2}
 Z =\;\; \sum_{\{\S\}} W(\S) \;\;=\;\; \sum_{\{\S\}} \prod_p \sum_{G_p}\; 
                             \Delta(S_p,G_p) \; V(G_p) \; A_{global}(\S)\;.
\eeq
The condition $\Delta(S_p,G_p) \ne 0
$ restricts the graph $G$ to consist of clusters,
i.e.\ divergence-free components compatible with the spin configuration.

The summation over spin configurations $\S$ can easily be done 
if a {\em reference spin configuration $\S_0$} exists (see also section \ref{ergodicity}),
in which all plaquette breakups $G_p$ with nonvanishing weight $V(G_p)$ 
are allowed (i.e. $W_p(S_{p0},G_p)>0$ whenever $V(G_p)>0$). 
Then {\em all} graphs with finite weight can be constructed from $S_0$.
By design, cluster flips do not change $W(S,G)$ when $A_{global}(S)$ is constant.
Each cluster can then be flipped independently and contributes a factor $2$.
With the weight choices from section \ref{ProbXXZ}, we see that for the AF region IV, 
the antiferromagnetically staggered configuration is such a reference configuration on bipartite lattices:
it allows all relevant breakups (vertical, horizontal, freezing of unequal spins) on any plaquette.
For region I (FM), the ferromagnetic spin configuration serves the same purpose on any lattice.

In these regions (as well as in region II), we can then sum over spin configurations in \eq{Z2}.
Without external weight $A_{global}$, we get
\beq{ZGraph}
 Z =\; \sum_{G=\union_p G^{ij}} \left(\prod_p w^{ij}\right)  \, 2^{N_c(G)} \;\;\ident\;\; \sum_{G} W(G) \;,
\eeq
where $N_{c}(G)$ is the number of clusters in $G$.
When $A_{global}$ is a product over contributions from each cluster (e.g.\ in case of a magnetic field),
the factor $2$ for each cluster is replaced by  $(A_{global}(\S(G)) + A_{global}(\SS(G)))$.

When there is {\em no} reference configuration, 
e.g.\ for region III (XY-like) or for antiferromagnets on non-bipartite lattices, 
or for different choices of breakup-weights,
we can still obtain a pure loop model.
Now there can be clusters which do not correspond to a continuous worldline configuration 
(i.e. the spin directions mandated by independently chosen breakups
 at different plaquettes within a cluster may contradict each other).
To remove graphs containing such clusters, we temporarily endow each loop with a direction, 
and introduce a constraint into the sum over graphs in \eq{ZGraph} enforcing compatibility 
of the loop directions of each cluster.

The mapping generalizes immediately to the anisotropic XYZ-like case, where the number 
of a priori breakup-possibilities per plaquette doubles, though they are graphically still the same
as in the XXZ-like case (see \fig{figBreakups}).

We have thus mapped all XYZ-like quantum spin and particle models, 
with any choice of breakup weights and in any dimension,
to a {\em loop} model,
in complete analogy with the Fortuin-Kasteleyn mapping \cite{KasteleynF69,FortuinK72}
of statistical mechanics.
This  mapping  is useful for analytical purposes (see above).
Note that for the  Heisenberg antiferromagnet, the loop model consists of antiferromagnetic 
selfavoiding polygons,
and  for the Heisenberg ferromagnet it has a similar graphical
representations as the worldlines themselves.
Note also that for a given physical model there are many different loop models,
corresponding to the different possible choices of breakup-weights.
Remarkably, the graph-decomposition \eq{WSGp}, 
and thus the transformation to a  loop model,
can also be written on an operator level
\cite{BrowerCW98} (see section \ref{Offdiagonal}). 
All observables can be measured in the loop representation \cite{BrowerCW98},
as correlation functions (``improved estimators'', see sections \ref{ImprEst}, \ref{Offdiagonal})
or as thermodynamic derivatives.

One can  perform  
{\em Monte Carlo simulations purely in the loop representation},
analoguously to Sweeny's method for the cluster representation of the Ising model \cite{Sweeny83}.
Indeed, closer inspection reveals that 
the Handscomb method for the ferromagnet is equivalent to a Monte Carlo in loop
representation with stochastic series expansion \cite{SouzaCL00,Handscomb62,Handscomb64,Lyklema82,LeeJN84,ChakravartyS82} !
For other models this has not yet been tried (but the method of section \ref{Merons} comes close).
In more than one spatial dimension, it is computationally more difficult than the loop algorithm
with graphs and spins, since one has to keep track of the number of clusters, which
can require traversing complete clusters for each local update,
unless it can be simplified by a binary tree search.

 \subsection{Single-Cluster Variant} \label{SingleLoop}
As in Swendsen-Wang Cluster updates, there are several ways to perform an update
of the worldline configuration $\S$ with the required detailed balance with respect
to $W(\S,G)$, \eq{WSGdetbal}.
There are two important approaches:
\begin{itemize}
\item [(i)] {\em Multi-Cluster Variant:} 
    Determine the whole graph (set of loops) $G$ and flip each cluster 
    (set of mutually glued loops)  in $G$
    with suitable probability (see section \ref{Loopsummary})
\item [(ii)] {\em Single-Cluster Variant} (Wolff-cluster) \cite{Wolff89a,EvertzLM93}
    Pick a site $(i_0,l_0)$ at random, and construct only the cluster which includes
    that site. This can be done iteratively, by following the course of 
    the loop through $(i_0,l_0)$
    until it closes, while determining the breakups (and thus the route of the loop(s))
    only on the plaquettes which are traversed.
    This corresponds to our initial loop-description in section \ref{outline}.
    At each plaquette at which a frozen breakup $G^{ii}$ is chosen, the current loop
    is glued to the other loop traversing this plaquette. That other loop 
    (and any loops glued to it) then also has to be
    constructed completely.
    Flip the complete cluster with probability $p_{flip}$ satisfying detailed balance
    with respect to  $A_{global}$   to get to a new spin configuration.
    Note that when $A_{global}=\mbox{\em const}$, we can choose $p_{flip}=1$ instead of \half.
\end{itemize}
Both approaches satisfy detailed balance in \eq{WSGdetbal}.
One may think of the single-cluster variant as if  all clusters had actually been 
constructed first, and then one of them chosen at random, by picking a site,
to make an update proposal.

The advantage of the single-cluster variant \cite{Wolff89a,Wolff89c,Sokal92}
is that by picking a random site, one is likely to pick a {\em large} cluster,
whose flip will produce a big change in the configuration and thus a large step
in phase space. This can reduce critical slowing down still further.
The effort (computer time) to construct the single-cluster is proportional to its length.
Normalized to constant effort, one finds indeed that
the single-cluster variant (and the corresponding ``Wolff-algorithm''
for Swendsen-Wang-like algorithms)
usually have even smaller dynamical critical exponents (see appendix \ref{CSD})
than the multi-cluster variant.
Note that improved estimators get a different normalization in the single-cluster variant
(see \eq{chimult} and below).
In some circumstances, the multi-cluster variant can still be advantageous overall, 
for example when employing parallel \cite{Mino91,Todo02a,Todo02b} or
vectorized \cite{Evertz93}  computers. It is also essential for the meron method (section \ref{Merons})
and e.g. four-point functions as improved estimators (section \ref{Offdiagonal}),
and it is easier to implement in continuous time.

The single-cluster method, on the other hand, generalizes into the worm and directed-loop
methods discussed in section \ref{Related}, which are applicable to any discrete model.

 \subsection{Arbitrary spatial dimension}\label{Dimensions}

There is vitually no change algorithmically in going to higher dimensions
\cite{EvertzLM93},
if one chooses to stay on a vertex-lattice.
Let us look at two spatial dimensions as an example.
The even/odd split of the Hamiltonian in \eq{split}
can be generalized to
\beq{splitgeneral}
 \hat{H} = \sum_\nu \hat{H}_\nu = \sum_\nu \sum_i \hat{H}_{i,i+\hat{\nu}}
\eeq
with a separate $\hat{H}_\nu$ for each direction of hopping 
(resp.\ spin coupling) in the Hamiltonian $\hat{H}$.
For a two-dimensional square lattice with nearest neighbor hopping
we thus get 4 parts $\hat{H}_\nu$, each the sum of commuting pieces
living on single bonds, in complete analogy with the one-dimensional case.

After the Trotter-Suzuki breakup, \eq{Zxxz}, these single bonds again develop into
shaded plaquettes. Each Trotter timeslice now has 4 subslices.
Locally on each shaded plaquette we have the {\em identical} situation
as in (1+1) dimension.
Thus the loop algorithm can be applied {\em unchanged}.
The  only thing that changes is the way that different plaquettes connect.
(Thus it is easy to write a loop-cluster program for general dimension.
This contrasts with the traditional local worldline updates, where a number of different
rather complicated updates are necessary \cite{MakivicD91} 
to achieve acceptable performance).
The same construction can be applied as long as the lattice and the Hamiltonian
admit a worldline representation in which commuting pieces of the Hamiltonian live on bonds.

 \subsection{Continuous time}\label{ContTime}
As one of the most important developments,
Beard and Wiese \cite{BeardW96} have shown that within the loop formulation,
one can directly take the time continuum limit $\dtau\TO 0$ in the Trotter-Suzuki 
decomposition, \eq{Zxxz}.
In fact, it turns out that one can also write the original spin model 
in operator language, directly in continuous time  
when $H$ has a countable basis (see section \ref{oper}) \cite{FarhiG92,AizenmanL90,AizenmanN94}.

In continuous time it is appropriate to describe worldlines 
by specifying the times $t$ at which a worldline jumps to a different site. 
This jump is now instantaneous.
%
\newfigh{figcont}{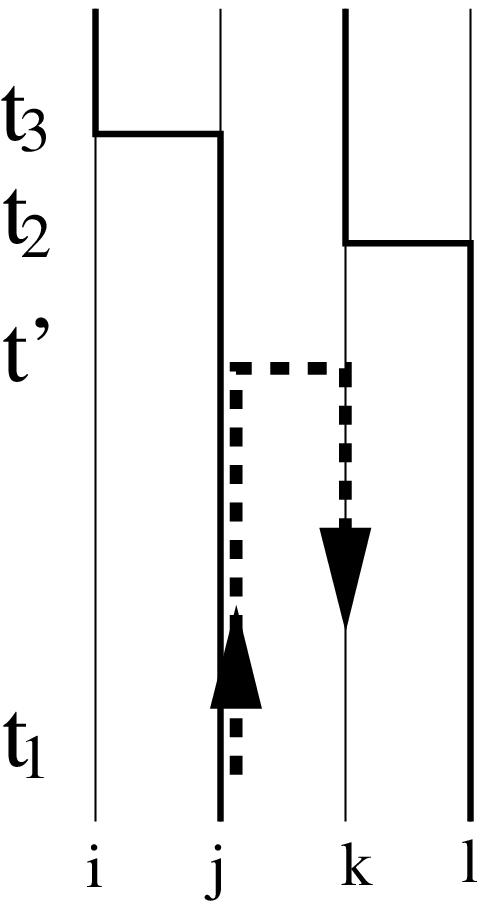}{5cm}{
Loop construction in continuous time.
Thick solid lines denote worldlines; the thick dashed line is a partial loop.
}
%

Let us look again at the simple case of a spin\ \half\ antiferromagnet.
Figure \ref{figcont} shows part of a worldline configuration.
In discrete time, this picture would be subdivided into plaquettes
of temporal extent $\Delta\tau$,
like \fig{figcheckerboard}.
On each such plaquette, the probability of a horizontal breakup is given by \eq{GAF}
                  \beqa{GAFa}
           p(\mbox{horiz.\ breakup})
                &=& \left\{
             \begin{array}{lllll}
                0,                           &\; \mbox{0 or 2 worldlines on plaquette}\;,\\
                \tanh(\frac{\Delta\tau}{2}J) &\; \mbox{1 straight worldline} \;,\\
                1,                           &\; \mbox{worldline jumps } \;,
             \end{array}
                \right.
         \eeqa
Now look at a specific lattice bond, e.g.\ $\langle jk\rangle$ in \fig{figcont}.
For any time interval in which the plaquette occupation on this bond does not change,
the breakup probability is constant.
Then the {\em breakup probability per time}
has a continuous time limit,
i.e. it becomes a constant {\em probability density}.
For lattice bond $\langle jk\rangle$ in \fig{figcont}
this is for example the case between times $t_1$ and $t_2$:
\beq{cl}\nonumber
         \frac{p(\mbox{horiz.\ breakup})}{\Delta\tau}
         ~~~~~\stackrel{\Delta\tau\TO 0}{\longrightarrow}~~~~~
          \frac{J}{2}
\eeq
Between times $t_2$ and $t_3$, the probability for a horizontal breakup
on this lattice bond from \eq{GAFa} is zero.
On plaquettes without horizontal breakup, there is a vertical one,
which means that loops will just continue in imaginary time without a jump.
A third case occurs on lattice bond $\langle ij\rangle$ 
at time $t_3$. Here the probability for a horizontal breakup from \eq{GAFa}
is 1.

The same pattern holds for the general case:
All breakup probabilities on plaquettes are either 0, 1, or proportional
to $\Delta\tau$ (for small $\Delta\tau$), and thus have a continuous-time limit.
Note that the probability densities are generated by the order $\Delta\tau$ of plaquette weights.
Thus they contain matrix elements of the Hamiltonian (or parts of it),
and no longer the exponential of $\hat{\cal{H}}$.

\ul{\em The multi-loop algorithm}, summarized in section \ref{Loopsummary},
therefore obtains a modified breakup-step in continuous time.
For each lattice bond:
\bit
\item[(a)] Identify each region of imaginary time in which the worldline
    configuration on this bond does not change.
    Randomly assign horizontal or diagonal breakups there with
    constant probability density, given by the continuous time limit of
    \eq{breakup}.
\item[(b)] At times $t$ where a worldline jumps across the lattice bond,
     there will be a non-vertical breakup with probability one. 
     For example, in case of the Heisenberg antiferromagnet, 
     this will always be a horizontal  breakup.
     In case of the XY-model, \eq{qTHREE} implies probabilty \ \half\ for both
     horizontal and diagonal breakups.
\eit    
The rest of the algorithm remains unchanged.
The technical implementation does however change completely.
In continuous time, one can no longer store plaquette configurations.
Instead, worldlines are specified by the {\em events} at which a worldline
jumps to a neighboring site.
It is useful to store a doubly linked list of such events for each site,
with each list-item containing
pointers to the preceeding and to the following event on the same site,
and a specification of the nature of the event including its time $t$.
(For the isotropic ferromagnet one needs only a singly linked list since all movement is
 forward in imaginary time).
The breakup step can then be performed for each lattice bond,
by following the lists for the corresponding two sites.
Breakups can, e.g.,  be inserted as a different kind of event into the same lists,
or be stored in separate lists.
Identification and flipping of clusters then involves manipulations of
these linked lists.

\ul{\em The single-loop variant} (section \ref{SingleLoop})
can also be performed in continuous time.
Instead of deciding breakups bond by bond, 
we now follow an individual loop-end along sites.

Choose a site $j$ at random. Start a loop at an arbitrary time $t_1$, 
moving (e.g.) upwards in time. An example is shown in \fig{figcont}.
The loop construction iterates the following procedure:

Determine the time interval $t_1 < t < t_2$ 
(equations are specified for moving upwards in time)
during which the worldline occupation of the neighboring sites does not change.
Technically this can be implemented by using linked lists of events
as above, and including additional pointers for each
event, e.g. to preceding events on neighboring sites.
For each such neighbor $k$ draw a random number $\tau_{jk}$ from the distribution 
$\lambda_{jk} \exp{(-\lambda_{jk}\tau_{jk})}$,
where $\lambda_{jk}=J_{jk}/2$ is the corresponding breakup probability density.
Now  move the loop-end on site $j$ up to time $t_1 + \min_k{(\tau_{jk})}$ and let it jump
to the corresponding site $k$ there%
\footnote{Alternatively one can use the sum of the rates $\lambda_{ij}$ to determine
 a transition time, and then decide where to move, according to the ratios of the $\lambda_{ij}$.}
.%
The situation is the same as in radioactive decay, with  decay constants
$\lambda_{jk}$ and neighboring sites  corresponding to  decay channels.

Moving in time-direction on site $i$ corresponds to choosing a vertical breakup 
for all infinitesimal ``plaquettes'' connecting site $i$ to its neighbors.
When $J_{jk}$ corresponds to a horizontal breakup,
the loop-end will move in the opposite time-direction at the new site.
When $J_{jk}$ corresponds to freezing, the loop branches and becomes a cluster of loops,
as usual.

In the standard loop-algorithm, loops are non-self-overlapping
and correspond to those of the pure loop-representation of the
simulated model (sections \ref{PureLoop} and \ref{oper}).
Thus in the above construction one has to exclude those 
temporal regions of neighboring sites which have already
been visited by the single loop (resp.\ cluster).
This constraint is removed for the so-called directed loops, section \ref{DirectedLoops}.

In case that all transition times $t_1 + \tau_{jk}$ exceed $t_2$, 
the loop-end stays at site $j$ and moves to time $t_2$.
If the worldline at site $j$ jumps at time $t_2$, the loop must also jump,
with the same probabilities as in the discrete time formulation.

Now iterate this procedure.%
\footnote{Note that when the transition probability per time is constant,  the stochastic 
          determination of a transition {\em time} can be interrupted and iterated arbitrarily
          without affecting the outcome. Thus the interruption at $t_2$ is allowed.}
Eventually the  loop closes and can be flipped as usual.

We see that in continuous time, the single loop method is technically 
more involved than the multi-loop algorithm.
It also lacks some of the improved estimators of the multi-loop method
(see sections  \ref{Offdiagonal},\ref{Merons}).
On the other hand it tends to have still better autocorrelations.
It can also be generalized to the directed-loop method (section \ref{DirectedLoops}).

The continuous time limit has several important advantages over the discrete time case.
It removes completely the systematic error from the Trotter breakup,
thus also removing the cumbersome need for calculations at several values of $\Delta\tau$
in order to extrapolate to $\Delta\tau =0$.
In addition, worldlines are  specified much more economically
by just specifying their transition times.
This helps especially  at low temperatures,
by strongly reducing the storage requirements for a simulation.
Longer range couplings imply larger sets of neighbors $j$ to treat at each step.
This is still cumbersome, but more economical than introducing extra Trotter slices.
The advantages of the loop algorithm are preserved.

Other approaches which work without Trotter approximation
are the worm algorithm (section \ref{Worms}) 
and  the stochastic series expansion (SSE) (sections \ref{operSSE},\ref{SSEoperupdate},
as well as directed loops (section \ref{DirectedLoops}).

 \subsection{Improved Estimators} \label{ImprEst}

In addition to the reduction of autocorrelations, the
combined representation \eq{FortKast}
allows a potentially drastic reduction of statistical errors by using so-called 
improved estimators \cite{Wolff89a,Wolff89c,Wolff90,WieseY94,AmmonEKTF98,Sweeny83}.
The Monte Carlo procedure provides us with a series of configurations $\S_i$.
For each such configuration, we construct a graph, with some $n_i$ clusters.
We can then  reach any state in a set ${\F}_i$ of $2^{n_i}$ worldline configurations by flipping 
a subset of the clusters. 
The probability for each of these configurations
is determined by the cluster flip probabilities $p_{flip}$. 
In the loop algorithm one configuration $\S_{i+1}$ will be chosen 
randomly according to these probabilities
as the next Monte Carlo configuration.
The standard thermal expectation value of an
observable $\cal O$ is calculated by averaging over the 
value of the observable in the configurations $\S_i$: 
\begin{equation}
        \langle \O\rangle  = \sum_{i} \O(\S_{i}).
\end{equation}

An improved estimator with 
$  \langle \O \rangle = \langle \O_{impr} \rangle $
can formally be defined as a weighted average 
of $\O$
over the states $\SS$ that can be reached from $\S$ with any valid Monte Carlo procedure.
With the loop algorithm we sum over the $2^{n_{i}}$ states $\SS\in{\F}_{i}$.
Since this is a sum over many states, it has reduced variance.
Ideally, it can be calculated completely and then depends only on the graph $G$  
\beq{OimprG}
  \O_{impr}(G) \;=\; \frac{\sum_{\SS}              W(\SS,G)        \O(\SS)      }
                                {\sum_{\SS}        W(\SS,G)                     }
                     \;=\; \frac{\sum_{\SS \in {\F}_{i}} A_{global}(\SS) \O(\SS)}
                                {\sum_{\SS \in {\F}_{i}} A_{global}(\SS)        },
\eeq
where we have used \eq{WSGdelta2}.
Note that $\O_{impr}(G)$ is the representation of $\hat{O}$ in the pure loop model 
(see sections \ref{PureLoop}, \ref{Offdiagonal}).
Alternatively, the sum in \eq{OimprG} can be evaluated stochastically,
\beq{imprEst}
  \O_{impr}(\S,G) \;=\; \sum_{\SS \in {\F}_{i}} \;p_G(\S\rightarrow\SS) \;\O(\SS) \;,
\eeq
where $p_G(\S\rightarrow\SS)$ is any probability that satisfies detailed balance for $A_{global}$
(and thus for $W(S,G)$); it need not be equal to the actual update probability. 
Usually it is a product over suitably chosen cluster flip probabilities $p_{flip}$.
We need to calculate this average over $2^{n_i}$ states in a time comparable to the
time needed for a single measurement. Fortunately that is often possible.

Especially simple improved estimators can  be found in the case
that $p_{\rm flip}={\Half}$ for all clusters. 
Then \eq{imprEst} simplifies to
\beq{impr}
\O_{impr} \;=\;  2^{-n_i}\sum_{\SS \in {\F}_{i}} \O(\SS),
\eeq
since all of the states in ${\F}_i$ now have the same probability $2^{-n_i}$.
In order to achieve  $p_{flip}=\frac{1}{2}$ when there is a nontrivial global weight $A_{global}$, 
we can use a probability function $p_G(\S\rightarrow\SS)$
that has some clusters fixed in a certain state, and then has a new flip probability of \half\  for all other clusters.
There are  many possibilities. 
One can for example fix a cluster in its present state with Metropolis probability,
or one can \cite{AmmonEKTF98} fix the state of a cluster with probability 
$  p_{\rm fix} = |2p_{\rm flip}-1|$, in its present state if $p_{\rm flip}<{\Half}$,
and in its flipped stated otherwise.
The improved estimators then contain the usual contributions from 
both states of the non-fixed clusters,
as well as contributions from the constant state of the fixed clusters \cite{AmmonEKTF98}.
In the following we assume for simplicity that no clusters are fixed.

\newcommand{\szsz}{\hat{S}^z_j \hat{S}^z_k} 
Let us calculate some useful improved estimators.
Consider as an example the spin correlation function 
$\langle \szsz \rangle$ between two spins 
at spacetime sites $j=({\bf r},\tau)$ and $k=({\bf r'},\tau')$.
Since each spin can be in one cluster only, the improved estimator \eq{impr}
is 
\beq{Oimpr2}
 4(\szsz)_{impr} = \left\{ \begin{array}{lll}
                                                \sigma_j\sigma_k ,&    
                           \mbox{\ \ if $j$ and $k$ are in the same cluster}, \\
     (1-2p_{\rm flip,j})\,(1-2p_{\rm flip,k}) \; \sigma_j\sigma_k, &
                           \mbox{\ \ otherwise},\\
  \end{array} \right.      
\eeq
where $\sigma_{j,k}=\pm 1$ are the current values of the worldline variables
($+1$ for a worldline, $-1$ for an empty site).
In case $p_{\rm flip}={\Half}$, the improved estimator is extremely simple:
\beq{Oimpr3}
 4(\szsz)_{impr} = \left\{ \begin{array}{lll}
     \sigma_j\sigma_k, & \mbox{\ \ if sites $j$ and $k$ are in the same cluster},\\
     0               , & \mbox{\ \ otherwise.}
                            \end{array} \right.      
\eeq
We see that the calculation of improved estimators of correlation functions
requires even {\em less} effort than for non-improved estimators.
Remarkably, the spin-spin correlation function corresponds to
the size distribution of the clusters.
In general one can compute $n$-point Greens functions, including off-diagonal ones, 
as improved estimators from the geometric properties of the clusters (section \ref{Offdiagonal}).
Note that for uniform correlations of the the Heisenberg FM 
and for staggered correlations of the Heisenberg AF, 
we always have $\sigma_j\sigma_k=+1$ in \eq{Oimpr3}.

The potential gain from using the improved estimator is easy to see when $\sigma_j\sigma_k=1$.
The expectation value is the same as for the unimproved estimator 
$\sigma_j \sigma_k = \pm 1$.
When $\langle \O \rangle$ is small (e.g.\ $\langle \O \rangle \sim \exp{(-R/\xi)}$
 at large distance $R=|{\bf r} -{\bf r'}|$), then the variance of $\O$ is 
\beq{var1}
  \langle \O^2 \rangle - \langle \O \rangle^2 \,=\, 1- \langle \O \rangle^2 \,\approx\, 1 \;,
\eeq
whereas the variance of $\O_{impr}$ is
\beq{var2}
  \langle \O_{impr}^2 \rangle - \langle \O_{impr} \rangle^2 \,=\,  
  \langle \O_{impr} \rangle -  \langle \O_{impr} \rangle^2 
  \,\approx\, \langle \O_{impr} \rangle \ident \langle \O \rangle \ll 1 \;.
\eeq
For a given distance R, this gain is largest at small correlation length $\xi$,
whereas the gain from reducing autocorrelations with the loop algorithm
is largest at large $\xi$.
On the other hand the non-improved estimator can have a sizeable amount of self-averaging at small $\xi$, 
so that the gain from using improved estimators as just a measurement tool can
be moderate in practice. 
See, however, sections \ref{Offdiagonal} and \ref{Merons} for other drastic effects
of using improved estimators.

Especially simple estimators can also be derived for magnetic susceptibilities.
The uniform magnetic susceptibility at vanishing  magnetic field
can be expressed as the sum over all correlation functions:
\begin{equation}
\langle\chi\rangle=\frac{\beta}{V}\left\langle
\left(\sum_{\bf r}{1\over M}\sum_{\tau}S^z_{{\bf r},\tau}\right)^2
\right\rangle,
\end{equation}
where $V$ is the spatial volume (number of sites), $M=2 d L_t$ is the number of time slices
in $d$ dimensions, and  $S^z_{{\bf r},\tau}=\pm \Half$.
This simplifies \cite{WieseY94} in the XXZ case, by using 
\beq{sus}
 \sum_\tau \frac{1}{M} \sum_{\bf r} S^z_{{\bf r},\tau}  \,=\,
  \sum_{\scr{(clusters c)}}  \sum_{(({\bf r},\tau) \scr{in } c)} \frac{1}{M} S^z_{{\bf r},\tau}  \,=\,
  \Half \, \sum_{\scr{clusters c}}  w_t(c) 
\eeq
to the sum of the square of the temporal winding numbers
$w_t(c)$ of the clusters $c$: 
\beq{chimult}
\langle\chi\rangle=\frac{\beta}{4V}\left\langle\sum_{\scr{clusters c}} w_t(c)^2\right\rangle
\label{EqChiImpr}
\end{equation}
In the single-cluster variant (section \ref{SingleLoop}), the sum over the clusters in
eq.~\ref{EqChiImpr} is also calculated stochastically. Since we
pick a single cluster with a probability $\frac{|c|}{M V}$, 
where $|c|$ is the cluster size and $M V$ is the number of sites in the space-time lattice,
we have to compensate for this extra factor and obtain 
 $      \langle \chi \rangle
        = \frac{\beta}{4V}
        \left\langle\frac{M V}{|c|}\, w_{t}(c)^{2}\right\rangle $.
Similarly, \eq{Oimpr3} implies that for an antiferromagnet with only horizontal breakups,
the {\em staggered} susceptibility corresponds to the sum over the squares of all cluster sizes, 
\beq{ChisImpr}
  \langle \chi_s \rangle 
       =   \frac{\beta}{V}  \left\langle 
                 \left(\sum_{\bf r} (-1)^{\bf r} {1\over M}\sum_{\tau}S^z_{{\bf r},\tau}\right)^2
                             \right\rangle
     \;=\; \frac{\beta}{4V M^2}  \left\langle \sum_{\scr{clusters c}} |c|^2 \right\rangle .
\eeq
Further improved estimators can be constructed, 
including cases with a sign problem \cite{AmmonEKTF98}.
Moreover, it was discovered \cite{ChandrasekharanW99}
that by clever use of the improved estimator for the fermion sign,
one can perform  fermionic simulations in a restricted class of models 
without  sign problem; see section \ref{Merons}.

 \subsection{Infinite Lattices and Zero Temperature} \label{sec:InfLattice}
The existence of a single cluster version (section \ref{SingleLoop})
and of improved estimators for
two-point functions which have support only on individual clusters,
allows for a surprising variant of the loop algorithm,
namely for simulations on borderless lattices,
which implies infinite size  ($L=\infty$)
and/or  exactly zero temperature ($\beta=\infty$),
while the simulation itself remains unchanged.
Evertz and von der Linden showed \cite{EvertzV01} that
one need only iteratively repeat the construction of 
a single cluster with {\em fixed geometrical starting point},
within a spin background of unlimited size which gets updated iteratively when the
clusters are flipped.
As iterations proceed, the single cluster updates will thermalize
the surroundings of the starting point, 
up to further and further distance (with probability proportional to the 
two-point function).
Thus the two-point function of the infinite size system becomes available,
converged to farther disctances in space and/or imaginary time, 
as the computation proceeds.
The infinite size data can be used as the asymptotic point in Finite Size Scaling.
It is especially valuable in systems for which the finite size behaviour is not known 
\cite{PleimlingH01,HenkelP01}.
The calculation of correct error bars for the resulting two-point function needs
special care \cite{EvertzV01}.
For a given distance, the computational time is, as usual, 
by far dominated by measurements, not by thermalization.

This method works whenever
the two-point function drops sufficiently quickly,
so that the corresponding susceptibility is finite,
e.g. in quantum spin systems with a gap.
We note that such a parameter range away from a phase transition is often the region of interest
when comparing simulation results to experimental measurements.


\begin{figure}[htb]
\begin{minipage}[t]{5cm}
  \epsfig{file=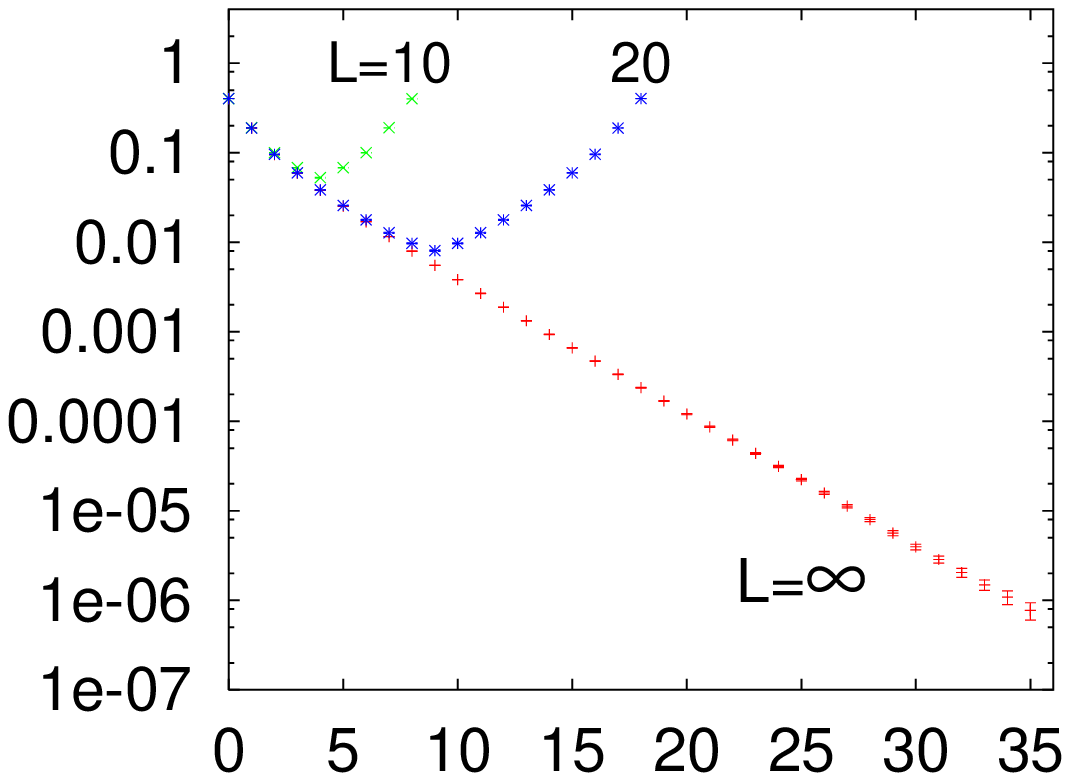,height=36mm} \\ 
  \epsfig{file=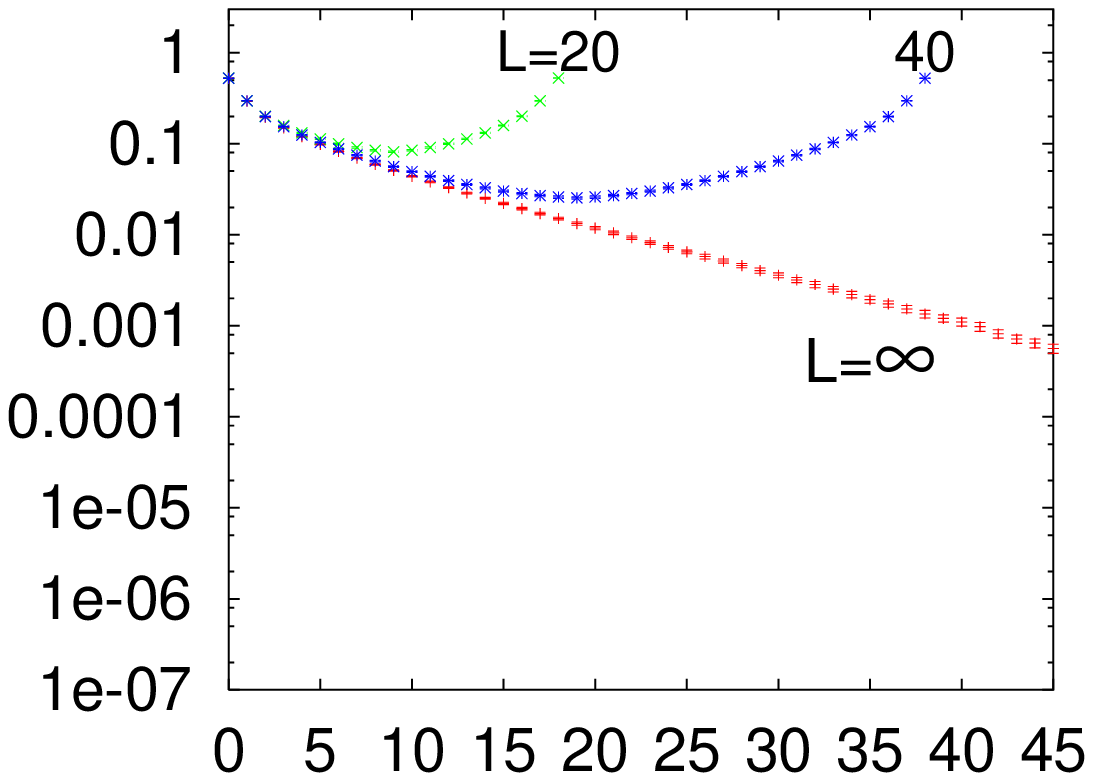,height=36mm}    
\end{minipage}
\begin{minipage}[t]{5cm}
  \epsfig{file=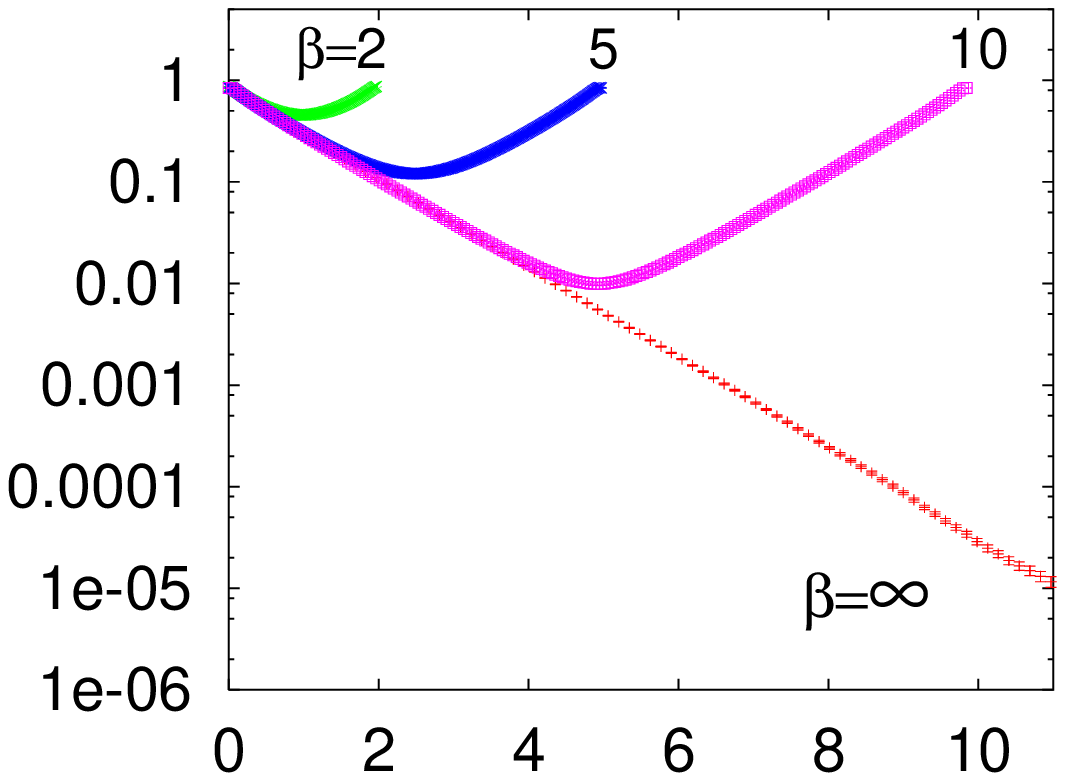,height=36mm}  
  \epsfig{file=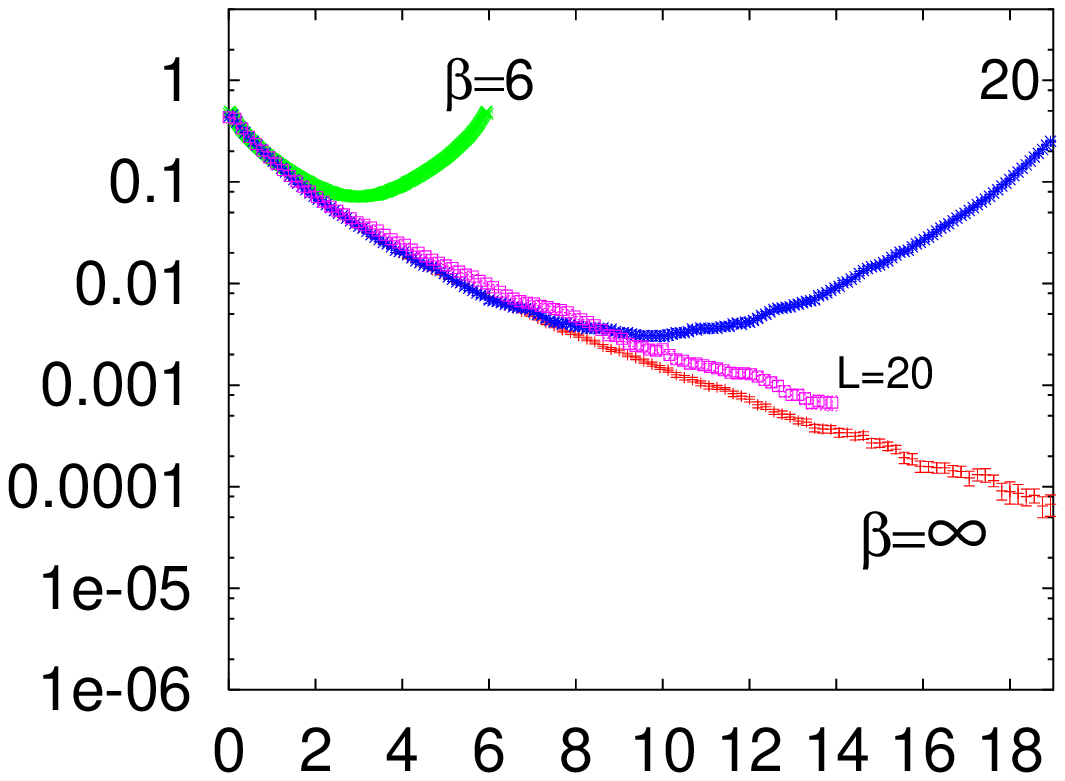,height=36mm}  
\end{minipage}
\begin{minipage}[t]{5cm}
  \epsfig{file=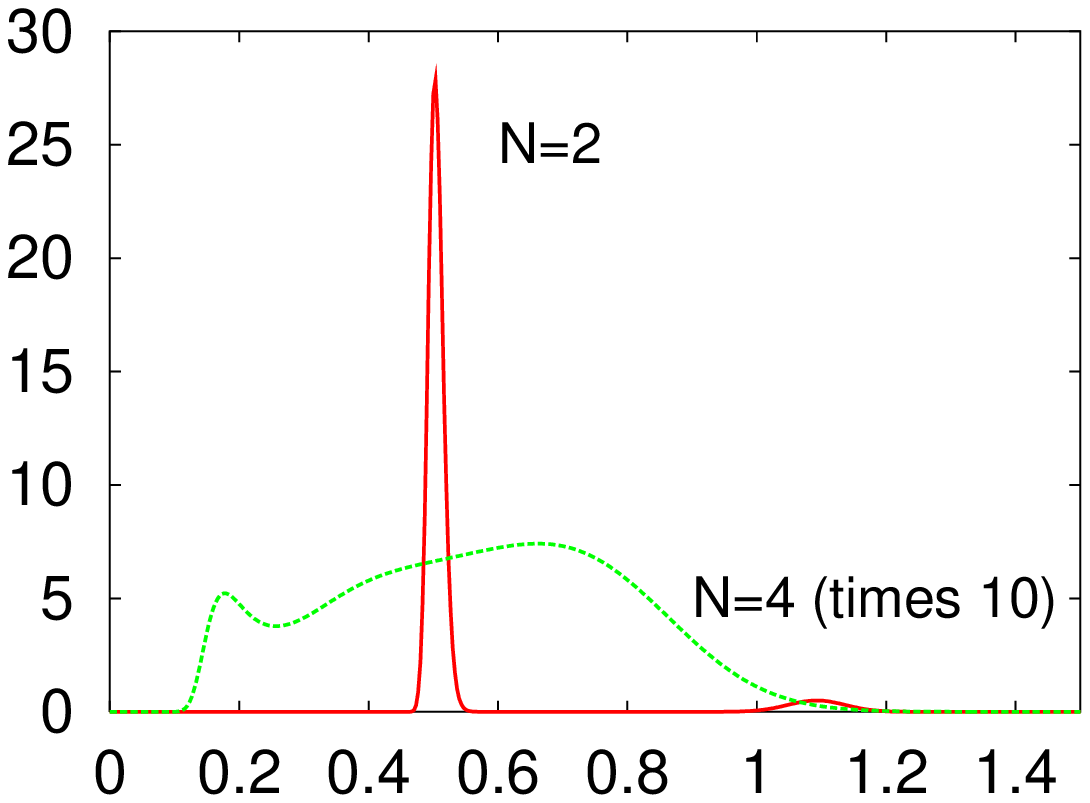,height=36mm}  
\end{minipage}                  %
     \caption{{\em Left}: Equal time staggered spatial correlation function 
              of isotropically coupled Heisenberg antiferromagnetic chains,
              at $q_\perp =\pi$ and $\beta=\infty$,
              for N=2 (top) and N=4 chains (bottom).~
              {\em Center:} 
              Greens functions 
              $\langle S(\vec{q},0) \, S(\vec{q},t) \rangle$
              at $\vec{q}=(\pi,\pi)$ for $L=\infty$, the infinite size system,
              with  $N=2$ (top) and $N=4$ (bottom).
              Results at $L=20$, $\beta=\infty$ have been added to exemplify finite size effects.
              {\em Right:}  
                            Spectrum $S(\vec{q},\omega)$ at 
                            $\vec{q}=(\pi,\pi)$ for $L=\infty$ and $\beta=\infty$.
              From ref.\ \cite{EvertzV01}.
             }
\label{fig:InfLattice} 
\end{figure}

%
As an example we show results for a  spin ladder system \cite{DagottoR96} 
with $N=2$ and $4$ legs for the isotropic antiferromagnet ($\lambda=1$).
The left side of \fig{fig:InfLattice} provides
results for 
the equal time staggered spatial correlation functions along the chains.
A fit to the infinite lattice result gives $\xi=2.93(2)$ for $N=2$ and $\xi=8.2(1)$ for $N=4$. 
The center part of \fig{fig:InfLattice} shows greens functions for $L=\infty$, 
the infinite size system.
Whereas finite temperature calculations give results periodic in imaginary time,
which have to be extrapolated, 
this approach provides the $\beta=\infty$ ($T=0$) result {\em directly}.
A fit to the exponential decay $G(\tau)\sim e^{-\tau\Delta}$ 
directly provides estimates for the gaps
$\Delta=0.5059(4)$ at $N=2$ and $\Delta=0.19(1)$ at $N=4$,
consistent with previous calculations.
Results for $L=20$ and $\beta=\infty$ are also shown,
to exemplify the effect of finite size systems.

Continuing the imaginary time greens function to real frequencies by the maximum entropy method
provides the spectra on the right side of \fig{fig:InfLattice}, in which the gaps, the single magnon peaks, 
and higher excitations for $N=4$ are clearly visible.

 \subsection{Performance} \label{Performance}
The most important advantages and limitations 
of the loop algorithm have already been summarized in the introduction.
Let us be more explicit here.
Further aspects of the performance are mentioned in the following sections.

{\em Autocorrelations: }
The biggest obstacle which the loop algorithm addresses 
are the long autocorrelation times 
of worldline algorithms with local updates, as discussed in the appendices (see \eq{growQ}).
They require a proportional increase in computer time,
so that simulations for large systems and/or low temperatures quickly become impossible.
The loop algorithm appears to remove these autocorrelations completely
in many cases (without magnetic field), like the
spin \half\ Heisenberg AF in any dimension, the two-dimensional spin \half\ XY-model,
and the spin $1$ Heisenberg chain.
For large systems and low temperatures this can save many orders of magnitude in
computer time.
As one striking example, see the gain in autocorrelation time for the one-dimensional
Hubbard model in \fig{figHubbard} in section \ref{FermionicModels}.

Autocorrelations and critical slowing down have been carefully determined
in the original loop algorithm paper \cite{EvertzLM93}
for the nonquantum six-vertex model, with the single-cluster variant of the 
loop algorithm.
In \cite{NovotnyE94}, a related study was done in which
spatial winding was allowed to vary, with similar results for autocorrelations.
In the massless phase (infinite correlation length) at
$\frac{a}{c}=\frac{b}{c}=\sqrt{2}$, the loop algorithm
completely eliminates critical slowing down, i.e.\ the autocorrelation times
are small and constant.
The dynamical critical exponent of the Monte Carlo method (see appendix \ref{CSD}) was
$z^{MC}_{int}\approx 0$ for all measured quantities, and $z^{MC}_{exp}=0.19(2)$ (or logarithmic dependence).
On the KT transition line, the exponential autocorrelation times are slightly larger
(up to $20$ on a $256^2$ lattice), with $z^{MC}_{exp}=0.71(5)$,
yet for the integrated autocorrelation times, which are relevant for MC errors,
we saw barely any autocorrelations in either case,
up to the largest lattices of size $256^2$.
Note that thus the dynamical critical exponent for the integrated autocorrelation time 
is zero here,
{\em different} from that for the exponential autocorrelation time.
Local updates, in contrast, indeed showed very long autocorrelation times,
and $z^{MC}=2.2(2)$, as expected.

Other studies have also seen very small 
integrated autocorrelation times for quantum systems, 
not significantly increasing with $L$ or $\beta$, 
with both the single- and the multi-cluster version of the loop-algorithm
for Heisenberg spin\ \half\ systems in 1d, 2d and on bilayers \cite{DorneichT01},
a spin-$1$ ladder \cite{TodoMYT01}, 
and for a \tJ chain \cite{AmmonEKTF98}.
Note that away from a critical point, integrated autocorrelation times can even decrease with 
increasing system size, due to self-averaging of observables \cite{DorneichT01}.

{\em Strong fields } (resp.\ chemical potentials) 
can however seriously impair the performance. 
They are discussed in section \ref{asymmetries} and \ref{TransverseField}.
See also section \ref{DirectedLoops}.

{\em Improved Estimators:}
The use of improved estimators (section \ref{ImprEst})
provides additional gains.
For example, in ref\ \cite{GrevenBW96} it has been possible to calculate the 
spin-spin correlation function (which in standard updates has large variance)
down to values of $10^{-5}$.

{\em Change of global quantities:} 
Since the loops are determined locally by the breakup decisions, 
they can easily, ``by chance'', wind around the lattice in temporal or in 
spatial direction. An example is given in \fig{figLoopProcedure}.
The flip of such a loop then changes a global quantity
(magnetization, particle number, spatial winding number).
(Of course one can also choose to restrict the simulation to part of 
the total phase space, e.g.\ the {\em canonical ensemble} by not allowing such flips).
This kind of configuration change is virtually impossible with standard local methods.
It has been used to investigate e.g.\ the KT transition in the quantum XY model \cite{HaradaK97,HaradaK98}.

{\em Freezing: }  
For the loop algorithm itself, apart from effects of global weights,
models which require finite freezing weights $w^{ii}$ could potentially  be difficult. 
The intuitive argument can easily be understood.
If two different loops meet at a ``frozen'' plaquette (i.e.\ one for which
the breakup $G^{ii}$ was chosen), they are glued together.
If this happens at overly many plaquettes, then the cluster of glued loops which must
be flipped together can occupy most of the lattice.
The flip of such a cluster is not an effective move in phase space.
It is basically equivalent to flipping all of the (few !) spins outside of that cluster.
As an example, in ref.\ \cite{EvertzLM93} we also investigated versions of the 
loop algorithm in which $w^{ii}$ was (unnecessarily !) chosen finite.
Sizeable autocorrelations were the result. 
Minimal freezing, on the other hand, appears not to be a problem.
(Ref.\ \cite{KohnoT97} includes freezing but also a large magnetic field).
As an example, note that as mentioned in section \ref{ProbXXZ} \cite{KawashimaG95b}, 
the limiting case $J_{x,y}\TO 0$ of the loop algorithm is the classical Swendsen-Wang
cluster algorithm, in which ``freezing'' is the {only} operation.
Yet this cluster algorithm also drastically reduces critical slowing down
in the corresponding classical models.
More general cases with minimal freezing have apparently not been tested.
For  alternatives to  freezing  see also sections \ref{operfreeze} and \ref{DirectedLoops}.

{\em Implementation: }
Implementation of the loop algorithm in imaginary time 
is actually considerably easier than for
local updates, which, especially in more than one dimension, require
rather complicated local updates \cite{MakivicD91}.

In section \ref{ContTime} it was explained how the
time continuum limit $\Delta\tau\TO\,0$ can be taken immediately in
the loop algorithm, eliminating the Trotter approximation, 
reducing storage and CPU-time,
and thus further extending the accessible temperature range.
The implementation within the stochastic series expansion appears to be 
even more efficient, because of the {\em discrete} time-like variable
used there. See section \ref{LoopSSE}.

The loop algorithm can be vectorized and parallelized 
similarly to the Swendsen Wang cluster algorithm (see e.g.\ \cite{Mino91,Evertz93}).
A vectorized version was used in ref.\ \cite{HaradaK97,HaradaK98}.
Vectorization or parallelization of the breakup process is trivial.
The computationally dominant part is to identify the resulting clusters.
This is equivalent to the well know problem of connected component labeling.
See, e.g.,  ref.\ \cite{EvertzM93}.
The optimal strategies are  different from the Swendsen Wang case,
because loops are linear objects.
Efficient parallelization has been discussed by Todo \cite{Todo02a,Todo02b}.
Each of $N_p$ nodes processes a slice of imaginary time of thickness $\beta/N_p$,
identifying the loops that close withing a slice.
The remaining unclosed loops are merged gradually by combining adjacent pairs
of slices and iterating this process in a binary tree fashion,
which produces  only logarithmic overhead.

 \section{Operator formulation of the loop algorithm}\label{oper}

A simple and straightforward derivation of the Loop Algorithm
can be given directly on the operator level, 
instead of working on the level of matrix elements, as we have done so far.
We will rewrite the Hamiltonian of our standard example,
the Heisenberg $XXZ$ model, in terms of {\em loop-operators},
which are equivalent to the breakups introduced previously.

We can then directly write $\tr \;\ebh$ as a continuous time path integral
over loop-operators and spin variables (worldlines).
Alternatively, we can express $\tr \;\ebh$ with the stochastic series expansion (SSE)
\cite{\SSElist}, 
arriving at a version of the complete loop algorithm within SSE.

Various parts of this formulation have appeared in the literature
in different guises,
especially in the operator formulation (on matrix element level) by Brower et al.\ \cite{BrowerCW98},
in the independent 
work by Aizenman and Nachtergaele on the  Heisenberg model \cite{AizenmanN94,Nachtergaele93}, 
in Sandvik's work on the interaction representation \cite{SandvikSC97}
and on ``operator loop updates'' for the isotropic AF \cite{Sandvik99b},
in the work by Harada and Kawashima \cite{HaradaK00},
and in connection with the meron approach \cite{Osborn00,ChandrasekharanO01,Chandrasekharan01a,ChandrasekharanCOW02}.

\subsection{Isotropic Antiferromagnet}\label{operAF}

The Hamiltonian of the spin {}\half{} Heisenberg XXZ model on a single 
lattice bond $\ij$,  without field, is given by \eq{HXXZ}
\beq{Hij}
 \hH_{ij} ~=~  
         J_{x}(\hat{S}^{x}_{{ i}}\hat{S}^{x}_{{ j}}+
         \hat{S}^{y}_{{ i}}\hat{S}^{y}_{{ j}})+
         J_{z} \hat{S}^{z}_{{ i}}\hat{S}^{z}_{{ j}}  \;.
\eeq
For the isotropic antiferromagnet $J_x = J_z = J > 0$, we rewrite
\beqa{Hijaf}
 -\frac{1}{J} \,\hH_{ij} \;+\;\frac{1}{4}

      &=& -\,\vS_i \vS_j \;+\;\frac{1}{4} \1

      &=&        - {1 \over 2} (\hat{S}^{+}_{{ i}} \hat{S}^{-}_{{ j}} +
        \hat{S}^{-}_{{ i}}\hat{S}^{+}_{{ j}}) 
        \;-\;\left(\hat{S}^{z}_{{ i}}\hat{S}^{z}_{{ j}} \;-\;\frac{1}{4}\right)\1

      &=& \frac{1}{2} \left(~
          -\myfig{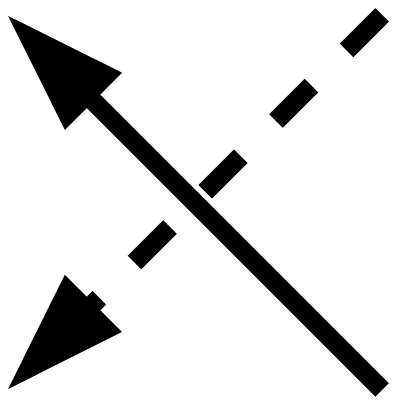} 
          -\myfig{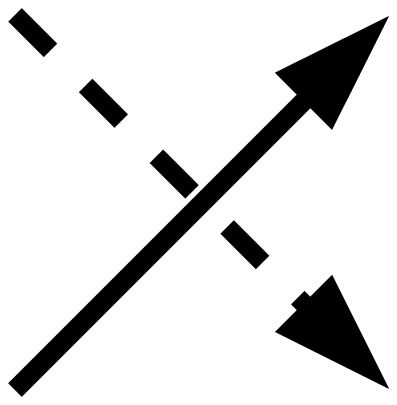} 
          +\myfig{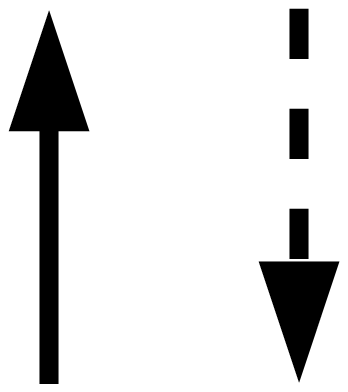} 
          +\myfig{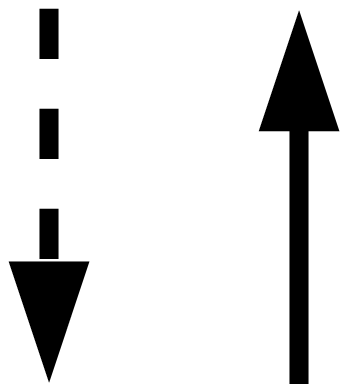}  ~\right)\2

      &=& \frac{1}{\sqrt{2}}\left(\, \ket{\up\down} - \ket{\down\up} \,\right) ~
          \frac{1}{\sqrt{2}}\left(\, \bra{\up\down} - \bra{\down\up} \,\right)  \;.
\eeqa
This operator acts towards the right, except for the third line, where we have given
a worldline-like picture for illustration,
to be interpreted as an operator acting towards the bottom.
We have added a constant $\frac{1}{4}$ to eliminate the contributions of parallel spins.
We see that 
{\em the bond operator $\hH_{ij}$ of the Heisenberg antiferromagnet
     is $-J$ times a \underline {singlet projection operator}}.\\

{On a bipartite lattice} we can change the sign of $J_x$ (see footnote \ref{bipartite}),
obtaining the operator $\tilde{\hH}_{ij}$, equivalent to $\hH_{ij}$, with
\beqa{Htijaf}
 -\frac{1}{J} \,\tilde{\hH}_{ij} \;+\; \frac{1}{4}


      &=& \frac{1}{\sqrt{2}}\left(\, \ket{\up\down} + \ket{\down\up} \,\right) ~
          \frac{1}{\sqrt{2}}\left(\, \bra{\up\down} + \bra{\down\up} \,\right)  \1

      &=& \frac{1}{2} \left(~
          \myfig{crossleft} 
          +\myfig{crossright} 
          +\myfig{leftup} 
          +\myfig{rightup}  ~\right)\2

      &=& \frac{1}{2} \left(~
          \myfig{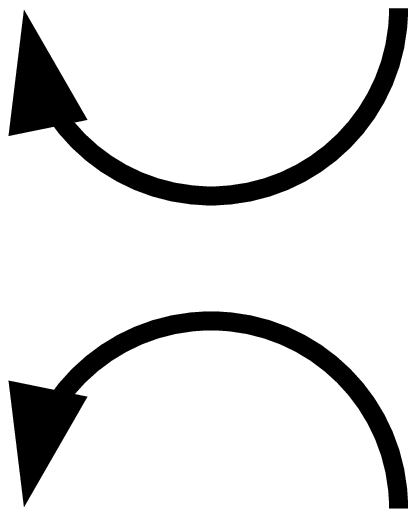} ~+~ \myfig{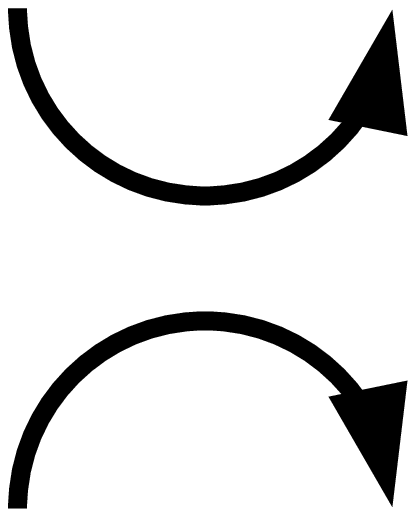} ~+~ \myfig{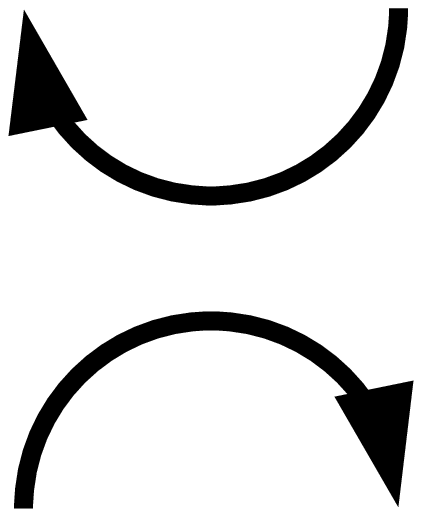} ~+~ \myfig{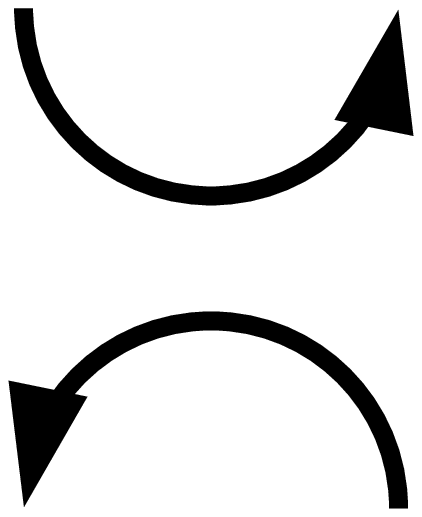}
          ~\right) \2

      &=:& \frac{1}{2} \myfig{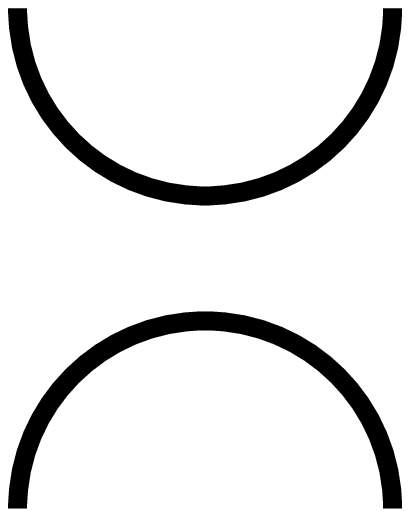}   ~~\equiv~~  \frac{1}{2}\; \hat{B}^h_{ij} ~.
\eeqa
On the third line we have again written a worldline-like picture. The arrows (spin directions)
remain the same on the fourth line, but we have now chosen to connect them differently,
while keeping continuity of arrows along the connecting lines.
As a result we see that the contributions to $\tilde{\hH}_{ij}$ with nonzero matrix 
elements are the same as those contributing to a horizontal loop-breakup.
Thus the horizontal breakup can be written as an {\em operator}
\beq{operh}
 \hat{B}^h_{ij} ~\equiv~ \myfig{breakh} ~=~ 
         \left(\, \ket{\up\down} + \ket{\down\up} \,\right) ~
         \left(\, \bra{\up\down} + \bra{\down\up} \,\right)  ~,
\eeq
with $\frac{1}{2}\hat{B}^h_{ij}$ a projection operator.
After sublattice rotation on a bipartite lattice, 
{\em the energy-shifted Heisenberg bond-operator is therefore the same as a horizontal breakup},
on an operator level.

In passing, we also note that the vertical breakup of the loop-algorithm
is just the identity operator
\beq{vert}
     \myfig{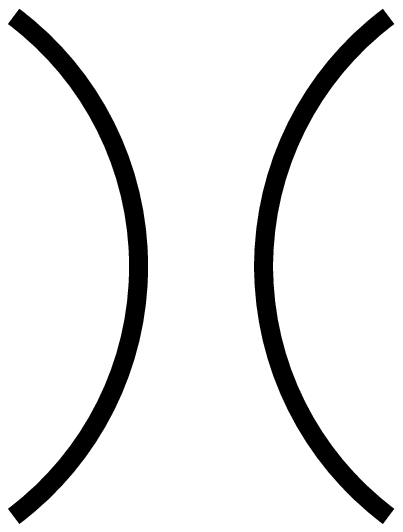} ~\equiv~ \hat{B}^v_{ij} ~\equiv~ \identity  ~.
\eeq

The partition function \eq{Zxxz} of the antiferromagnet ($J>0$) on a bipartite lattice becomes
\beq{Zafoper}
 Z = \tr\, e^{-\beta \hat{H}} ~=~  \; e^{\beta \frac{J}{2} \sum_{\ij} 
                                      (\hat{B}^h_{ij} -\frac{1}{2})} ~.
\eeq

\subsection{Isotropic Ferromagnet}\label{operFM}
The ferromagnet can be treated in the same way.
Now $J_x = J_z = J < 0$, and 
\beqa{Hijfm}
 -\frac{1}{|J|} \,\hH_{ij} ~+~\frac{1}{4} 

      &=& \vS_i \vS_j \;+\;\frac{1}{4} \1

      &=&         {1 \over 2} (\hat{S}^{+}_{{ i}}\hat{S}^{-}_{{ j}} + 
        \hat{S}^{-}_{{ i}}\hat{S}^{+}_{{ j}})
        \;+\;\left(\hat{S}^{z}_{{ i}}\hat{S}^{z}_{{ j}} \;+\;\frac{1}{4}\right)\1

      &=& \frac{1}{2} \left(~
           \myfig{crossleft} 
          +\myfig{crossright} 
          +\myfig{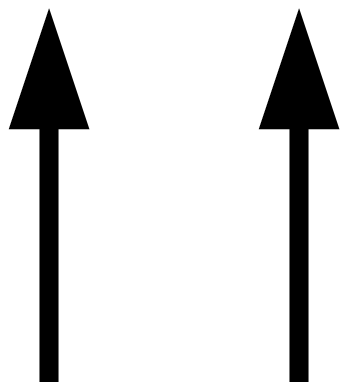} 
          +\myfig{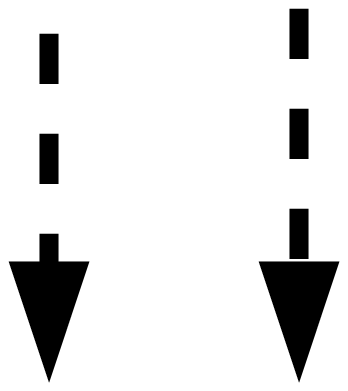}  ~\right)\2

      &=& \frac{1}{2} \left(~
           \myfig{crossleft} 
          +\myfig{crossright} 
          +\myfig{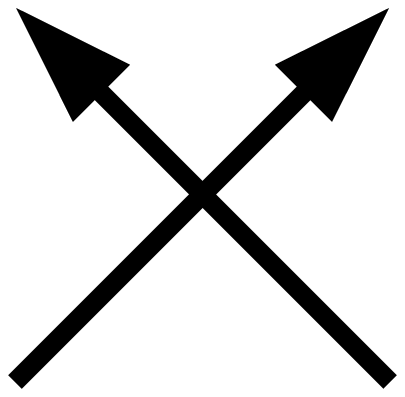} 
          +\myfig{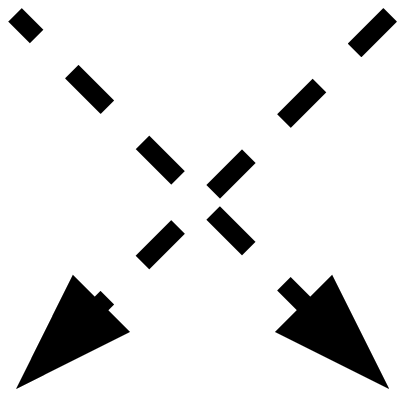}  ~\right)\2

      &=& \frac{1}{2} \myfig{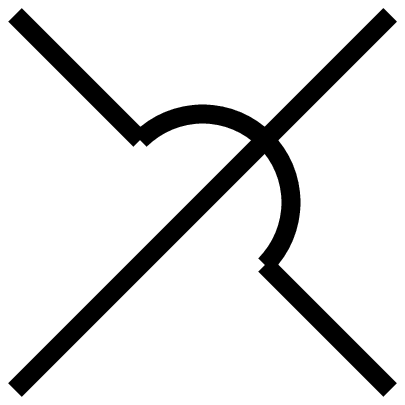}   ~~\equiv~~  \frac{1}{2}\; \hat{B}^d_{ij} ~.
      
\eeqa

Here the constant $\frac{1}{4}$ was chosen to eliminate the contributions of antiparallel spins.
On the third line we have again written a worldline-like picture. The arrows (spin directions)
remain the same on the fourth line, but we have again chosen to connect them differently,
while keeping continuity of arrows along the connecting lines.
As a result we see that for the isotropic ferromagnet on any lattice,
the operator $\tilde{\hH}_{ij}$, after an energy-shift,
is proportional to the {\em operator} $\hat{B}^d_{ij}$ for a {diagonal} loop-breakup,
which indeed is the permutation operator.

\noindent
The partition function \eq{Zxxz} of the ferromagnet on any lattice becomes
\beq{ZafoperFM}
 Z = \tr e^{-\beta \hat{H}} ~=~  \; e^{\beta \frac{|J|}{2} \sum_{\ij} 
                                      (\hat{B}^d_{ij} -\frac{1}{2})} ~.
\eeq
Note that the difference between antiferromagnet and ferromagnet is connected to requiring 
positivity of the final exponent for the partition function, leading
to horizontal breakups for the antiferromagnet and diagonal ones for the ferromagnet.

\subsection{Anisotropy}\label{operaniso}
To treat models with anisotropy $\Delta=J_z/|J_x|$, we can use the operator identities
\beqa{Hzz}
 4\; S_i^z S_j^z 

             &=& \myfig{breakd} ~-~ \myfig{breakh}  ~,
\eeqa
which follows by subtracting \eq{Htijaf} from \eq{Hijfm}, and
\beqa{Hpm}
    2\left( \hat{S}^{+}_{{ i}}\hat{S}^{-}_{{ j}} + \hat{S}^{-}_{{ i}}\hat{S}^{+}_{{ j}} \right)
  &=& \myfig{breakd} ~+~ \myfig{breakh} ~-~ \identity ~.
\eeqa

\noindent
For the {antiferromagnet} 
on a bipartite lattice we get
\beqa{Hafoperanbi}
    -\frac{1}{|J_x|} \tilde{\hat{H}}_{ij} 
        &=& \frac{1+\Delta}{4}\myfig{breakh} ~+~\frac{1-\Delta}{4}\myfig{breakd} ~-~\frac{1}{4} ~.
\eeqa

\noindent
The anisotropic {ferromagnet} on any lattice is given by the same equation, with $\Delta<0$.
These results are equivalent to  \eq{qTHREE}.
They provide positive weights when $|\Delta|\le 1$.

\subsubsection*{Freezing and alternatives}\label{operfreeze}
For $\Delta>1$, i.e.\ the Ising-like regions of parameter space,
the operator formulation leads to, e.g., the following approaches: 
\bit 
 \item[(1)] One possibility is to use $(S_i^z S_j^z \pm \frac{1}{4})$
            as an operator, with weight $\Delta-1$.
            Then on bonds where this operator acts, 
            the weight of a configuration with AF (or FM) neighboring spins is zero,
            i.e.\ forbidden. This amounts to an operator formulation of {\em freezing}
            of the opposite spin orientation (see section \ref{weights}).
 \item[(2)] Alternatively, one can use $S_i^z S_j^z$ as an operator,
            and proceed similarly to the case of merons (section \ref{Merons}).
            The flip of a loop connected to some other loop by an odd number
            of such operators will result in a sign flip for the
            total weight of the configuration, and therefore in a combined contribution of
            zero to the total weight. Thus only configurations with an even number
            of such connections contribute, while keeping loops independent.

 \item[(3)] An interesting different alternative has been developed by Otsuka \cite{Otsuka01}:
            He treats $(\Delta-1)S_i^z S_j^z$ by introducing Hubbard-Stratonovich variables,
            which act locally like a magnetic field. Then loop flips remain independent,
            with a flip probability that depends on the Hubbard-Stratonovich configuration.
\eit

\subsection{Treating $\ebh$ as a continuous imaginary time path integral}\label{operpath}
We can obtain the  path integral in continuous imaginary time
by directly employing a 
Poisson process representation of $\ebh$, as specified
rigorously by Aizenman and Nachtergaele (section 2.2 of reference \cite{AizenmanN94}).
For related work see \cite{AizenmanL90,FarhiG92}.
Rephrased for our context the statement refers to a Hamiltonian of the form
\beq{NH}
 \hat{H} = - \sum_b J_b\, \hat{h}_b
\eeq
with bonds $b$, nonnegative couplings $J_b$, and selfadjoint operators $\hat{h}_b$,
which acts on a finite system.
Then $\ebh$ can be expressed as a Poisson integral:

\beqa{Nachtergaele}
 e^{-\beta \hat{H}}
  &= e^{\beta \sum_b J_b} \lim_{\Delta t\to 0} 
     \left({\prod}_b e^{(-J_b+J_b \hat{h}_b) \Delta t}\right)^{\beta/\Delta t}\1

  &= e^{\beta \sum_b J_b}\lim_{\Delta t\to 0}
     \left({\prod}_b \{(1-J_b\Delta t) +J_b  \hat{h}_b\Delta t\}\right)^{\beta/\Delta t}\1

  &= e^{\beta \sum_b J_b}\int\rho(d\omega)\prod^* \hat{h}_b ~,
\eeqa

where  $\prod^* \hat{h}_b$ is a time ordered product of bond operators $\hat{h}_b$,
$\omega$ is the bond configuration,
and $\rho(d\omega)$ is a poissonian probability measure with density $\prod_b J_b dt$.
Thus, we get a random countable collection of
time-indexed bonds which occur independently in disjoint
regions of spacetime.
This is a non-commutative version of the usual power series
expansion of the exponential function.

\subsubsection*{Loop operator representation}\label{operpathloop}
If we choose to represent the antiferromagnet as a sum over horizontal loop operators,
\beq{Hop}
 \hat{H} ~+~ const ~~=~~ -\frac{J}{2} \;\sum_\ij \;\hat{B}^h_{ij} ~,
\eeq
then the theorem states that we can represent $\ebh$ as stochastic integral
over configurations of loop operators $\hat{B}^h_{ij}$
which appear with constant probability density $J/2$ in continuous imaginary time.
In between these operators there is just the identity operator, to which
we have been referring as a vertical breakup.
Quantizing in $S^z$ basis, one obtains continuous lines between the loop operators.
Finally, taking the trace of $\ebh$ provides periodic boundary conditions 
in imaginary time 
for the resulting configuration of horizontal loop operators and lines.
\newfigh{loopconttime}{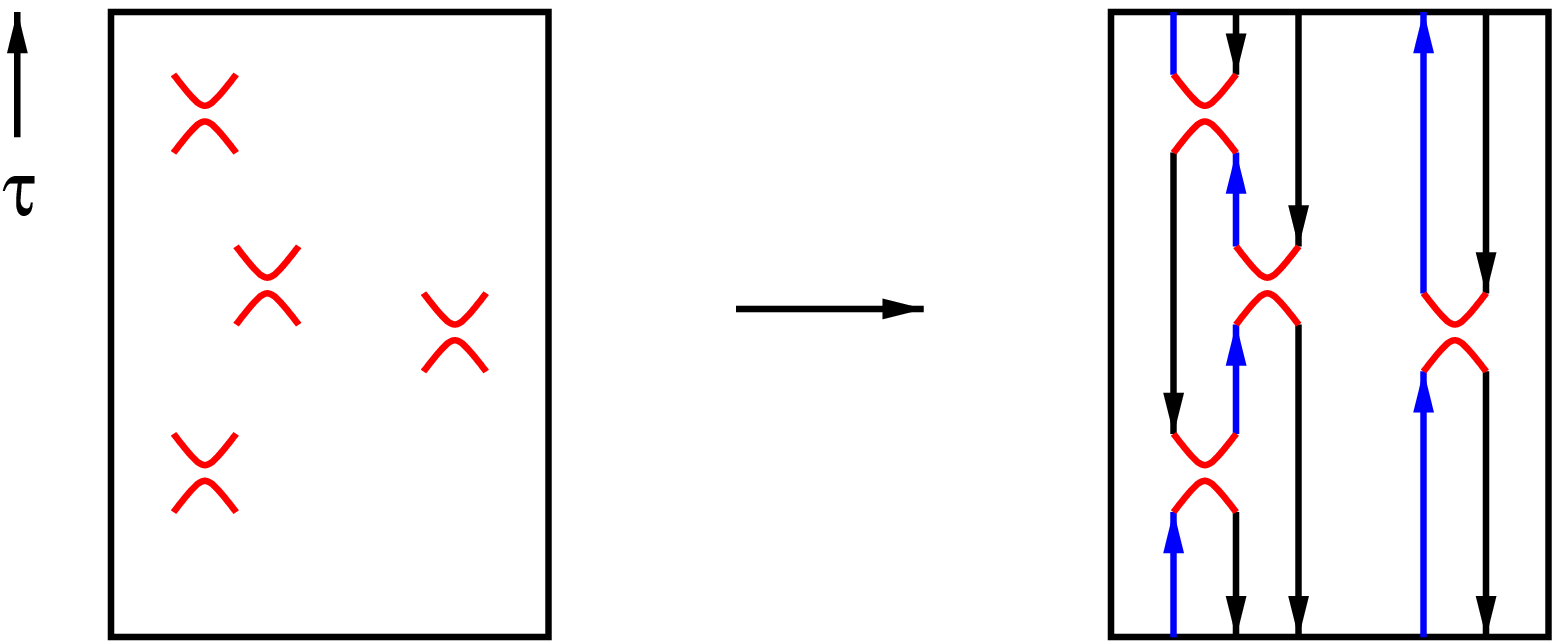}{5 cm}{
    Configuration of horizontal loop operators (left),
    and a compatible loop and (world)line configuration (right).}
Since these lines do not start anywhere, they close to form loops,
with two possible orientations of arrows (spins) on each loop.
When the spin points upwards, these lines are worldlines.
From \eq{Htijaf} we see that all matrix elements are unity, so the total weight of the
configuration (once stochastically chosen) is unity.

We have arrived at the combined loop-operator and spin {\em representation} of the
Heisenberg model in continuous time.
From this representation we can immediately derive the loop {\em algorithm}
as a Quantum Monte Carlo procedure by switching back and forth between
(i) choosing a new operator configuration compatible with the current
    spin configuration, with probability density $J/2$ for each operator,
and (ii) choosing a new spin configuration compatible with the current operator
configuration. This is the same procedure as that derived in section \ref{ContTime}.

When we sum over all possible spin configurations, two for each loop,
we obtain the pure-loop representation of the Heisenberg model
(see section \ref{PureLoop}).

\subsubsection*{Standard hopping representation}\label{operpathhop}
We can also stay in the standard representation \eq{Hij},
in which the operators $S^+S^-$ correspond to hopping of worldlines,
and $4S^z_i S^z_j \pm 1$ is diagonal, both with matrix elements unity or zero.

Then we obtain a standard worldline representation, 
but {\em directly in continuous time}.
Thus the existence of a continuous time representation is independent
 of whether one uses a loop representation \cite{FarhiG92}.
Here one can perform, e.g.,  something like the usual local updates of standard worldline Monte Carlo,
e.g.\ by proposing to shift an existing hopping operator $S^+S^-$ by some time $\Delta\tau$,
uniformly chosen within a suitable range.

\subsection{Treating $\ebh$ by stochastic series expansion}\label{operSSE}
The stochastic series expansion of an operator $\ebh$,
was introduced by Anders Sandvik \cite{\SSElist}.
It is related to the Handscomb method \cite{Handscomb62,Handscomb64,Lyklema82,LeeJN84,ChakravartyS82}
for the Heisenberg model,
but also applicable to many other models.
We assume again that 
$\hat{H}=\sum_{\nu,b}  \hat{H}_{\nu,b}$ 
is a sum of operators $\H_{\nu,b}$ living on bonds $b$.
(Site terms can also be brought into this form).
SSE proceeds from the power series
\beqa{SSEeq}
 \ebh &=& \sum_{n=1}^\infty \;\frac{\beta^n}{n!} \;(-\H)^n \1

      &=& \sum_{n=1}^\infty \;\frac{\beta^n}{n!} \;(-\H_1 -\H_2 -...)(-\H_1 -\H_2 -...)... \1

      &=& \sum_{n=1}^\infty \;\frac{\beta^n}{n!} \;\sum_{j_1,j_2,...,j_n} 
                        (-\H_{j_1})(-\H_{j_2}) ... (-\H_{j_n}) ~,

\eeqa
where $j$ is a shorthand for $(\nu,b)$.
The  third line is obtained by expanding the product in the second line, 
resulting in a sum with one sub-Hamiltonian $\H_{j_k}$ from each of the $n$ factors.
For a fixed value of $n$, we obtain a sum over {sequences} of $n$ operators $\H_{j_k}$.
There is now a discrete index $l=1..n$ with ``time-like'' properties, 
since it describes a sequence of Hamiltonians.
At each value of $l$, only one of the sub-Hamiltonians acts.
The partition function becomes
\beq{Zsse}
  Z ~=~ \sum_\alpha \,\sum_{n=0}^\infty \,\sum_{S_n} \,\frac{\beta^n}{n!} \;
         \bra{\alpha} \,\prod\, (-\H_{(\nu,b)_i}) \, \ket{\alpha}
\eeq
where $S_n$ is a sequence $((\nu,b)_1,(\nu,b)_2,...,(\nu,b)_n)$ of operator indices.
We insert $\hS^z$ eigenstates for $\ket{\alpha_l}$ between all operators.
The result is again a worldline-like
representation of $\tr\;\ebh$, this time with a discrete index space instead of
continuous imaginary time.

Let us emphasize that the stochastic series expansion, as presented so far,
is a {\em different treatment of $\ebh$}, independent of 
the choice of sub-Hamiltonians and 
of any Monte Carlo procedure.
We also note that the continuous time representation \eq{Nachtergaele},
with a time ordered set of sub-Hamiltonians, 
and the stochastic series expansion are indeed closely related.
The biggest difference is the discreteness of index space in case of SSE.
This discreteness increases the efficiency of SSE with respect to continuous time.

\subsubsection*{Standard hopping representation}\label{operSSEhop}

If we choose the standard hopping representation \eq{Hij},
then {\em local} updates of the operator configuration,
as specified in the original SSE papers, 
are the natural choice.

\subsubsection*{Loop operator representation}\label{operSSEloop}

If instead we choose a loop operator representation like \eq{Hop},
then we arrive again at a combined worldline/loop-operator representation
of $\tr\;\ebh$.
Switching back and forth between choosing operators and choosing worldlines
results in the loop algorithm,
this time in  discrete index space.
For the isotropic Heisenberg antiferromagnet and ferromagnet, 
a variant of this procedure was 
introduced by Sandvik as ``operator loops with deterministic updates'' 
\cite{Sandvik99b,HeneliusS00},
developed from a different approach (section \ref{SSEoperupdate}). 
From the present formulation 
we see that indeed the complete loop algorithm can be applied within SSE,
including anisotropic cases and generalizations.

\subsection{Loop algorithm in SSE}\label{LoopSSE}
%
\newfigh{figSSE}{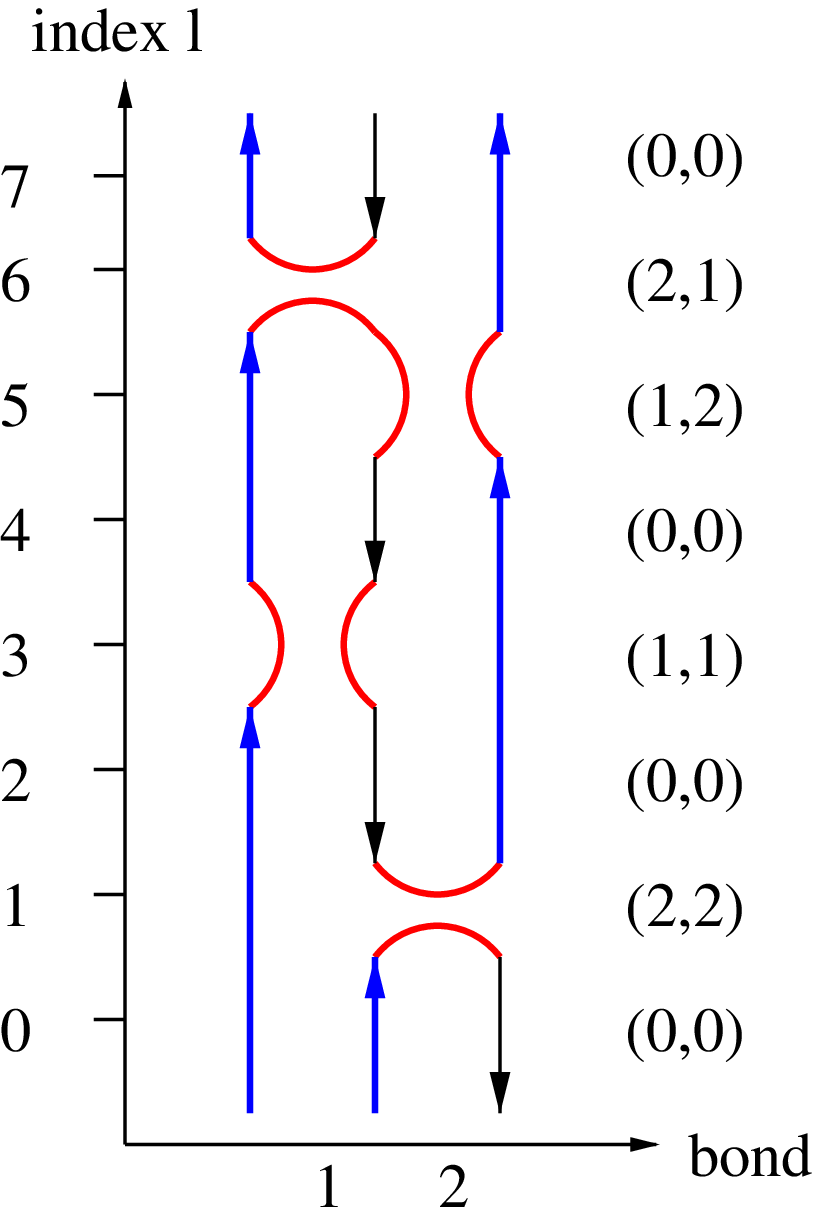}{7cm}{
Part of an SSE configuration in loop-operator representation.
The diagonal operator $\identity$ is drawn as a vertical breakup.
A compatible spin-configuration is denoted by the arrows.
The vertical axis marks the propagation index $l$,
the horizontal axis numbers bonds $b$.
The sequence $(\nu,b)$ is given on the right,
with $\nu=1$ ($\nu=2$) denoting a vertical (horizontal) breakup.
}%
%
Let us describe a possible implementation of the loop algorithm with SSE in more detail. 
The continuous time version was described in section \ref{ContTime}.
For further information, we refer to the literature on SSE 
\cite{\SSElist,Sandvik99b,DorneichT01,SyljuasenS02}.

We describe the case of an anisotropic Heisenberg model on a bipartite lattice.
From \eq{Hafoperanbi} we get (with $J_x=1$ as unit of energy)
\beqa{Hsse}
 -\hat{H}_{ij} 

       &:=&  - \frac{1}{2} (\hat{S}^{+}_{{ i}}\hat{S}^{-}_{{ j}} +
                            \hat{S}^{-}_{{ i}}\hat{S}^{+}_{{ j}})
        \;-\;  \Delta \; \hat{S}^{z}_{{ i}}\hat{S}^{z}_{{ j}} 
        \;+\; \frac{1}{4}  \;+\; \en\, \identity
        \1
       &=& \frac{1+\Delta}{4}\myfig{breakh} ~+~\frac{1-\Delta}{4}\myfig{breakd} 
            ~+~\en\, \identity \1
      &=:&  J_h  \hat{B}^h ~+~ J_d \hat{B}^d ~+~ \en\, \hat{D}
\eeqa
with $|\Delta|\le 1$, $\en > 0$ 
and a ``diagonal operator'' $\hat{D}=\identity$.
We have added a constant to $\H_{ij}$ to ensure positivity of matrix elements.
The constant $\en$ might be varied to further improve convergence \cite{SyljuasenS02}.

It is advantageous to extend the SSE sequence to a fixed length $L$ by introducing
additional ``empty'' unit operators $\h^0=\identity$. Then \cite{Sandvik97}
\beq{ZMsse}
 Z ~=~ \sum_\alpha\,\sum_{S_L} \,\frac{\beta^n (L-n)!}{L!} 
         \bra{\alpha} \,\prod_{i=1}^L (-\H_{(\nu,b)_i}) \, \ket{\alpha}
\eeq
with $n$ the number of non-unit operators in the sequence $S_L$.
This can be done without approximation in practice, since the average length of
the operator string is just the total energy of the system, as seen by differentiating
with respect to $\beta$, and its fluctuations are finite in any finite run.
A value for $L$ with a safe margin can be determined during thermalization.

The SSE configuration is specified by $S_n$, 
which can be written as a sequence of tuples $(\nu,b)$, with $(0,0)$ denoting $\h^0$.
When the operators $\H_(\nu,b)$ are ``non-branching'', it suffices to specify
the initial state $\alpha(0)$.
Otherwise (as for horizontal loop-operators), the ``worldline configuration'' $\ket{\alpha(l)}$ 
has to be stored, too.
A picture of an SSE configuration is shown in \fig{figSSE}. 

The simulation begins with, e.g., a configuration of all down spins
and $S_n$ consisting of only empty operators.

An overall Monte-Carlo-update consists of three parts.
First, empty operators $\h^0$ and diagonal operators operators $\hat{D}$
are exchanged with Metropolis probability.
\beqa{DIAGsse}
 p(\,(0,0)_l\rightarrow(1,b)_l\,) ~=~ 
       \min\left(1,\frac{N_b\,\beta\, \bra{\alpha(l)}\hat{D}_b\ket{\alpha(l)}}%
                                                       {L-n}\right) \1
 p(\,(1,b)_l\rightarrow(0,0)_l\,) ~=~ \min\left(1,\frac{L-n+1}%
                                          {N_b\,\beta\, \bra{\alpha(l)}\hat{D}_b\ket{\alpha(l)}}%
                                                       \right)
\eeqa
where $N_b$ is the total number of bonds and $\nu=1$ stands for the diagonal operator $\hat{D}$.
This step changes the expansion order $n$ by $\pm 1$.
It is attempted consecutively for each $l=1,...,L$.

In the second step, the ``breakup decision'',
diagonal operators $\hat{D}$ and loop-operators $\hat{B}^h$, $\hat{B}^d$ are exchanged.
Now only the ``vertices'' with non-empty operators matter.
As in the case of continuous time, one needs a doubly linked list
connecting each of the four ``legs'' of each vertex with the corresponding leg of the 
next (resp.\ previous) vertex. 
The information on spins can be stored with the vertices.

As usual at each vertex, the local spin configuration allows two of the three breakups,
vertical (operator $\hat{D}$), horizontal $\hat{B}^h$ or diagonal  $\hat{B}^d$.
The choice is made at each vertex with heat-bath probability.
Thus on an antiferromagnetic vertex (like plaquettes $2^\pm$ in fig.\ \ref{figplaquettes}),
the horizontal breakup in our example is chosen with probability
\beq{breakupHsse}
\frac{ 1+\Delta}{4\en \,+\, (1+\Delta)}
\eeq
and on a vertex with spin-flip (like plaquettes $3^\pm$)
it is chosen with probability
\beq{breakupHFsse}
\frac{ \frac{1}{4}(1+\Delta)}{\frac{1}{4}(1-\Delta) \,+\, \frac{1}{4}(1+\Delta)}
     ~=~ \frac{1}{2}(1+\Delta) ~.
\eeq

The third step consists as ususal of identifying and flipping loops.
Nontrivial diagonal operators as well as external fields can influence
the flip probability.
Note that on sites on which no operators act there is a straight loop.

The ``deterministic operator-loops'' previously constructed \cite{Sandvik99b}
for isotropic ferro- and antiferromagnet
are somewhat different. They use a different diagonal operator in \eq{Hsse}
which forbids ferromagnetic vertex configurations.
The generalization of the loop algorithm just outlined has lower autocorrelations,
as recently shown by Syljuasen and Sandvik \cite{SyljuasenS02}.

We have described the multi-cluster loop-algorithm.
In the {\em single cluster version}, loop construction starts 
at a randomly chosen vertex-leg, and decides breakups iteratively until the loop closes.
(This construction generalizes to the directed loops in section \ref{DirectedLoops},
 which are better suited, e.g.,  for large magnetic fields.)

{\em Measurements} of static observables are straightforward \cite{SandvikSC97}.
For example, the total energy is given by 
$ E = \-\frac{1}{\beta} \,\langle n \rangle $
and the heat capacity by
\beq{CHsse}
 C_V ~=~ \langle n^2 \rangle \;-\; \langle n \rangle^2 \;-\; \langle n \rangle ~.
\eeq
Time-dependent correlations are also available \cite{Sandvik97}.
An imaginary time separation $\tau$ corresponds to a binomial 
distribution of propagation distances $\Delta l$,
and correlators are given by 
\beq{CorrSSE}
 \langle \hat{A}(\tau) \,\hat{B}(0) \rangle ~=~
     \left\langle \sum_{\Delta l=0}^n \, \left( n\atop {\Delta l} \right) 
                                      \, \left( \frac{\tau}{\beta} \right)^{\Delta l} 
                                      \, \left(1-\frac{\tau}{\beta}\right)^{n-\Delta l} 
                   C_{\hat{A}\hat{B}}(\Delta l) \,\right\rangle
\eeq
with
\beq{CABsse}
 C_{\hat{A}\hat{B}}(\Delta l) ~=~ \frac{1}{n+1}\,\sum_{l=0}^n \,\hat{B}(l+\Delta l)\, \hat{A}(l) ~.
\eeq
Improved observables are available as usual.
An efficient method to measure time-dependent Greens functions has been 
provided by Dorneich and Troyer \cite{DorneichT01}.

\subsection{Discussion}\label{operdisc}

We have seen that the issue of representing $\tr\;\ebh$ by a (continuous or discrete)
imaginary time path integral, or by SSE,
is independent of the choice of representation of $\H$,
e.g.\ in terms of traditional hopping operators, or in terms of loop operators,
and independent of the resulting possible update procedures.
The loop operator representation of $\H$ and the full loop algorithm
can be applied both in the imaginary time path integral and in SSE.
Additional operators in $\H$ can be treated similarly 
(see sections \ref{operfreeze},\ref{TransverseField}).
Because of its discrete index space, 
the stochastic series expansion appears to be somewhat more efficient
computationally than continuous imaginary time.

 \section{Generalizations}\label{Generalizations}
So far we have purposely restricted ourselves to XYZ-like
models in order to simplify the presentation.
This covered both the case of spin \half\ quantum spin models,
where we have inserted eigenstates 
$\left|S^z_{il}\right>=\left|\pm 1\right>$
(or eigenstates along a different quantization axis) 
and models of fermions or hard core bosons, where we have inserted occupation 
number eigenstates \cite{Scalettar-WL99}.
We have developed the formalism for the general anisotropic XYZ-like 
(eight-vertex-like) case.
We have computed explicit update probabilities for all XXZ and most XYZ-like cases.

Let us now describe further generalizations, several of which are immediate.
For all generalizations, with a slight modification for
continuous time (section \ref{ContTime}),
it remains true that {\em locally} on the vertices we have a situation like in the 
six- (or eight-) vertex model, 
so that the loop formulation, 
on the level of matrix elements or on the level of operators,
described above can be applied directly.

 \subsection{Long range couplings}\label{longrange}
Hopping or spin-spin interactions beyond nearest neighbor can be handled
by the same approach as higher dimensions,
namely by introducing extra parts $\hat{H}_\nu$ in the split of the Hamiltonian,
i.e.\ extra ``bonds'' on the lattice, with a corresponding set of 
shaded plaquettes living on separate Trotter time subslices.

When the number of additional couplings is large,
this approach becomes impractical.
It is somewhat less cumbersome 
in the continuous time version of the loop algorithm (section \ref{ContTime}).
A stochastic approach by Luijten and Bl\"ote to extend cluster simulations to such cases
may be helpful here \cite{LuijtenB95}.

 \subsection{Bond disorder, diluted lattices, and frustration}\label{disorder}

Bond disorder refers to spatial variations in the spin couplings $J_{ij}$
(resp.\ hopping strengths $t_{ij}$ and/or density-density couplings $V_{ij}$).
This modifies the loop-construction probabilities {\em locally},
making them plaquette-dependent. Otherwise nothing changes !
(One does need to check whether ergodicity is still achieved.)

The same is true for diluted lattices, which can be viewed as a case of bond disorder
in which the coupling vanishes completely on some bonds.

The situation is different for frustrated couplings whence some matrix elements become
negative and cannot be transformed to a positive representation.
Whereas the loop algorithm itself is unaffected,
this produces a second type of sign problem 
which needs to be handled in the same way as the fermion sign problem (section \ref{FermionicModels}).
If the strength and/or frequency of frustrated matrix elements, as well as system size and $\beta$ are not too large,
this sign problem can remain manageable \cite{AmmonEKTF98,MiyaharaTJU98}.
In general, however, it has so far precluded simulations of strongly frustrated models.
It is possible to find improved estimators for frustrated couplings \cite{AmmonEKTF98},
to  alleviate this sign problem.
Note that for non-bipartite lattices, the breakup probabilities have to
be modified to ensure ergodicity of the algorithm, as has been discussed in section \ref{ergodicity}.
This modification might introduce additional freezing, which, however, is
dwarfed by the sign problem.

Promisingly, the meron-strategy outlined in section \ref{Merons} can also be applicable 
to frustrated systems.
Indeed, this way Henelius and Sandvik have succeeded in simulating 
a ``semi-frustrated'' Heisenberg quantum spin system without sign problem \cite{HeneliusS00},
though at present  other frustrated model.
In special cases, one can also find a new basis without sign problem,
e.g. for a  Trellis lattice with special range of couplings \cite{Nakamura97,NakamuraO97}.

 \subsection{Asymmetric Hamiltonians: Magnetic field, chemical potential}\label{asymmetries}
Asymmetric weights can be caused by a (uniform or random) magnetic field in $\hat{S}^z$-direction
or equivalently a chemical potential,
or by other non ``particle-hole-symmetric'' terms,
like e.g.\ softcore bosons.
Large diagonal fields  seriously affect the performance of the loop algorithm.
Transverse fields, however, are easily handled (see section \ref{TransverseField}).

Within the algorithm presented so far, 
asymmetric parts of the Hamiltonian need to be taken into the global weight $A_{global}$,
such that $W^{plaq}$ is symmetric with respect to a flip of all four spins on each shaded
plaquette. 
Then they contribute to the flip probabilities $p_{flip}$ of the clusters.

A magnetic field $h$ (or chemical potential $\mu$) affects only clusters
which change the number of worldlines, i.e.\ which wind around the lattice
in temporal direction.
The acceptance rate for the flip of such clusters is proportional to $\exp(-\beta h n_w)$,
where $n_w$ is the temporal winding number of the cluster.
At sufficiently {\em large values of $\beta h$} 
the acceptance rate becomes very small on average and results in an 
\ul{\em exponential slowing down} of the simulation.
This was indeed observed \cite{ZhangKCG95,KashurnikovPST99,DorneichT01}.
For such large  diagonal fields, 
the  ``worm-algorithm'',
and especially the general method of ``directed loops'',
discussed in section \ref{Related}, are much better suited.
When comparing computational effort for equal error bars 
for one- and two-dimensional Heisenberg systems, 
the threshold in comparison to SSE worms (see section \ref{SSEoperupdate}) 
was $\beta h \gsim 3$  \cite{DorneichT01}.
When comparing to the worm-algorithm for a Heisenberg-chain,
the threshold was at $\beta h \gsim 5$ \cite{KashurnikovPST99}.

On the other hand, at small fields $\beta h \lsim 1$ it appears to be advantageous to
use the loop algorithm with global acceptance step \cite{DorneichT01}
to achieve equal errors with smaller computational effort.

To minimize the acceptance problem, one should normally choose $A_{global}$
such that its fluctuations are minimized. 
Somewhat surprisingly, it has also been reported \cite{OnishiNKM99} that supplementing the
cluster updates with explicit global flips of worldlines helps considerably,
even at rather larger values of $\beta h$. 
An interesting variant of the loop algorithm which appears to work well for very large diagonal fields
was introduced by Syljuasen \cite{Syljuasen00}.

Let us  note that
the \ul{\em canonical ensemble} (or ensemble of constant magnetization) is simulated by disallowing
the flip of clusters which would change the number of worldlines.
It is also possible, and can reduce autocorrelations, to allow the number of worldlines to fluctuate, 
and afterwards treat each subset of configurations with a certain number of worldlines
as a canonical simulation \cite{RoosM99,NishinoORYM99}.

 \subsection{Transverse field}\label{TransverseField}
\newcommand{\Sx}{$\hat{S}^x$}
There is a different way to treat magnetic fields and  chemical potentials,
which can be simulated without difficulties.
One can change the axis of quantization
to turn a magnetic field into the x-direction.
The field operator 
 $\hat{S}^x = \frac{1}{2}  \;(\!\begin{array}{cc} {\scriptstyle 0 \, 1} \\[-1ex] {\scriptstyle 1 \, 0} \end{array} \!)$
then flips the direction of a spin.
It can act at any spacetime point $(il)$.
Worldlines now become piecewise continuous,
with spin flips whereever \Sx\ acts.
For the isotropic Heisenberg model, the physics is invariant under this rotation.

Rieger and Kawashima \cite{RiegerK99}
have shown how to treat such source operators in a 
cluster algorithm%
\footnote{Note that with sources, local updates allow the change of global quantities and 
          become ergodic too. See also section \ref{Related}.}.
They investigated the Ising model in a transverse field, for which $J_x=0$
and all worldlines are thus straight.
Here we apply their method to the loop algorithm.

We enlarge our graph to consist not only of breakups but also of the information,
for each spacetime site, whether \Sx\ acts there or not.
\Sx\ occurs in matrix elements like
$ W_h \ident \langle + | \exp{(\Delta\tau h \hat{S}^x)} |-\rangle = \sinh{(\frac{\Delta\tau}{2} h)}$.
Its presence or absence in a graph can be treated with Metropolis probability%
\footnote{I.e.\ {\em propose} creation of new operators with $p=W_h$, 
          and  deletion of existing ones with $p=1$.}%
: Tentatively put additional operators \Sx\ with probability%
\footnote{In continuous time \cite{RiegerK99} this become constant probability $W_h/\Delta\tau$ per time.}
$p=W_h$ 
on any lattice site where there is no \Sx\ yet. 
After choosing breakups as usual we now have {\em segments} of loops 
between successive occurences of sources \Sx. 
Flip each segment%
\footnote{When there is freezing, we similarly have to flip subgraphs of clusters.}
with probability \half\ 
(modified with any remaining external weight $A_{global}$).
Finally, remove the sources between segments of equal orientation.
This completes one update.
A generic framework for such tentative updates has been provided in ref.\ \cite{EvertzEV01}.

Now the magnetic field (chemical potential) can be of arbitrary size,
without acceptance problems (section \ref{asymmetries}).
We can measure off-diagonal Greens functions  in \Sx\ 
directly in the worldline representation, as well as in the loop representation (section \ref{Offdiagonal}).
When the original model is not isotropic, one can still rotate the magnetic field
into  the $x$ direction, obtaining an anisotropic XYZ model (see end of section \ref{Setup}).
The loop algorithm for this model (section \ref{ProbXXZ})
involves breakups for which the arrow direction changes on both loop segment, 
which is equivalent to the original breakups plus two sources.

The procedure just described works in the ferromagnetic case.
For an {\em antiferromagnet}, we have to perform a rotation
$\hat{S}^{x,y} \rightarrow -\hat{S}^{x,y}$
on one sublattice to achieve positive matrix elements in \eq{matrixelements}.
This minus sign stays with the source operators \Sx, multiplies
the configurational weight $W(\S)$%
\footnote{I thank Matthias Troyer for this observation.}
and results in a sign problem.
Chandrasekharan, Scarlet, and Wiese \cite{ChandrasekharanSW99,CoxGHSW99} 
showed that this sign problem
can be removed completely with the meron approach described in section \ref{Merons}.
To include sources, they use a method similar to that of Rieger and Kawashima.
They introduce additional time slices and employ the Swendsen-Wang
algorithm \cite{SwendsenW87,Sokal92} to set or unset bond-variables (as part of loops)
which force equal spins along the additional timelike lattice bonds. This subdivides
the loop-clusters and is equivalent to the Metropolis procedure described above.
They show that for large fields, the algorithm with transverse fields 
performs orders of magnitude better than the original one with diagonal fields.

The strategy described here, namely to include operators \Sx\ stochastically, resembles their occurence
in the stochastic series expansion (section \ref{operSSE}). 
Indeed, one can also separate other operators from the Hamiltonian and include
them stochastically in the same manner as \Sx.
(See section \ref{operfreeze}).

 \subsection{Higher Spin representations}\label{HigherSpins}
The loop algorithm for spin models has so far been formulated for the
spin-\half\ case.
One way to extend it to higher spin representations would be \cite{EvertzLM93}
to use the corresponding vertex representation (19-vertex model for spin-1)
and to try the same formalism as for spin-\half.

Kawashima and Gubernatis have successfully employed a different approach \cite{KawashimaG95a,KawashimaG94}.
They write higher spin representations as a product of spin-\half\ representations,
with a projection operator onto the proper total spin.
They arrive at new ``shaded plaquettes'', 
between the different spin-\half\ representations at each space-time site.
Locally on each plaquette, the situation looks again like a six- or eight- vertex model.
By the same approach, Kawashima also treated the anisotropic XYZ case for general spins
\cite{Kawashima95}.
(Here the number of different graphs quickly proliferates).
This generalization was successfully tested on an antiferromagnetic Heisenberg chain
with $S=1$, finding complete removal of autocorrelations \cite{KawashimaG94}.

Since the Hamiltonian commutes with the projection operator, 
the procedure can be simplified \cite{HaradaTK98,TodoKT98,TodoK01}
by projecting onto total spin $S$ only at time zero.
For general spin $S$ \cite{TodoKT98,TodoK01} there are again 
$2S$ spin-variables per site.
Breakups with the usual probabilites (section \ref{ProbXXZ})
are allowed between any pair of spin variables from neighboring sites.
Projection is achieved at time zero by symmetrizing over permutations 
of the worldline variables at each site, subject to worldline continuity.
This procedure immediately generalizes to continuous time 
(section \ref{ContTime}) \cite{HaradaTK98,TodoKT98}.

Harada and Kawashima \cite{HaradaK02} recently improved this approach in the framework of directed loops 
by adjusting the algorithm to reflect a stochastic mapping 
from the space of $2S$ spin-\half\ variables
to a single variable $S^z\in \{1\dots 2S+1\}$ (see section \ref{DirectedLoops}).

 \subsection{Off-diagonal operators}\label{Offdiagonal}
\newcommand{\spsm}{\hat{S}_i^+ \hat{S}_j^-}
\newcommand{\avspsm}{$\langle \spsm \rangle$}
Brower, Chandrasekharan, and Wiese showed \cite{BrowerCW98} 
that one can obtain  off-diagonal greens functions,
especially two-point functions, 
by measuring properties of the loop clusters, i.e.\ improved estimators,
instead of properties of the worldline configurations.
We first discuss the off-diagonal two-point function \avspsm\ between spacetime sites $i,j$.
It corresponds to a configuration with an additional propagator from $i$ to $j$,
i.e.\ with a partial worldline. Such a configuration never occurs during the simulation.
Thus, in a regular worldline Monte-Carlo with continuous worldlines, 
off-diagonal Greens functions cannot be measured at all%
\footnote{One can measure equal-time correlation functions by introducing 
          open b.c.\ on one timeslice \cite{HirschSSB82}.}.
Instead, one would have to perform a separate simulation with fixed sources for each pair of sites $(i,j)$,
and somehow measure and adjust the normalization 
$Z_{\scr{sources}\, i,j}/Z_{\scr{no sources}}$. 
This is not feasible with the usual worldline method.
Note, however, that off-diagonal two-point functions can be measured 
directly in the extended ensemble of worm methods (section \ref{Related}),
both with imaginary time and with SSE.
They are also directly accessible when there are transverse fields 
(section \ref{TransverseField}).

In the loop algorithm, \avspsm\ is easy to measure even without transverse fields.
Let us look at a given configuration of loops.
If spacetime sites $i$ and $j$ are connected by a loop, we could  
flip the partial loop from $i$ to $j$ (or from $j$ to $i$, depending on the current spins at $i,j$)
to obtain the desired propagator, and thus a contribution to \avspsm.
Note that we do not actually need to perform such a partial loop flip as a Monte Carlo update,
we can just virtually do it to perform the measurement as an improved estimator (section \ref{ImprEst}).
When there is no external asymmetry $A_{global}$, the proper flip probability 
for use in the improved estimator is $\Half$ for the partial loop,
and the improved estimator for $\spsm$ is 
\beq{spsm}
 4(\spsm)_{impr} = \left\{ \begin{array}{lll}
                                 1\cdot\phi_{ij}, & \mbox{\ \ if the sites $i$,$j$  are on the same loop},\\
                                 0,               & \mbox{\ \ otherwise.}
                            \end{array} \right.      
\eeq
(For the antiferromagnet a phase factor $\phi_{ij}=(-1)^{|{\bf r}_i - {\bf r}_j|}$
appears due to the sublattice rotation (footnote \ref{bipartite}); $\phi_{ij}=1$ otherwise.)
This resembles the improved estimator \eq{Oimpr3} for the diagonal correlation function.
There the criterion for a finite contribution is that  the two sites are located 
on the same {\em cluster}, 
whereas here it is the same {\em loop}.
For uniform correlations of the isotropic Heisenberg ferromagnet 
and for staggered correlations of the antiferromagnet,
the improved estimators \eq{Oimpr3} and \eq{spsm} are identical, directly reflecting spin rotation invariance.
Note that by using the improved estimator, one can {\em simultaneously} measure $\spsm$ 
in a given cluster configuration for all pairs of sites $i,j$.
When there is an external weight $A_{global}$, the improved estimator \eq{spsm} has to be modified,
as described in section \ref{ImprEst}.

Brower, Chandrasekharan, and Wiese \cite{BrowerCW98} show that improved estimators
exist for general Greens functions, by developing an {operator} description 
of loop clusters.
In their formulation, each specific plaquette breakup 
is written as a local transfer-matrix operator $\cal T$
that evolves worldline spins across the plaquette in time direction.
The operator is specified by its matrix elements, which are linear combinations of 
Kronecker-deltas for the spin values, reflecting the specific breakup.
For an explicit representation in terms of spin operators, see section \ref{oper}.
The important point here is the {existence} of such an operator.
A given graph $G$ then corresponds to an operator $\hat{M}_G$,
with $\tr_{spins} \hat{M}_G = 2^{(\scr{\em number of clusters in } G)}$ for the fieldless case.
The partition function is  
 $ Z = \sum_G \,w(G)\, \tr \hat{M}_G$ 
(see \eq{ZGraph}), where $w(G)$ is the product over plaquettes 
of the breakup-weights $w^{ij}$ occuring in $G$.
In this operator language in graph space, we can now also write expectation values:
\beq{OpExp}
 \langle \hat{O} \rangle = \frac{1}{Z} \sum_G w(G) \tr\left(\hat{O}\hat{M}_G\right)
                         = \langle \frac{\tr\hat{O}\hat{M}_G}{\tr\hat{M}_G}\rangle_{Monte Carlo},
\eeq
where the trace is over spin configurations, and the last average is taken
over the cluster configurations
generated in a Monte Carlo simulation.
For the two-point Greens function, $\hat{O}$ consists of two source terms, and we get 
the improved estimator \eq{spsm}. 
Similar results obtain for general n-point Greens functions, where now contributions
can come from more than one cluster.
A four-point function $\langle S^+_i S^-_j S^+_k S^-_l \rangle$, for example,
gets contributions from   sites $i,j,k,l$  which are (1) all on the same loop,
or (2) on two different loops, with pairs of sites $(i,j),(j,k)$ or $(i,l),(j,k)$ 
being on the same loop.
For some further details we refer to refs.\ \cite{BrowerCW98,AlvarezG00}.
We also note, that ostensibly four-point properties
can sometimes be reduced to two-point functions, as shown for the Drude-weight 
in \cite{AlvarezG00,AlvarezG02}.

 \subsection{Fermionic Models}\label{FermionicModels}
%
Fermionic models can be treated in the same way as hard core bosons \cite{Scalettar-WL99},
with the addition of a fermionic sign for each permutation of worldlines \cite{HirschSSB82}. 
The Monte Carlo simulation is performed with the modulus $|W(\S)|$ of the configuration weight,
and observables are determined as 
\beq{signobs}
  \langle \O \rangle_W = \frac {\langle \O \; \mbox{\em sign}\rangle_{|W|}}{\langle \mbox{\em sign}\rangle_{|W|}} ,
\eeq
where {\em sign} is the sign of $W(\S)$. 
Since%
\footnote{ The exponential form results since negative contributions to $\langle \mbox{\em sign} \rangle$
           originate in finite spacetime regions more or less independently,
           from negative matrix elements and/or winding of worldlines.
           More formally \cite{ChandrasekharanW99},
           $\langle \mbox{\em sign} \rangle = \exp{(-\beta V \Delta f)}$ 
           corresponds physically to the difference $\Delta f$ in free energy between the fermionic model
           and the ``bosonic'' model $|W|$, whose nature depends on the quantization and weight function $W$
           that was chosen.
         }
 $\langle \mbox{\em sign} \rangle \sim \exp{(-\mbox{\em const} \cdot \beta \cdot\mbox{\em Volume})}$,
simulations have so far been restricted mostly to one-dimensional unfrustrated models
and to small or very low-doping higher-dimensional or frustrated systems.
The identical sign problem also occurs for fermion simulations with the loop algorithm 
(first performed in ref.\ \cite{Wiese93}).
By clever use of improved estimators,  a solution of the sign problem 
was found for a  restricted class of models (see section  \ref{Merons}).
This class does, however, 
not so far include the standard Hubbard model nor the \tJ model.
For these models, several generalizations of the loop algorithm have been developed:

\EMM{The Hubbard Model}
can be viewed as consisting of two systems of tight binding fermions, each mapping
to an XXZ-model (plus fermion sign), coupled by the Hubbard interaction
$U \sum_i n_i^\uparrow n_i^\downarrow$.
It can therefore immediately be simulated by employing a loop algorithm for
each of the XXZ models, and taking the Hubbard interaction
as well as the chemical potential $\mu \sum_i (n_i^\uparrow + n_i^\downarrow)$  
into the global weight $A_{global}$, \eq{WS}.
However, for large values of $|U|$ this procedure is not very efficient,
since the global weight will fluctuate too strongly,
resulting in small acceptance rates, especially for the flips of large loops.

Kawashima, Gubernatis, and Evertz \cite{KawashimaGE94} therefore added an additional
new type of loop-update, called loop-exchange, which flips between spin-up and spin-down,
leaving unoccupied sites in the worldline lattice unchanged.
These loops move upwards in time on spin-up sites, and downwards on spin-down sites.
The breakup probabilites can be constructed with the formalism of section \ref{algorithm}.
The loops were chosen to change direction (i.e.\ use a horizontal breakup) when spin-up
and spin-down worldlines meet. 
Flips of these loops  are not affected by the Hubbard interaction nor
by the chemical potential, and can therefore always be accepted.

In the 1D Hubbard model
this additional type of loop updates eliminated all remaining autocorrelations in the
Hubbard simulations. 
An example is shown in \fig{figHubbard}.%
%
\newfigh{figHubbard}{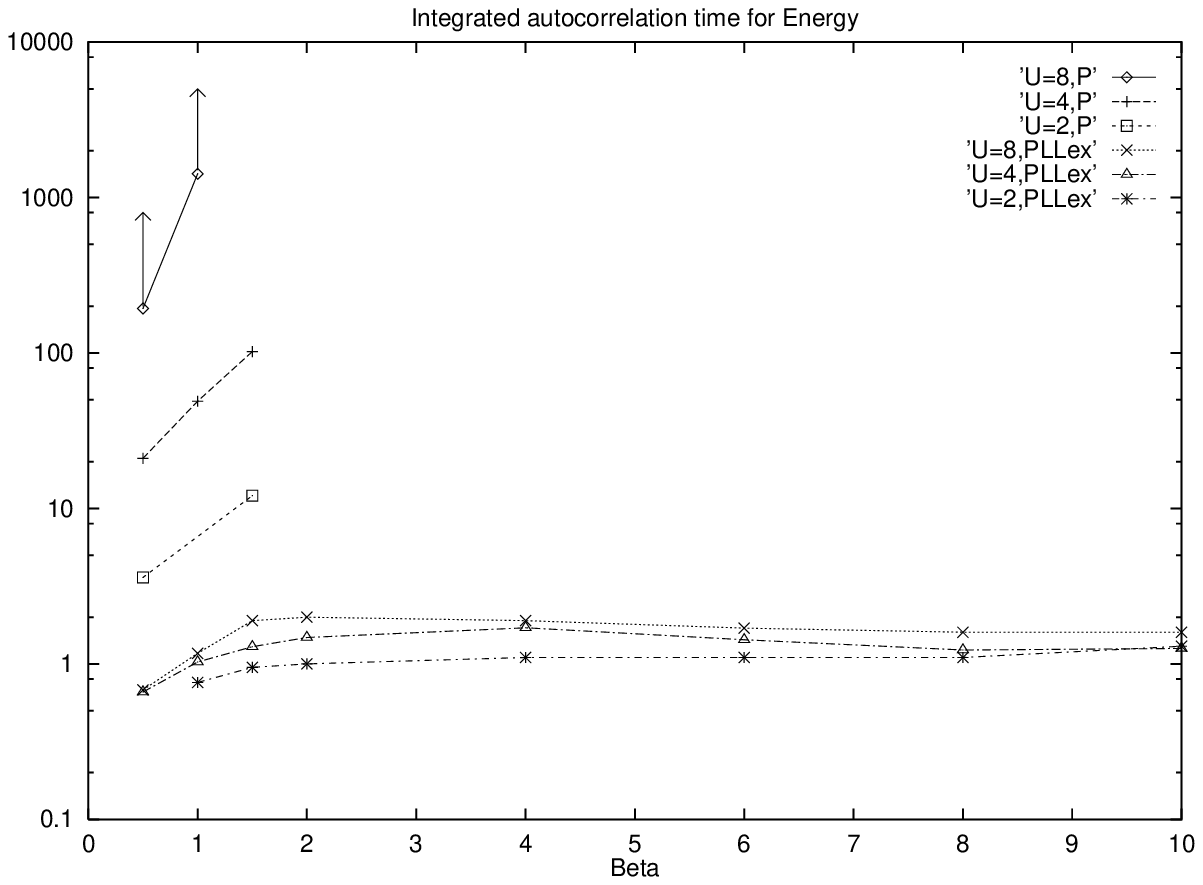}{7cm}{
Comparison of the loop algorithm and local updates for the 1D Hubbard model (32 sites),
adapted from ref.\ \cite{KawashimaGE94}.
The figure shows the integrated autocorrelation time $\tau^E_{int}$ 
(which is proportional to the computer time required for a given accuracy)
for the total energy E, on a logarithmic scale.
The autocorrelation times for other quantities behave similarly.
The upper three curves are for local updates, the others for loop updates
including loop exchange. In both cases, $U=2$, $4$, and $8$, with autocorrelation times
increasing with growing $U$.
With {\em local updates}  the autocorrelation times are very large
already at small $\beta$, and grow rapidly when $\beta$ or $U$ are increased
(roughly consistent with $\tau\sim \beta^z U^x$ with $z\lsim 2$ and $x\gsim 3$).
For $U=8$ these simulations did not converge, and only lower bounds for $\tau$ are shown.
At $\beta\gsim 1.5$ none of the simulations with local updates converged.
With  {\em loop updates} the situation improves drastically. All autocorrelation 
times are less than about $2$.
Thus at large $\beta$ and $U$ several (likely many) orders of magnitude
in computer time are saved. 
}%
%
For the loop updates, the autocorrelation times remain smaller than $2$ at all temperatures.
No slowing down is visible at all.
For local updates, on the other hand, the autocorrelation times in \fig{figHubbard}
show the expected rapid increase
(see appendix \ref{CSD}), consistent with $t\sim \beta^{\mbox{\normalsize $z$}}$.
They are orders of magnitude larger than the loop-autocorrelation times,
already at small $\beta$ and even for the energy as an observable,
which as a locally defined quantity is expected to converge relatively fast
in a local algorithm.
Beyond $\beta\gsim 1.5$, the local Monte Carlo did not converge anymore.
The autocorrelation times are expected to continue to grow.
Note that the autocorrelation times for the local algorithm will additionally grow 
like $1/(\Delta\tau)^2$ for improved Trotter discretization,
whereas the loop algorithm does not suffer from this effect, and can moreover
be implemented directly in continuous time.

A similar loop-update method was independently developed 
and recently employed by Sengupta, Sandvik, and Campbell \cite{SenguptaSC02}
within the SSE formalism for  a Hubbard model including nearest neighbor Coulomb repulsion.
The authors also show how to apply the parallel tempering technique to such simulations.

\EMM{For the \tJ model}, a number of  generalizations of the loop algorithm have 
been developed
\cite{Kawashima96,AmmonEKTF98,BrunnerM98,BrunnerAM99}.
Here we have three kinds of site-occupation: spin-up, spin-down, and empty.
The model can be simulated by a divide and conquer strategy \cite{Kawashima96,AmmonEKTF98}, 
using three different types of loop updates.
In the first update, the empty sites in the worldline configuration 
are left untouched. The remaining sites with spin-up or spin-down can be updated
with a loop algorithm very similar to that for the spin-\half\ Heisenberg antiferromagnet.
In the second update, sites with spin-up are left untouched, and updates between
spin-down and empty sites are done with a loop algorithm (which now looks 
similar to that for hard core particles). 
In the third update, loops on spin-up and empty sites are constructed.
All three kinds of loops fall within the XXZ case discussed in section \ref{ProbXXZ}.
The combined algorithm is indeed a working cluster algorithm for the \tJ model,
ergodic, removing autocorrelations, and providing improved estimators which further
reduce statistical errors. 
It was successfully 
used in ref.\ \cite{AmmonEKTF98} on single \tJ chains and on two and three coupled \tJ chains.
In ref.\ \cite{VebericPE00} it was used to study thermodynamic and diamagnetic properties of the square lattice \tJ model
with up to 2 holes, including cases with spin-anisotropy.

Brunner and Muramatsu \cite{BrunnerM98} use another representation of the \tJ model, in which only
holes are fermionic. They appear in bilinear form and can only live on spin-down sites.
In this representation, two kinds of loop updates therefore suffice.
The authors study the stability of the Nagaoka state on two-dimensional lattices of size
up to $10\times 10$ at small $J$, with up to two holes, reaching temperatures
down to $\beta t=2500$ for a single hole  and $\beta t=150$ for 2 holes at $J=0$,
much lower than previously attainable with other methods.

Brunner, Assaad, and Muramatsu developed a different 
very interesting method \cite{BrunnerAM99}.
In the same representation, they simulate just the spin degrees of freedom (pure Heisenberg model),
and can then calculate dynamical properties of a single hole in imaginary time 
by {\em exactly} integrating out the bilinear fermion degrees of freedom. 
The method has been applied to several 1- and 2-dimensional cases \cite{BrunnerAM99,BrunnerAM00,BrunnerCAM01}.

\EMM{For relativistic gauge theories},  Wiese et al.\ \cite{Wiese98} 
have developed the so-called quan\-tum-link formalism. 
They employ an additional artificial ``time'' direction,
for which they can  apply variants of the loop algorithm,
e.g.\ for $U(1)$ gauge theory, aiming at efficient simulations of fermionic lattice QCD.

In a recent different approach to the sign problem \cite{Lee02}, Lee observed that
the typical distance which fermions ``walk'' is only of order $\sqrt{\beta t}$.
In this approach fermion permutations are restricted to finite zones,
and results extrapolated in the size of these zones.

 \subsection{Fermion simulations without sign problem: The meron method}\label{Merons}
Chandrasekharan and Wiese \cite{ChandrasekharanW99} discovered an exciting possibility 
to overcome the fermion sign problem for a restricted class of models,
by clever use of improved estimators.
They treat   the fermionic version of the
spin \half\ antiferromagnetic Heisenberg model, i.e. spinless fermions with 
nearest neighbor repulsion $V = 2t$, with Hamiltonian
\beq{HV}
 H=-t \sum_{\langle ij\rangle} (c^+_i c_j + c^+_j c_i) 
    +V \sum_{\langle ij\rangle} (\hat{n}_i-\frac{1}{2})(\hat{n}_j-\frac{1}{2}), 
\eeq
and the usual relations 
$J_{x,y} =2t$, $J_z=V$,  $\hat{n_i}-\frac{1}{2}=\hat{S}_i^z$, and  $n_i=0,1$
between particle language and spin language.

Then there are only vertical and horizontal breakups (section \ref{Recipe}),
no diagonal ones.
In this case one can write down an improved estimator for the fermion sign
in which the contribution from each cluster is {\em independent}.
Namely, the fermion sign of the worldline configuration changes by a {\em factor}
$ (-1)^{1+n_w+n_h/2} $ when flipping a cluster,
where $n_w$ is the temporal winding number of the cluster,
and $n_h$ is the number of horizontal moves in the cluster.
They denote by {\em ``meron''} a cluster for which this factor is $-1$.
Without chemical potential, each cluster can flip independently with propability \half.
Each meron then contributes a factor $((+1) + (-1))/2 =0$ to the improved estimator for the sign,
which is therefore  zero whenever there is at least one meron.
When there is no meron, the improved estimator is $+1$, since for the Heisenberg antiferromagnet
{\em any} allowed loop configuration can be constructed from the 
reference staggered worldline configuration
(see section \ref{ImprEst}), which has $\mbox{\em sign}=+1$, 
and since without merons, all configurations
contributing to the improved estimator have the same sign.
Thus,
\beq{signimpr}
 (\mbox{\em sign})_{impr} = \left\{  \begin{array}{lll}
                               1, & n_{meron}=0 \\ 
                               0, & n_{meron}>0 
                                     \end{array} \right.
\eeq
has no negative contributions. In a standard simulation, it is however dominated by contributions 0,
and still has an exponentially bad variance.

Suppose  we want to measure staggered two-point functions like 
$\O = (-1)^{i-j} (n_i-\frac{1}{2})(n_j-\frac{1}{2})$.
To use \eq{signobs}, we determine the improved estimator \eq{Oimpr3} for 
$(\O \cdot \mbox{\em sign})$.
We can evaluate it by using the reference configuration.
It is equal to $\frac{1}{4}$ in two cases: 
(i) when there are no merons and $i$ and $j$ are on the same loop;
(ii)when there are two merons and $i$ is located on one, $j$ on the other meron;
otherwise it is zero.
\newcommand{\vol}{\scr{Vol}}
For the susceptibility 
    $\chi = \frac{\beta}{\vol} \; \langle \hat{O}^2 \rangle $
of the staggered occupation
       $\hat{O} = \sum_{\bf i} (-1)^{\bf i} (\hat{n}_i - \frac{1}{2})$
%
%
%
we use \eq{ChisImpr} for the improved estimator and get 
\beq{chimeron}
 \langle \chi \rangle =\; \frac{\langle (\chi \cdot \mbox{\em sign})_{impr} \rangle}{\langle \mbox{(\em sign)}_{impr} \rangle}
                    \;=\; \frac{\beta}{4 M^2 \mbox{Vol}}
                          \frac{\langle \delta_{n_{meron},0}  \sum_{\scr{(clusters c)}} \, |c|^2 
                                        + 2\delta_{n_{meron},2}\, |c_1|\,|c_2|\rangle}
                               {\langle \delta_{n_{meron},0} \rangle}
\eeq
which has contributions only from the zero- and two-meron sector.
Here the meron-contribution obtains a factor $2$ since sites $i$ and $j$ can be on different merons two ways.

Chandrasekharan and Wiese point out that one can therefore restrict simulations
to $n_{meron}\le 2$, thereby drastically reducing the visited phase space.
This is achieved by starting with the reference worldline configuration
and with completely vertical breakups.
The breakups are then updated with the usual probabilities {\em plaquette by plaquette}, 
subject to the additional constraint that the total number of merons in a cluster configuration can at most be 2.
Each plaqutte update requires that one follows the partial loops attached to that
plaquette to determine their meron nature.
Since the typical loop size increases with growing spatial and temporal correlation 
length, this step increases the computational cost, but at most by a factor
proportional to space-time volume%
\footnote{This cost has been reduced to logarithmic by a binary tree search \cite{Osborn00}
}.%
The resulting loop clusters are flipped (e.g.) after all plaquettes have been updated,
to get to a new worldline configuration (the sign of which is irrelevant).

The denominator in \eq{chimeron} can be viewed as the remaining effect of the
fermionic sign. This denominator would up to now still be very small.
This can be avoided by {\em reweighting} the simulation so that roughly half of the
generated cluster configurations have zero merons. The corresponding reweighting factor
has to be cancelled when computing expectation values \cite{ChandrasekharanW99}.
All n-point functions can be computed similarly by including $O(n)$ merons.
Overall, the total computational effort per sweep is proportional to between $V_t$ and $V_t^2$,
where $V_t$ is the spacetime volume, with strictly positive estimators and 
no exponential sign problem remaining.
The sign problem for this model is thus solved !

For the meron approach to work we need a multi-cluster method with   two stringent  conditions.
Loop-flips must be independent in their effect on the worldline-sign.
For the $tV$-model this means that there cannot be any diagonal breakups, 
so we need $V\ge 2t$; 
and there must be a reference worldline configuration in the sense of section \ref{ImprEst},
with {\em positive} worldline sign.
We also need to ensure ergodicity.

These restrictions unfortunately exclude most interesting models at present,
like the standard Hubbard and \tJ models.
Nevertheless, valuable progress has been made.
Chandrasekharan \cite{Chandrasekharan99} developed a meron method for free spinless fermions 
at a large chemical potential. In $d=2$, this implies a filling of $2\%$ at $\beta=3$, becoming less
with larger $\beta$. This method uses only free fermions.
Chandrasekharan and Osborn \cite{Osborn00,ChandrasekharanO01,Chandrasekharan01a} 
succeeded in a meron method for a non-standard  type of Hubbard model,
with unusual correlated hopping,
constructed by requiring that the clusters for up- and down-spins should be identical;
then the reference configuration automatically has a positive sign.
They studied an attractive Hubbard model of this class  on lattices as large as $128\times 128$
and find critical behaviour that is well described by a Kosterlitz-Thouless transition.
The method is also available for repulsive models with limited chemical potential.
For a review, see ref.\ \cite{ChandrasekharanCOW02}.

The meron approach can also be applied to other sign problems.
Chandrasekharan, Scarlet, and Wiese \cite{ChandrasekharanSW99} used it 
successfully for antiferromagnetic Heisenberg 4-leg ladders
in a transverse uniform magnetic field (see section \ref{TransverseField}).
The same technique was used for the $tV$ model in a staggered field \cite{Chandrasekharan00,ChandrasekharanO00a}.
Henelius and Sandvik succeeded in applying the meron technique to a 
``semi-frustrated'' Heisenberg quantum spin system.
They also discuss details of the implementation and optimal reweighting.

 \section{Related Methods}\label{Related}
%
The main limitation of the loop algorithm is its exponential slowing down
for diagonal magnetic fields (or chemical potentials) $\beta h\gsim 3$.
This problem can be overcome by ``worm-like'' methods,
which extend phase space by source operators
and perform {\em local} updates which eventually combine to a permissible 
update in the original ensemble.
We first describe the original worm algorithm,
then the so-called 'operator-loop updates' in SSE,
for which one limiting case are ``deterministic operator loops''.
Recently, the very capable worm-like method
of ``directed loops'' has been constructed.
In limiting  cases it becomes similar to
single-loop construction in the loop algorithm.
It is briefly described in section \ref{DirectedLoops}.

 \subsection{Worm Algorithm}\label{Worms}

In the standard worldline formulation with local updates it is almost impossible to
compute single particle Greens functions like e.g.\ 
$\langle a(x_0,t_0) \, a^\dagger(x_1,t_1)\rangle$ (where $a^\dagger, a$ are creation
and annihilation operators).
In order to do so, one would have to introduce sources at $(x_0,t_0)$
and $(x_1,t_1)$ explicitely, 
with a partial worldline between these two points,
and to perform a separate simulation for each such 
pair of coordinates (see also section \ref{Offdiagonal}).

A very elegant solution to this problem was  provided by 
Prokof'ev, Svistunov, and  Tupitsyn \cite{ProkofevST98b}.
Their method can be viewed from the perspective of single loop construction.
{\em During} that construction, there is a partial loop with two open ends.
Flipping this partial loop would result in a partial worldline, i.e.\ a propagator 
between two sources, just as desired.
Thus every step in a single-loop construction can be taken to provide
a configuration for the  measurement of Greens functions (see also section \ref{Offdiagonal}).

Prokof'ev et al.\ turn this observation around and explicitely construct a single 
propagator with two ends (``Worm'') 
in continuous time, related to the way  a single loop would be constructed.
The Monte Carlo moves are thus {\em local} in space and in time 
(within a constant neighborhood, see section \ref{ContTime}).
Each local step  provides a new configuration to the measurement of Greens functions.
When the sources meet and annihilate (equivalent to the closing of a single loop),
contact is made to the sourceless partition function, thus providing the
correct normalization. 
Prokof'ev et al. supplement these moves by additional moves corresponding to the flip of 
small closed loops, in order to make the simulation faster.
The Worm-algorithm is available for any spin-magnitude. 
It is ergodic in the same way that the loop algorithm is.
It is expected to have a small dynamical critical exponent \cite{KashurnikovPST99};
On a one-dimensional chain, $z$ was measured to be close to zero \cite{KashurnikovPST99}.

A very important advantage of the {\em local} updates is
that {\em all} interactions in the Hamiltonian, like e.g.\ magnetic fields,
can be taken into account in each step, without encountering prohibitively small acceptance rates.
This is in contrast to the loop algorithm itself, which has to put unsuitable interactions
into the global weight $A_{global}$
(see section \ref{asymmetries}).

Prokof'ev et al.\ 
have applied the method very successfully to the 1D Bose Hubbard model 
\cite{ProkofevST98b,ProkofevS98b,KashurnikovKS98} 
with soft core bosons%
, also with disorder, 
to the 2d \tJ model with a single hole \cite{MishchenkoPS01},
to the 3d Hubbard model in an optical trap \cite{KashurnikovPS02},
as well as to  other problems including
\cite{GoeppertGPS98,ProkofevS98a,KaganKKPS98,ProkofevS99,KashurnikovMTK99}.

The worm-method has also been adapted to classical spin models \cite{ProkofevS01}.

 \subsection{SSE with ``operator loop updates''}\label{SSEoperupdate}
Sandvik developed a worm-like method called
``operator loop updates'' \cite{Sandvik99b} within
the stochastic series expansion.
It can be seen as a special case of the new directed loops discussed below,
for which the new standard  solution is  still more efficient.
It has been used to study a variety of models with fields and other asymmetries,
like two-dimensional hardcore bosons with chemical potential and/or next-near neighbor repulsion
\cite{HebertBSSTD01,SchmidTTD01,BernadetBMSTD01,BernadetBTD02}
and numerous others
\cite{SandvikC99,WesselH01,WesselOH01,Sandvik00,WesselNSH00,Sandvik99a,Sandvik02,%
      Sandvik01a,ShevchenkoSS00,HeneliusS00,SenguptaSC02}.

In the special case of $h=0$ and $\Delta=\pm 1$, operator loop updates become
``deterministic'': Once a set of vertices has been constructed in the first 
update step of the SSE procedure (see section \ref{LoopSSE}),
each vertex then becomes a horizontal (resp.\ diagonal) breakup with probability 1, so that 
the loop-configuration is already fixed.
This method is similar to (but not quite the same as) the loop algorithm 
reviewed here. 
It has been used in a sizeable number of studies mentioned in section \ref{Applications}.

 \subsection{Directed Loops}\label{DirectedLoops}

Syljuasen and Sandvik recently developed the worm-like method of
directed loops, both in SSE and in continuous time \cite{SyljuasenS02},
for very general Hamiltonians.
Like in the original worm method, a single ``loop''
is constructed by propagating a source-term (``loop-head'')
until it meets the second source.
Updates are performed locally, and spins are flipped immediately.
The source term can be a creation or an annihilation operator.

At each vertex (in the language of SSE) which the loop-head enters
through some entry-leg,
there are \`a priori  five possibilities:
(i) The loop-head can exit through one of the other three legs.
One of these is usually forbidden, unless the Hamiltonian contains 
pair-creation/annihilation operators;
(ii) it can ``bounce'', i.e. exit through the same leg through which it entered,
thus undoing the last spin flip and starting to {backtrack},
or (iii) it can just stop,  in case the Hamiltonian contains source terms \cite{DorneichT01}.

Detailed balance is constructed in the enlarged phase space of local spin configurations
and local {\em directed} loop paths, 
similar to the extended phase space of the loop algorithm
in the Kandel-Domany formalism, section \ref{KDframework}.
Weights $W(s,l_1,l_2)$ are assigned to vertices with spin configuration $s$,
entry leg $l_1$ and exit leg $l_2$, such that
\beq{DloopsW}
  W(s) ~=~ \sum_{l_2} \, W(s,l_1,l_2)
\eeq
where $W(s)$ is the usual weight of the spin configuration $s$.
Detailed balance for an overall spin-update, after the loop closes, requires
\beq{DLDB}
p(s\rightarrow s') \,W(s) ~=~  p(s' \rightarrow s)\, W(s') ~.
\eeq
This can be ensured by demanding that the local weights are symmetric under 
reversal of the loop-path
\beq{DLDBloc}
 W(s,l_1,l_2) ~=~ W(s',l_2,l_1) ~.
\eeq
Other constraints are symmetries of the weights with respect to
spatial and temporal reflections.
A very desirable criterion is minimization of the bounce-probabilities.

For the XXZ model in a field, Syljuasen and Sandvik provide such a solution.
(See also \cite{HaradaK02}.)
Within a whole region $|\Delta| + \frac{h}{d} \le 1$ 
including finite magnetic field, they achieve zero bounce probability.
For the XY-model ($\Delta=0$) this includes all fields up to the saturation field.
Continuous imaginary time for directed loops is implemented in a similar way as in the loop algorithm.

Directed loops work very well.
Syljuasen and Sandvik show that this method
works better than the previously used worm-like ``operator loop update'' 
\cite{Sandvik99b} (not identical to the loop algorithm reviewed here)
which has larger bounce probabilities.
It shows very small autocorrelation times in zero field and
in any magnetic field for one- and two-dimensional Heisenberg models.
The more complex simulations using continuous imaginary time show  lower autocorrelation
times than those using SSE, which contain a parameter $\en$ that needs to be adjusted.
Measured critical exponents for integrated autocorrelation times are small:
$z=0$ for a bilayer Heisenberg model, $0.25$ for the 3D model, and about $0.75$ for a 1D chain.

In ref.\ \cite{SyljuasenS02} it is pointed out that directed loops differ 
from the worm algorithm of section \ref{Worms},
which show much larger autocorrelations in a magnetic field \cite{KashurnikovPST99}.

It is interesting to compare worms and loops in cases with similar move-probabilities at vertices.
At $h=0$, $|\Delta|\le 1$,
the directed loop method has the same probabilities for traversing
a vertex as the loop algorithm described in this review.
Yet the methods are not the same: When the loop-head reaches a vertex a second time,
the exit decision in a worm-method is in general independent of the previous route at this vertex,
unlike the breakup-decision of the loop algorithm.
Therefore the worm can self-overlap (even if it does not bounce),
and a set of worms does not subdivide the lattice into clusters.
(The ``deterministic operator loops''\cite{Sandvik99b}
 although originally devised as worms, are an exception.)
One consequence is that improved estimators are not available,
except those obtained from the presence of sources.
Therefore methods like the meron-method
or the infinite-lattice method, which intrinsically use improved estimators,
(i.e. the loop-representation of a model), are not available in a worm-approach.

An adaptation of the directed loop method to Spin $1$ was provided 
by Bergkvist et al \cite{BergkvistHR02}
and applied to the random bond Heisenberg chain.
Within the framework of directed loops, 
Harada and Kawashima recently developed a method to simulate
general spin $S$ systems \cite{HaradaK02}.
They stochastically map the usual system of symmetrized $2S$ spin-\half\ variables to 
``coarsened''  variables with only $2S+1$ values.
For these coarsened variables, the directed loop-probabilities were worked out 
directly, so that the original $2S$ variables are not needed,
and the simulation can be performed in a single variable.

 \section{Some Applications}\label{Applications}
%
We briefly point to some of the applications of the loop algorithm
to show in which ways it has been used in practice.
For worm-like-methods see section \ref{Related}. 
Other applications have been mentioned in previous sections.
Most recent calculations have been performed with the
continuous time loop algorithm; 
many also with the slightly different ``deterministic'' SSE operator loops
\cite{SandvikC99,WesselH01,WesselOH01,Sandvik00,WesselNSH00,Sandvik99a,Sandvik02,%
Sandvik01a,ShevchenkoSS00,HeneliusS00,SenguptaSC02} (see section \ref{SSEoperupdate}).

{\em Spin \half\ isotropic Heisenberg antiferromagnets} have been investigated in many studies.
For variations of this model, the loop algorithm has been particularly valuable.
%
It has for example allowed high precision calculations 
of the critical exponents of a  quantum critical point \cite{UedaTKSL96,TroyerKU96,TroyerIU97,TroyerI97} 
in a 2D depleted system.
System sizes up to one million sites (at $\beta J=5.5$) \cite{KimT98}
and temperatures down to $\beta J=1000$ (for 2500 sites) \cite{KatoTHKMT00} have been accessible.
No sign of critical slowing down has been reported in the calculations for this model
without magnetic fields.

The even/odd structure \cite{DagottoR96} 
and correlation lengths of spin ladders  
\cite{GrevenBW96,FrischmuthAT96,FrischmuthHSR97,SyljuasenCG97,ChandrasekharanSW99,KimB00,EvertzV01} 
and coupled ladders \cite{KimBKLESY99,TworzydloDZ98,TworzydloOvZ99},
including  quantum phase transitions  in 
three dimensionally coupled chains and ladders \cite{TroyerZU97,WesselOH01}
have been investigated in detail,
with a very extensive recent study in ref.\ \cite{JohnstonETal00b}.
%
Very precise studies of the finite size scaling of the 2D system  \cite{KimT98,BeardBGW98,KimLT97,Beard00} 
have been able to extract the asymptotic infinite lattice low temperature behavior,
with correlation lengths up to 350,000, and to test
the predictions from chiral perturbation theory. 
Similar studies have been performed for layered 2D systems \cite{YinTC98,KorotinEATK99,ShevchenkoSS00}.
Universality of the KT transition temperature in a bilayer system with
small magnetic field has been shown \cite{TroyerS98}.

Dimerized systems have been studied extensively 
\cite{JohnstonETal00a,OnishiM00,MatsumotoYTT01,NakamuraT01,NakamuraT02}.
Random systems have become accessible at sufficiently low temperatures and large
system sizes, including the necessary averaging over disorder realizations.
Bond disorder in chains \cite{FrischmuthS97,TodoKT98,FrischmuthSAT99,AmmonS99,NishinoORYM99,NishinoOYM00},
bond dilution in 2d systems \cite{Sandvik99a,YasudaTMT01a,YasudaTMT02}
and coupled layers \cite{Sandvik02},
as well as random nonmagnetic impurities in ladders \cite{IinoI96,MiyazakiTOUY97,GrevenB98},
coupled ladders \cite{ImadaI97}, and two-dimensional systems 
\cite{KatoTHKMT00,Sandvik00,TodoTK01,YasudaTHKMT01,Sandvik01a,WesselNSH00,YasudaTMT01b} 
have been investigated.
%
Frustrated models with a sign problem have been simulated by making use
of the increased precision \cite{MiyaharaTJU98,AmmonEKTF98,JohnstonETal00b}
and by removing the sign problem in a semifrustrated case\cite{HeneliusS00}.
Spin chains with quantum phonons have been investigated 
both in first \cite{OnishiM02} and second quantization \cite{SandvikC99,KuhneL99,RaabLUK01}.
Transport properties have recently become accessible in spin chains \cite{AlvarezG01,AlvarezG02}.

{\em For the quantum XY-model}, accessibility of winding number fluctuations
has allowed high precision studies of the Kosterlitz-Thouless transition
(including a model with site disorder \cite{ZhangKCG95})
via the jump in helicity \cite{HaradaK97,HaradaK98}.  

{\em With Ising anisotropy,} magnetic field driven transitions could be studied
in two and three dimensions for moderately large fields \cite{KohnoT97}.
Griffiths-McCoy singularities have become accessible for the random transverse Ising
model (section \ref{TransverseField}) in two dimensions \cite{RiegerK99,IkegamiMR98,PichYRK98}.

{\em For higher spin representations}, 
correlations in one \cite{KimGWB98,TodoK01} and two \cite {HaradaTK98,Beard00} dimensions 
and for ladders \cite{TodoMYT01} 
have been analyzed with very high precision,
including computation of the spin-$1$ Haldane gap to 5 digits \cite{TodoK01}.
The effects of bond dimerization \cite{KohnoTH98,HagiwaraNKKNST98,OnishiM01,MatsumotoYTT01}
including a new order parameter \cite{NakamuraT01,NakamuraT02},
spatially anisotropic coupling \cite{KimB00,MatsumotoYTT01},
bond disorder \cite{TodoKT98,Nishiyama98,TodoKT00}, 
spin-\half\ impurities \cite{RoosM99,WesselH01} in $S=1$ chains,
and additional biquadratic interaction \cite{HaradaK00,HaradaK01}
have been investigated,
as well as a quantum phase transition from site dilution on the square lattice \cite{KatoTHKMT00}.

The \tJ model has been simulated 
\cite{Kawashima96,AmmonEKTF98,BrunnerM98,BrunnerAM99,VebericPE00,VebericPE02,BrunnerAM00,BrunnerCAM01}
in 1 and 2 dimensions with up to 2 holes, 
examining thermodynamic, spectral, and magnetic properties,
including two-hole \tJ model simulations at previously unaccessibly low temperatures.
The Hubbard model has been simulated in 1d \cite{KawashimaGE94}
and recently in detail with nearest neighbor Coulomb interaction and SSE in 2d \cite{SenguptaSC02},
A non-standard Hubbard-like model, suitable for merons, has been  shown to have a KT transition
\cite{Osborn00,ChandrasekharanO01,Chandrasekharan01a,ChandrasekharanCOW02}.
Merons were also used to study the $tV$ model and its phase transitions
\cite{ChandrasekharanW99,ChandrasekharanCHW00,Chandrasekharan00,ChandrasekharanO00a,CoxH00}.

Other applications include 
the roughening transition in classical spin models \cite{HasenbuschP97,HasenbuschMP96},
spin-orbit coupling \cite{FrischmuthMT99,MilaFDT99},
relativistic gauge theory \cite{Wiese98} (see section \ref{FermionicModels}),
hardcore bosons with disorder \cite{ZhangKCG95},
the physics of spins in ordered and fluctuating striped systems 
\cite{TworzydloOvZ99,TworzydloDZ98,ZaanenOKNT01}
and a new method to calculate the free energy of a quantum system \cite{TroyerWA02}.

 \section{Conclusions}\label{Conclusions}
%
The loop algorithm and its generalizations have opened up exciting new opportunities.
Many of them remain to be investigated.
A summary of advantages and limitations of the loop approach has been given 
in the introduction.
For models in which it can be applied
without sign problem 
and without overly big global weights,
it offers large benefits.
Some examples were given in \fig{fig:InfLattice}, section \ref{sec:InfLattice},
and in \fig{figHubbard}, section \ref{FermionicModels}.
For models with sizeable fields, Directed Loops have become available.
Last, but not least, the mapping to a combined spin and loop model
that is the basis of the loop algorithm 
is intruiging  on the theoretical side.

\section*{Acknowledgements}
I am indebted to Mihai Marcu for his long standing friendship and collaboration.
Without him the loop algorithm would not exist.
I am grateful to W. Hanke, M. Imada, D.J. Scalapino, and W. von der Linden 
for their support and ongoing interest.
I would like to thank K. Harada, N. Kawashima, A. Sandvik, S. Todo, M. Troyer, and U.-J. Wiese,
who have played an especially important and decisive role 
in the application and generalization of the loop algorithm,
for numerous discussions,
and them as well as B. Beard, W. Koller, and D. Veberic for helpful comments 
on the first edition of this review.
I thank the Austrian Science fund for support under project FWF P15834.

\section*{Appendices} \addcontentsline{toc}{section}{Appendices}
        \def\thesubsection    {\Alph{subsection}}

\makeatletter  
       \@addtoreset{equation}{subsection}
       \renewcommand\theequation{\thesubsection.\@arabic\c@equation}
\makeatother

For correct Monte Carlo simulations it is essential,
yet often neglected, that convergence and statistical errors are properly determined.
To facilitate this task we provide a prescription.
The requirements of detailed balance and ergodicity are briefly summarized 
in Appendix \ref{MonteCarlo}.
Appendix \ref{CSD} discusses autocorrelations and their increase in 
physically interesting situation, 
which can drastically increase the necessary simulation times. 
The loop algorithm was designed to overcome this problem.
Autocorrelations, especially very large ones, can easily can be overlooked
and can be a serious problem in practice, causing (even drastically) incorrect results.
Appendix \ref{Errors} then describes how to properly ensure convergence 
and how to calculate correct error estimates.
%

\subsection{Detailed Balance and Ergodicity}\label{MonteCarlo} 
%
There are several excellent reviews of the Monte Carlo approach, 
e.g.\ in refs.\ \cite{Sokal92,Binder,LandauBinder-book00,BinderL01}.
Here we briefly summarize some properties which we need elsewhere.
The Monte Carlo procedure in classical statistical physics allows 
stochastic evaluation of expectation values
\beq{expvalue}
   \langle \O \rangle = \frac{1}{Z}\; \sum_{\S\in \{\S\}} \,\O(\S) \,W(\S)
\eeq
with respect to the partition function
$  Z= \sum_{\S\in \{\S\}}  \,W(\S) $
and the phase space $\{\S\}$, 
by generating a Markov chain of configurations
$
 \S_{(1)},\, \S_{(2)},\, \S_{(3)},\, ...
$, 
which is distributed like $W(\S)$.
Therefore one can compute $\langle \O \rangle$ from a sample of configurations
\beq{MCaverage}
 \langle \O \rangle = \lim_{n\goesto\infty} \,\frac{1}{n} \sum_{i=1}^n \O(\S_{(k+i)}) \;.
\eeq
(In practice, the first $k>\tauexp$ configurations should be discarded
 to allow ``thermalization'' into the Boltzmann distribution. See appendix \ref{Errors}.) 
A set of sufficient conditions to achieve this distribution is
\begin{itemize}
\item[(1)] {\em Detailed Balance:} The transition probability 
         $0\le p(\S_{(i)}\TO \S_{(i+1)}) \le 1$ 
         of the Markov chain satisfies
      $
          W(\S)\; p(\S\TO\SS) \,=\, W(\SS)\; p(\SS\TO \S) 
      $.
\item[(2)] {\em Ergodicity:}
      Every configuration $\S\element\{\S\}$ can be reached from every other configuration
      with finite probability in a finite number of steps.
\end{itemize}
Solutions for detailed balance are for example the Metropolis probability \cite{MetropolisRRTT53}
\beq{Metropolis}
   p(\S\TO\SS) = \max(1,\frac{W(\SS)}{W(\S)})
\eeq
and the heat bath like probability 
\beq{Heatbath}
   p(\S\TO\SS) = \frac{W(\SS)}{W(\S)+W(\SS) + \mbox{\em const}} \;.
\eeq

It is often advantageous, as it is for the loop algorithm,
to split the weight $W(\S)$ into two parts:
\beq{splitMC}
  W(\S) = W_1(\S) \,\cdot\, W_2(\S) \;.
\eeq

Let $p_{1}(\S\TO\SS)$  be a transition probability
that satisfies detailed balance
with respect to $W_{1}$.
We get a Monte Carlo procedure for $W$ by 
using $W_2$ as a ``filter'' to accept or reject $\SS$.
More precisely:
First  apply $p_1$ to propose a Markov step $\S\TO\SS$.
Then  decide with a probability $p_{accept}(\S\TO\SS)$ whether
to take $\SS$ as the next configuration in the Markov chain. Otherwise keep $\S$.
Here $p_{accept}$ only needs to satisfy detailed balance 
between $\S$ and $\SS$ with respect to $W_2$,
\beq{p2}
  W_2(\S) \;\, p_{accept}(\S\TO\SS) = W_2(\SS) \;\, p_{accept}(\SS\TO \S) \;.
\eeq
One can easily see that the overall update satisfies 
detailed balance with respect to $W$. 
For $p_{accept}$ we can for example 
choose the heatbath probability
$  p_{accept}(\S\TO\SS) = {W_2(\SS)}/({W_2(\S) + W_2(\SS)})$.
Ergodicity has to be shown separately for the overall procedure.

\subsection{Autocorrelations and Critical Slowing Down}\label{CSD}

Successive configurations $\S_{(1)},\S_{(2)},\S_{(3)},...$ in the Markov chain
of a Monte Carlo configuration are correlated.
Here we discuss the corresponding autocorrelation times, which can be extremely large.
For other treatments of this topic, 
see e.g.\ references \cite{AllenTildesley-book87,Sokal92,Binder,BinderL01}.
We follow refs.\ \cite{Sokal92} and \cite{KawashimaGE94} in slightly simplified form.
Define a Monte Carlo average from $n$ measurements 
\beq{Obar}
 \boldbar{\O} := \frac{1}{n} \,\sum_{i=1}^{n} \,\O_{(i)} 
\eeq
and define the autocorrelation function for the observable $\O$
\newcommand{\NN}{\frac{1}{n}}
\beqa{COO}
       &C_{\O\O}(t) &:=&  \langle \O_{(i)} \O_{(i+t)} \rangle
                       -\langle \O_{(i)} \rangle \langle \O_{(i+t)} \rangle \1
\approx&\boldbar{C}_{\O\O}(t) &:=&  
     \NN\sum_{i=1}^n \left\{ \left( \O_{(i)}   - \NN\sum_{i=1}^{n} \, \O_{(i  )})\right)
                          \; \left( \O_{(i+t)} - \NN\sum_{i=1}^{n} \, \O_{(i+t)} \right)\right\} \;,
\eeqa
with the normalized version
\beq{Gamma}
 \Gamma_{\O\O}(t) := \frac{C_{\O\O}(t)}{C_{\O\O}(0)}  \;.
\eeq
Typically, $\Gamma_{\O\O}(t)$ is convex and will decay exponentially at large $t$
like $e^{-|t|/\tau}$.
Define the {\em exponential autocorrelation time} for the observable $\O$
by this asymptotic decay
\beq{tauexpO}
 \tau^\O_{exp} := \limsup_{t\TO\infty} \frac{t}{-\log|\Gamma_{\O\O}(t)|} \;.
\eeq
This is the relaxation time of the slowest mode in the Monte Carlo updates which couples to $\O$.
The slowest overall mode is
\beq{tauexp}
 \tau_{exp} := \sup_{\O} \left( \tau^\O_{exp} \right) 
\eeq
and corresponds to the second largest eigenvalue of the Markov transition matrix.
(The largest eigenvalue is 1 and has the Boltzmann distribution as eigenvector).

If the $\O_{(i)}$ were statistically independent, then the error estimate of $\boldbar{\O}$
would be 
$\sigma/\sqrt{n}$ 
with 
\beq{sigma}
\sigma^2 = \frac{n}{n-1} \boldbar{C}_{\O\O}(0) \;.
\eeq
Instead, the statistical error of $\boldbar{\O}$ is controlled by the 
{\em integrated autocorrelation time}
\beq{tauint}
 \tau^\O_{int} := \frac{1}{2} \,+\, \sum_{t=1}^\infty \Gamma_{\O\O}(t)
\eeq
and becomes 
$\sigma_{int}/\sqrt{n}$ with \cite{BinderL01}
\beq{sigmaint}
 \sigma^2_{int} \,\simeq\, 2 \tau^\O_{int} \; \boldbar{C}_{\O\O}(0)  \;\;\;\;\mbox{for}\; n\gg\tau^\O_{int} \;.
\eeq
{\em Therefore a Monte Carlo run of $n$ measurements effectively contains only 
$n/(2\tau^\O_{int})$ independent samples for measuring $\langle \O\rangle$.}
If $\Gamma_{\O\O}(t)$ is a single exponential $e^{-t/\tau^\O_{exp}}$,
i.e.\ if only a single mode of the Markov transition matrix couples to $\O$,
then $\tau^\O_{int}=\tau^\O_{exp}$,
otherwise
$\tau^\O_{int}<\tau^\O_{exp}$.

In simulations of classical statistical systems, 
autocorrelation times typically grow like
\beq{grow}
 \tau^\O_{int,exp} \sim \min(L,\xi)^{z^{MC}_{int,exp}(\O)} \;,
\eeq
where $L$ is the linear size of the system, $\xi$ is the physical correlation length
in the infinite volume limit at the same couplings,
and $z^{MC}$ is called the (Monte-Carlo) dynamical critical exponent.
In general, $z^{MC}$ depends on the observable $\O$, and 
$z^{MC}_{int}(\O)\ne z^{MC}_{exp}(\O)$.
(Note that $\tau^\O_{exp}$ is a correlation time, whereas $\tau^\O_{int}$ resembles
the corresponding ``susceptibility''; they will in general have different critical behavior.)
For local unguided updates, 
including the case of local updates in the determinental formalism,
one has so far always found 
\beq{zlocal}
z^{MC,local}\,\gsim\, 2 \;.
\eeq
The intuitive reason is that changes in a configuration have to spread over a distance
$\min(L,\xi)$ in order to provide a statistically independent configuration.
With local updates, this spread resembles a random walk with step size one \cite{HohenbergH77},
which needs $r^2$ steps to travel a distance $r$.
For local overrelaxed updates (see ref.\ \cite{Sokal92} and section \ref{Worms}),
$z^{MC}$ can be as small as 1.

In nonrelativistic quantum simulations, space and imaginary time are asymmetric.
With local Monte Carlo updates one can expect 
\beq{growQ}
 \tau \,\sim\, \left\{\max\left(\; \min(L,\xi) \,,\; 
                    \frac{1}{\Delta\tau}\min(\beta,\frac{1}{\Delta}) \;\right)
                     \right\}^{\mbox{\large$z^{\scr{MC}}$}} \;,
\eeq
where now $L$ and $\xi$ are spatial lengths,
$\beta/{\Delta\tau} = L_t$ is the temporal extent of the lattice, 
$\Delta$ is the energy gap, and $1/{(\Delta\Delta\tau)}$ is the temporal correlation 
length.
Again we need to distinguish $\tau^\O_{int}$ from $\tau^\O_{exp}$,
and again $z^{MC}\,\gsim\, 2$ for local updates.

Close to phase transitions ($\xi\TO\infty$) or at low temperatures 
and small gaps ($\Delta\TO 0$),
the autocorrelation times of local algorithms, with $z^{MC}\gsim 2$, will grow very fast.
Away from a phase transition, at finite $\xi$ and $\Delta$, $\tau$ should reach a constant
value for large $L$ and $\beta$.
In addition, one needs to take the limit $\Delta\tau\TO 0$. 
For algorithms in discrete time, this results in a large factor
$1/(\Delta\tau)^{(z+1)}$ in required computer time. (We get ``$(z+1)$'' since each MC sweep 
has to update $\beta/\Delta\tau$ timeslices).
With the loop algorithm, on the other hand, critical slowing down often disappears: $z^{MC}\approx 0$,
and in continuous time, the factor $1/\Delta\tau$ is replaced by a constant of order 1.

\subsection{Convergence and Error Calculation}\label{Errors}
%
Since MC measurements are correlated (see appendix \ref{CSD}),
it is not at all trivial to calculate correct statistical errors,
or even to ensure convergence of a MC simulation (see also \fig{figHubbard}).
For error calculations, there are two good strategies in practice,
{\em binning} and {\em Jackknife} (or the related bootstrap method), which can also be combined.
Let us emphasize that to ensure convergence, 
it is \underline{\em indispensable} to begin simulations on extremely small systems
and at unproblematic parameters (e.g.\ high temperature),
and to slowly increase system size, while monitoring autocorrelations
through  binning and/or a thorough analysis of the time series and its 
autocorrelation function $\Gamma$ for all observables to be measured
(preferrably for all suspect observables).
Otherwise, if one starts on too big a system,
the simulation can all too easily be locked in some region of phase space,
{\em without} any detectable signal in measured autocorrelations,
but with possibly entirely wrong results.
In practice a very good instrument for detecting moderately long autocorrelations is
to just look at time series of various observables.
This way one can notice long time scales with far less statistics than necessary in the
analysis of autocorrelation functions.

Remarkably, it has been discovered that it is possible to perform Monte Carlo simulations with
{\em exact} convergence, generating completely independent configurations from the exact desired
Boltzmann distribution, although with rather large computational effort. 
Such an  {\em Exact Monte Carlo} was
developed as ``Coupling From The Past'' (CFTP) by Propp and Wilson \cite{ProppW96}.
An extension has  been proposed for for Swendsen-Wang-like cluster algorithms by M. Huber 
\cite{ProppW96},
but appears to be too slow. 

\medskip
{\bf Binning:} (We follow ref.\ \cite{KawashimaGE94}).
Group the $n$ measurements $\O(i)\ident \O_{(i)}$ into $k$ bins of length
$l=n/k$ with (e.g.) $l=2,4,8,...$ .
Compute $k$ bin averages
\beq{binav}
\boldbar{\O}_b(l) := \frac{1}{l} \sum_{i=(b-1)l+1}^{bl} \,\O(i) \;\;,b=1,..,k
\eeq
and the variance of these averages
\beq{binsigma}
\sigma^2(l) := \frac{1}{k-1} \sum_{b=1}^k \left( \boldbar{\O}_b(l) - \boldbar{\O} \right)^2 \;.
\eeq
This variance should become inversely proportional to $l$ as the bin size $l$
becomes large enough, whence the $\boldbar{\O}_b(l)$
as a function of $b$ become statistically independent \cite{AllenTildesley-book87}.

The expectation value of the quantity
\beq{tauintOl}
 \tau_{int}^\O(l) := \frac{l\sigma^2(l)}{2\sigma^2} \;,
\eeq
where $\sigma$ is given by \eq{sigma},
grows monotonically in $l$.
When statistical independence is approached, 
$\tau_{int}^\O(l)$ approaches the integrated autocorrelation time $\tau^\O_{int}$ from below. 
The converged asymptotic value of $\tau^\O_{int}(l)$ (or rather its expectation value)
can therefore be used in \eq{sigmaint}
to compute the actual statistical error of $\boldbar{\O}$.
Note that $\tau_{int}^\O(l)$ will start to fluctuate at large $l$,
since for finite number of measurements the number of bins $k$ becomes small.

{\bf Convergence: }
If $\tau_{int}^\O(l)$ 
does not converge, then its expectation value at the largest available $l$ is a {\em lower bound}
for $\tau^\O_{int}$, giving a {\em lower} bound for the error of $\boldbar{\O}$.
In that case the MC run has not converged, and the data cannot be used to deduce physical results
for $\langle \O\rangle$.
Convergence of $\tau_{int}^\O(l)$ is a {\em prerequisite} for using the MC results.
Since $\tau^\O_{int}$ varies for different observables $\O$,
$\tau_{int}^\O(l)$ may have converged for some $\O$, and not for others.
This is a dangerous situation, since the very slow modes
visible in the nonconverged observables may be relevant for
the apparently converged observables, too.
Moreover, before starting measurements, the Monte Carlo configuration must be
allowed to {\em thermalize}, i.e.\ to approach the Boltzmann distribution.
It can be shown that the thermalization (from an arbitrary starting configuration)
is governed by the overall exponential autocorrelation time $\tau_{exp}$,
i.e.\ the very largest time scale in the simulation.
The thermalization time needs to be a reasonably large multiple of $\tau_{exp}$.
Therefore it is necessary to have at least an upper bound on $\tau_{exp}$ available.
If an insufficient time is spent on thermalization, then the MC averages $\boldbar{\O}$
contain a systematic bias, and will converge more slowly.
A good approach in practice is to spend $10-20\%$ of the total simulation time on thermalization,
which still does not noticeably increase statistical errors.

An unfortunate problem in practice is that simulations
may be started on an overly big system, for which -- unbeknownst to the simulator --
there are huge autocorrelation times.
Then it may happen that within any feasible MC run, these large time scales remain
invisible, so that the MC run appears to have converged, whereas in reality it has
barely moved in phase space and the results may be completely wrong.
(Take for example the simulation of a simple Ising model with  a local algorithm
at low temperature. The total magnetization takes an exponentially large time
to change sign. It may never do so  during the simulation, and may appear converged
at a large finite value, whereas the true average magnetization for a finite system is zero.)
Unless an ``Exact Monte Carlo'' method \cite{ProppW96} 
can be used, 
the only way to avoid this problem is apparently to begin simulations on extremely
small systems and away from problematic parameter regions, so that convergence
is guaranteed by brute force.
Slowly increasing system size, while measuring autocorrelation times,
one can ensure that autocorrelations do not get out of hand.
Note that this approach does not require much additional computer time,
since simulations on small systems will be fast.

One rather sensitive and simple instrument to detect some autocorrelations
long before they are visible in a binning analysis 
is to simply plot the MC evolution $\O(i),\,i=1,...,n$ graphically
and to look for long correlations by eye.

{\bf Autocorrelation function: }
A quantitative analysis of autocorrelations beyond $\tau^\O_{int}$,
e.g.\ in order to calculate $\tau^\O_{exp}$,
requires calculation of the autocorrelation function $\Gamma_{\O\O}(t)$.
This is feasible only when $\boldbar{\O}$ has converged.
Contrary to claims in the literature, one {\em cannot} reliably 
extract the integrated autocorrelation time $\tau^\O_{int}$ from $\Gamma_{\O\O}(t)$
by neglecting it beyond a ``window'' $t<W$ selfconsistently determined
from the slope of $\Gamma_{\O\O}(W)$.
Typically (but simplified \cite{Sokal92}), 
we have
\beq{GammaSpectrum}
\Gamma_{\O\O}(t) = \sum_j \,c^\O_j \,e^{-t/\tau_j} \;, 
\eeq
with contributions from all eigenmodes of the Markov transition
matrix, unless they are orthogonal to $\O$ (whence $c^\O_j=0$).
Therefore $\tau^\O_{int}\approx \sum_j c^\O_j \tau_j$
can get sizeable contributions from very large time scales $\tau_j$,
even when they couple only with small matrix elements $c^\O_j$ and are therefore not
visible at small times $t$.
This does indeed commonly happen in practice.
A more reliable procedure is the following:
Ensure convergence of $\boldbar{\O}$. Calculate $\Gamma_{\O\O}(t)$ for $t<t_{max}$,
where $t_{max}$ is chosen as large as possible while $\Gamma_{\O\O}(t)$ remains well
above zero within error bars for all $t<t_{max}$.
Compute  estimates for $\tau^\O_{exp}$ and its matrix element $c^\O_{exp}$ 
from the (hopefully) asymptotic decay of $\Gamma_{\O\O}(t)$.
Calculate $\tau^\O_{int}$ from \eq{tauint} by summing $t$ up to the order of $\tau^\O_{exp}$
and computing the remainder of $\tau^\O_{int}$ from the asymptotic form of \eq{GammaSpectrum},
$\Gamma_{\O\O}(t)\sim c^\O_{exp} e^{-t/\tau^\O_{exp}}$.
Of course, even this procedure will fail if the MC run is too short to show the largest
autocorrelation times.

{\bf Jackknife: }
A binning-type analysis is a prerequisite for checking convergence.
It also produces values for the autocorrelation times $\tau^\O_{int}$.
However, standard error propagation becomes rather cumbersome for nonlinear quantities, like
correlation functions in simulations with a sign-problem.
It is much easier to compute errors  with the jackknife procedure \cite{jackknife}.
We give a brief recipe. 
Split the measured values $\O(i)$ into $k$ groups of length $l=n/k$.
To obtain the {\em asymptotic} error, $l$ must be 
significantly larger than the relevant autocorrelation time $\tau^\O_{int}$.

Now perform the complete, possibly highly nonlinear, analysis of the MC-data $k+1$ times:
first with all $l\cdot k $ data, leading to a result ``$R^{(0)}$'',
then, for $j=1,..,k$, with all data except those in bin $j$
(i.e.\ pretend that bin $j$ was never measured), leading to values ``$R^{(j)}$''.
Then the overall result $R$ is 
\beqa{jackknife}
     R             &=& R^{(0)} - \mbox{\em Bias}, \mbox{\ \ where} \\
   \mbox{\em Bias} &=& (k-1) \, (R^{av} - R^{(0)}) ,\\
     R^{av}        &=& \frac{1}{k}\, \sum_{j=1}^k R^{(j)} \;,
\eeqa
with statistical error 
\beq{Jackknifeerror}
 \delta(R) =  (k-1)^{1/2} \left( \frac{1}{k}
                            \sum_{j=1}^k (R^{(j)})^2 -  (R^{av})^2 \right)^{1/2} \;.
\eeq
In this procedure, error propagation is {\em automatic}.
In each of the $k+1$ analyses, almost the full set of data is used, avoiding problems
in the usual analysis like  instabilities of fits.
It is also possible to combine Jackknife and binning by repeating the Jackknife procedure
for different bin lengths, to check for convergence and to 
compute integrated autocorrelation times and asymptotic errors according to \eq{tauintOl}.

\pagebreak


\begin{thebibliography}{100}

\bibitem{EvertzLM93}
{\sc Evertz, H.~G.}, {\sc Lana, G.}, and {\sc Marcu, M.}, 1993, {\em Phys. Rev.
  Lett.\/}, 70, 875, cond-mat/9211006

\bibitem{EvertzM92}
{\sc Evertz, H.~G.} and {\sc Marcu, M.}, 1993, {\em Nucl. Phys. B\/}, S30, 277,
  hep-lat/9211047

\bibitem{EvertzM94}
{\sc Evertz, H.~G.} and {\sc Marcu, M.}, 1994, {\em Quantum Monte Carlo Methods
  in Condensed Matter Physics\/}, edited by M.~Suzuki, chap. Overcoming
  Critical Slowing Down in Quantum Monte Carlo Simulations, 65 (World
  Scientific)

\bibitem{talk91}
The loop algorithm was first presented in a talk by the present author at {\em
  Lattice 91}, Tsukuba, Japan

\bibitem{Scalettar-WL99}
{\sc Scalettar, R.~T.}, 1999, {\em Quantum Monte Carlo Methods in Physics and
  Chemistry\/}, edited by M.~P. Nightingale and C.~J. Umrigar, chap. World-Line
  Quantum Monte Carlo (Kluwer)

\bibitem{SandvikK91}
{\sc Sandvik, A.~W.} and {\sc Kurkij{\"a}rvi, J.}, 1991, {\em Phys. Rev. B\/},
  43, 5950

\bibitem{Sandvik92}
{\sc Sandvik, A.~W.}, 1992, {\em J. Phys. A\/}, 25, 3667

\bibitem{Sandvik97}
{\sc Sandvik, A.~W.}, 1997, {\em Phys. Rev. B\/}, 56, 11678

\bibitem{KawashimaG95a}
{\sc Kawashima, N.} and {\sc Gubernatis, J.~E.}, 1995, {\em J. Stat. Phys.\/},
  80, 169, cond-mat/9502065

\bibitem{AizenmanN94}
{\sc Aizenman, M.} and {\sc Nachtergaele, B.}, 1994, {\em Comm. Math. Phys.\/},
  164, 17, cond-mat/9310009

\bibitem{Nachtergaele93}
{\sc Nachtergaele, B.}, 1994, {\em Probability Theory and Mathematical
  Statistics (Proceedings of the 6th Vilnius Conference)\/}, edited by
  B.~Grigelionis et~al., 565--590, cond-mat/9312012, ~VSP/TEV,
  Utrecht-Tokyo-Vilnius, cond-mat/9312012

\bibitem{BrowerCW98}
{\sc Brower, R.}, {\sc Chandrasekharan, S.}, and {\sc Wiese, U.~J.}, 1998, {\em
  Physica A\/}, 261, 520, cond-mat/9801003

\bibitem{Sandvik99b}
{\sc Sandvik, A.~W.}, 1999, {\em Phys. Rev. B\/}, 59, R14157, cond-mat/9902226

\bibitem{HaradaK00}
{\sc Harada, K.} and {\sc Kawashima, N.}, 2001, {\em J. Phys. Soc. Jpn.\/}, 70,
  13, cond-mat/0011346

\bibitem{SwendsenW87}
{\sc Swendsen, R.~H.} and {\sc Wang, J.~S.}, 1987, {\em Phys. Rev. Lett.\/},
  58, 86

\bibitem{BeardW96}
{\sc Beard, B.~B.} and {\sc Wiese, U.~J.}, 1996, {\em Phys. Rev. Lett.\/}, 77,
  5130

\bibitem{RiegerK99}
{\sc Rieger, H.} and {\sc Kawashima, N.}, 1999, {\em Europ. Phys. J.\/}, B 9,
  233, cond-mat/9802104

\bibitem{CoxGHSW99}
{\sc Cox, J.}, {\sc Gattringer, C.}, {\sc Holland, K.}, {\sc Scarlet, B.}, and
  {\sc Wiese, U.~J.}, 2000, {\em Nucl. Phys. B-Proc. Suppl.\/}, 83-4, 777,
  hep-lat/9909119

\bibitem{ChandrasekharanSW99}
{\sc Chandrasekharan, S.}, {\sc Scarlet, B.}, and {\sc Wiese, U.-J.},
  cond-mat/9909451, Meron-Cluster Simulation of Quantum Spin Ladders in a
  Magnetic Field

\bibitem{ChandrasekharanW99}
{\sc Chandrasekharan, S.} and {\sc Wiese, U.~J.}, 1999, {\em Phys. Rev.
  Lett.\/}, 83, 3116, cond-mat/9902128

\bibitem{ChandrasekharanCHW00}
{\sc Chandrasekharan, S.}, {\sc Cox, J.}, {\sc Holland, K.}, and {\sc Wiese,
  U.~J.}, 2000, {\em Nucl. Phys. B\/}, 576, 481, hep-lat/9906021

\bibitem{Chandrasekharan99}
{\sc Chandrasekharan, S.}, 2000, {\em Nucl. Phys. B\/}, Proc.\ Suppl.\ 83, 774,
  hep-lat/9909007

\bibitem{Chandrasekharan00}
{\sc Chandrasekharan, S.}, 2000, {\em Chin. J. Phys.\/}, 38, 696,
  hep-lat/0001003

\bibitem{ChandrasekharanO00a}
{\sc Chandrasekharan, S.} and {\sc Osborn, J.~C.}, 2000, {\em Phys. Lett. B\/},
  496, 122, hep-lat/0010036

\bibitem{ChandrasekharanO00b}
{\sc Chandrasekharan, S.} and {\sc Osborn, J.}, 2000, {\em Springer Proc.
  Phys.\/}, 86, 28

\bibitem{CoxH00}
{\sc Cox, J.} and {\sc Holland, K.}, 2000, {\em Nucl. Phys. B\/}, 583, 331,
  hep-lat/0003022

\bibitem{Osborn00}
{\sc Osborn, J.}, 2001, {\em Nucl. Phys. B\/}, Proc.\ Suppl.\ 94, 864,
  hep-lat/0010097

\bibitem{ChandrasekharanO01}
{\sc Chandrasekharan, S.} and {\sc Osborn, J.~C.}, cond-mat/0109424, 2001,
  Kosterlitz-Thouless Universality in a Fermionic System

\bibitem{Chandrasekharan01a}
{\sc Chandrasekharan, S.}, 2001, Novel Quantum Monte Carlo Algorithms for
  Fermions, hep-lat/0110018, proceedings of the "Quantum Monte Carlo" meeting
  (Trento, Italy, July 3-6, 2001)

\bibitem{ChandrasekharanSW01}
{\sc Chandrasekharan, S.}, {\sc Scarlet, B.}, and {\sc Wiese, U.-J.},
  hep-lat/0110215, 2001, From Spin Ladders to the 2-d O(3) Model at Non-Zero
  Density, (To be published in computer physics communications)

\bibitem{ChandrasekharanCOW02}
{\sc Chandrasekharan, S.}, {\sc Cox, J.}, {\sc Osborn, J.}, and {\sc Wiese,
  U.-J.}, cond-mat/0201360, 2002, Meron-Cluster Approach to Systems of Strongly
  Correlated Electrons

\bibitem{HeneliusS00}
{\sc Henelius, P.} and {\sc Sandvik, A.~W.}, 2000, {\em Phys. Rev. B\/}, 62,
  1102, cond-mat/0001351

\bibitem{KawashimaG94}
{\sc Kawashima, N.} and {\sc Gubernatis, J.}, 1994, {\em Phys. Rev. Lett.\/},
  73, 1295

\bibitem{Kawashima95}
{\sc Kawashima, N.}, 1996, {\em J. Stat. Phys.\/}, 82, 131, cond-mat/9506075

\bibitem{HaradaTK98}
{\sc Harada, K.}, {\sc Troyer, M.}, and {\sc Kawashima, N.}, 1998, {\em J.
  Phys. Soc. Jpn.\/}, 67, 1130, cond-mat/9712292

\bibitem{TodoKT98}
{\sc Todo, S.}, {\sc Kato, K.}, and {\sc Takayama, H.}, 1998, {\em Computer
  Simulation Studies in Condensed Matter Physics XI\/}, edited by D.~P. Landau
  et~al., Springer Proceedings in Physics, cond-mat/9803088

\bibitem{TodoK01}
{\sc Todo, S.} and {\sc Kato, K.}, 2001, {\em Phys. Rev. Lett.\/}, 87, 047203,
  cond-mat/9911047

\bibitem{KimGWB98}
{\sc Kim, Y.~J.}, {\sc Greven, M.}, {\sc Wiese, U.~J.}, and {\sc Birgeneau,
  R.~J.}, 1998, {\em Eur. Phys. J. B\/}, 4, 291, cond-mat/9712257

\bibitem{HaradaK01}
{\sc Harada, K.} and {\sc Kawashima, N.}, 2002, {\em Phys. Rev. B\/}, 65,
  052403, cond-mat/0109431

\bibitem{KawashimaGE94}
{\sc Kawashima, N.}, {\sc Gubernatis, J.~E.}, and {\sc Evertz, H.~G.}, 1994,
  {\em Phys. Rev. B\/}, 50, 136, cond-mat/9403082

\bibitem{Kawashima96}
{\sc Kawashima, N.}, 1996, {\em Computer Simulations in Condensed Matter
  Physics IX\/}, edited by D.~P. Landau et~al., Springer Proceedings in Physics

\bibitem{AmmonEKTF98}
{\sc Ammon, B.}, {\sc Evertz, H.~G.}, {\sc Kawashima, N.}, {\sc Troyer, M.},
  and {\sc Frischmuth, B.}, 1998, {\em Phys. Rev. B\/}, 58, 4304,
  cond-mat/9711022

\bibitem{BrunnerM98}
{\sc Brunner, M.} and {\sc Muramatsu, A.}, 1998, {\em Phys. Rev. B\/}, 58,
  R10100, cond-mat/9707108

\bibitem{BrunnerAM99}
{\sc Brunner, M.}, {\sc Assaad, F.~F.}, and {\sc Muramatsu, A.}, 2000, {\em
  Eur. Phys. J. B\/}, 16, 209, cond-mat/9904150

\bibitem{SenguptaSC02}
{\sc Sengupta, P.}, {\sc Sandvik, A.~W.}, and {\sc Campbell, D.~K.}, 2002, {\em
  Phys. Rev. B\/}, 65, 155113, cond-mat/0102141

\bibitem{WieseY94}
{\sc Wiese, U.~J.} and {\sc Ying, H.~P.}, 1994, {\em Z. Phys. B-Condens.
  Mat.\/}, 93, 147, cond-mat/9212006

\bibitem{ProkofevST98b}
{\sc Prokof'ev, N.}, {\sc Svistunov, B.}, and {\sc Tupitsyn, I.}, 1998, {\em J.
  Exp. Theor. Phys.\/}, 87, 310, cond-mat/9703200, cond-mat/9703200

\bibitem{SyljuasenS02}
{\sc Syljuasen, O.~F.} and {\sc Sandvik, A.~W.}, cond-mat/0202316, 2002,
  Quantum Monte Carlo with Directed Loops

\bibitem{TroyerKU96}
{\sc Troyer, M.}, {\sc Kontani, H.}, and {\sc Ueda, K.}, 1996, {\em Phys. Rev.
  Lett.\/}, 76, 3822

\bibitem{TroyerIU97}
{\sc Troyer, M.}, {\sc Imada, M.}, and {\sc Ueda, K.}, 1997, {\em J. Phys. Soc.
  Jpn.\/}, 66, 2957, cond-mat/9702077

\bibitem{TroyerI97}
{\sc Troyer, M.} and {\sc Imada, M.}, 1997, {\em Computer Simulations in
  Condensed Matter Physics X\/}, edited by D.~P. Landau et~al., Springer
  Proceedings in Physics, cond-mat/9703049

\bibitem{Trotter59}
{\sc Trotter, H.~F.}, 1959, {\em Proc. Am. Math. Soc.\/}, 10, 545

\bibitem{Suzuki76}
{\sc Suzuki, M.}, 1976, {\em Prog. Theor. Phys.\/}, 56, 1454

\bibitem{Baxter-book89}
{\sc Baxter, R.~J.}, 1989, {\em Exactly Solved Models in Statistical
  Mechanics\/} (New York: Academic)

\bibitem{KawashimaG95b}
{\sc Kawashima, N.} and {\sc Gubernatis, J.~E.}, 1995, {\em Phys. Rev. E\/},
  51, 1547

\bibitem{KosterlitzT73}
{\sc Kosterlitz, J.~M.} and {\sc Thouless, D.~J.}, 1973, {\em J. Phys. C\/}, 6,
  1181

\bibitem{Lieb67}
{\sc Lieb, E.~H.}, 1967, {\em Phys. Rev. Lett.\/}, 18, 1046

\bibitem{LiebW72}
{\sc Lieb, E.~H.} and {\sc Wu, F.~Y.}, 1972, {\em Phase Transitions and
  Critical Phenomena\/}, edited by C.~Domb and M.~S. Green, vol.~1, 331
  (Academic)

\bibitem{Suzuki-book94}
{\sc Suzuki, M.} (ed.), 1994, {\em Quantum Monte Carlo Methods in Condensed
  Matter Physics\/} (World Scientific)

\bibitem{Sokal92}
{\sc Sokal, A.~D.}, 1992, {\em Quantum Fields on the Computer\/}, edited by
  M.~Creutz, chap. Bosonic Algorithms (World Scientific), (Available
  electronically via ref.\ \cite{ProppW96})

\bibitem{KandelD91}
{\sc Kandel, D.} and {\sc Domany, E.}, 1991, {\em Phys. Rev. B\/}, 43, 8539

\bibitem{KasteleynF69}
{\sc Kasteleyn, P.} and {\sc Fortuin, C.}, 1969, {\em J. Phys. Soc. Jpn.\/},
  26(Suppl.), 11

\bibitem{FortuinK72}
{\sc Fortuin, C.} and {\sc Kasteleyn, P.}, 1972, {\em Physica\/}, 57, 536

\bibitem{RoosM99}
{\sc Roos, P.} and {\sc Miyashita, S.}, 1999, {\em Phys. Rev. B\/}, 59, 13782,
  cond-mat/9812397

\bibitem{NishinoORYM99}
{\sc Nishino, M.}, {\sc Onishi, H.}, {\sc Roos, P.}, {\sc Yamaguchi, K.}, and
  {\sc Miyashita, S.}, 2000, {\em Phys. Rev. B\/}, 61, 4033, cond-mat/9906426

\bibitem{EvertzHMPS91}
{\sc Evertz, H.~G.}, {\sc Hasenbusch, M.}, {\sc Marcu, M.}, {\sc Pinn, K.}, and
  {\sc Solomon, S.}, 1991, {\em Phys. Lett.\/}, 254B, 185

\bibitem{EvertzHMPS92}
{\sc Evertz, H.~G.}, {\sc Hasenbusch, M.}, {\sc Marcu, M.}, {\sc Pinn, K.}, and
  {\sc Solomon, S.}, 1992, {\em Int. J. Mod. Phys. C\/}, 3, 235

\bibitem{HasenbuschLMP92}
{\sc Hasenbusch, M.}, {\sc Lana, G.}, {\sc Marcu, M.}, and {\sc Pinn, K.},
  1992, {\em Phys. Rev. B\/}, 46, 10472

\bibitem{HasenbuschMP92}
{\sc Hasenbusch, M.}, {\sc Marcu, M.}, and {\sc Pinn, K.}, 1992, {\em Nucl.
  Phys. B\/}, Proc.\ Suppl.\ 26B, 598, hep-lat/9207019

\bibitem{HasenbuschMP94}
{\sc Hasenbusch, M.}, {\sc Marcu, M.}, and {\sc Pinn, K.}, 1994, {\em Physica
  A\/}, 208, 124, hep-lat/9404016

\bibitem{KondevH96}
{\sc Kondev, J.} and {\sc Henley, C.~L.}, 1996, {\em Nucl. Phys. B\/}, 464,
  540, cond-mat/9511102,

\bibitem{HasenbuschP97}
{\sc Hasenbusch, M.} and {\sc Pinn, K.}, 1997, {\em J. Phys. A-Math. Gen.\/},
  30, 63, cond-mat/9605019

\bibitem{HasenbuschMP96}
{\sc Hasenbusch, M.}, {\sc Meyer, S.}, and {\sc Putz, M.}, 1996, {\em J. Stat.
  Phys.\/}, 85, 383, hep-lat/9601011

\bibitem{Kondev97}
{\sc Kondev, J.}, 1997, {\em Int. J. Mod. Phys. B\/}, 11, 153, cond-mat/9607181

\bibitem{Sweeny83}
{\sc Sweeny, M.}, 1983, {\em Phys. Rev. B\/}, 27, 4445

\bibitem{SouzaCL00}
{\sc de~Souza, A.~J.~F.}, {\sc Costa, U.~M.~S.}, and {\sc Lyra, M.~L.}, 2000,
  {\em Phys. Rev. B\/}, 62, 8909, cond-mat/0004176

\bibitem{Handscomb62}
{\sc Handscomb, D.~C.}, 1962, {\em Proc. Cambridge Philos. Soc.\/}, 58, 594

\bibitem{Handscomb64}
{\sc Handscomb, D.~C.}, 1964, {\em Proc. Cambridge Philos. Soc.\/}, 60, 115

\bibitem{Lyklema82}
{\sc Lyklema, J.~W.}, 1982, {\em Phys. Rev. Lett.\/}, 49, 88

\bibitem{LeeJN84}
{\sc Lee, D.~H.}, {\sc Joannopoulos, J.~D.}, and {\sc Negele, J.~W.}, 1984,
  {\em Phys. Rev. B\/}, 30, 1599

\bibitem{ChakravartyS82}
{\sc Chakravarty, S.} and {\sc Stein, D.~B.}, 1982, {\em Phys. Rev. Lett.\/},
  49, 582

\bibitem{Wolff89a}
{\sc Wolff, U.}, 1989, {\em Phys. Rev. Lett.\/}, 62, 361

\bibitem{Wolff89c}
{\sc Wolff, U.}, 1989, {\em Phys. Lett.\/}, 228B, 379

\bibitem{Mino91}
{\sc Mino, H.}, 1991, {\em Computer Physics Communications\/}, 66, 25

\bibitem{Todo02a}
{\sc Todo, S.}, 2002, {\em Computer Simulation Studies in Condensed Matter
  Physics XV\/}, edited by D.~P. Landau et~al., Springer Proceedings in Physics

\bibitem{Todo02b}
{\sc Todo, S.}, 2002, {\em Prog. Theor. Phys. Suppl.\/}, 145, 188

\bibitem{Evertz93}
{\sc Evertz, H.~G.}, 1993, {\em J. Stat. Phys.\/}, 70, 1075

\bibitem{MakivicD91}
{\sc Makivic, M.} and {\sc Ding, H.-Q.}, 1991, {\em Phys. Rev. B\/}, 43, 3562

\bibitem{FarhiG92}
{\sc Farhi, E.} and {\sc Gutmann, S.}, 1992, {\em Ann. Phys. (N.Y.)\/}, 213,
  182

\bibitem{AizenmanL90}
{\sc Aizenman, M.} and {\sc Lieb, E.~H.}, 1990, {\em Phys. Rev. Lett.\/}, 65,
  1470

\bibitem{Wolff90}
{\sc Wolff, U.}, 1990, {\em Nucl. Phys. B\/}, 334, 581

\bibitem{EvertzV01}
{\sc Evertz, H.~G.} and {\sc von~der Linden, W.}, 2001, {\em Phys. Rev.
  Lett.\/}, 86, 5164, cond-mat/0008072

\bibitem{PleimlingH01}
{\sc Pleimling, M.} and {\sc Henkel, M.}, hep-th/0103194, 2001, Anisotropic
  scaling and generalized conformal invariance at Lifshitz points

\bibitem{HenkelP01}
{\sc Henkel, M.} and {\sc Pleimling, M.}, 2002, {\em Comp. Phys. Commun.\/},
  147, 419, cond-mat/0108454

\bibitem{DagottoR96}
{\sc Dagotto, E.} and {\sc Rice, M.}, 1996, {\em Science\/}, 271, G18

\bibitem{NovotnyE94}
{\sc Novotny, M.~A.} and {\sc Evertz, H.~G.}, 1994, {\em Quantum Monte Carlo
  Methods in Condensed Matter Physics\/}, edited by M.~Suzuki (World
  Scientific)

\bibitem{DorneichT01}
{\sc Dorneich, A.} and {\sc Troyer, M.}, 2001, {\em Phys. Rev. E\/}, 64,
  066701, cond-mat/0106471

\bibitem{TodoMYT01}
{\sc Todo, S.}, {\sc Matsumoto, M.}, {\sc Yasuda, C.}, and {\sc Takayama, H.},
  2001, {\em Phys. Rev. B\/}, 64, 224412, cond-mat/0106073

\bibitem{GrevenBW96}
{\sc Greven, M.}, {\sc Birgeneau, R.~J.}, and {\sc Wiese, U.~J.}, 1996, {\em
  Phys. Rev. Lett.\/}, 77, 1865, cond-mat/9605068

\bibitem{HaradaK97}
{\sc Harada, K.} and {\sc Kawashima, N.}, 1997, {\em Phys. Rev. B\/}, 55,
  11949, cond-mat/9702081

\bibitem{HaradaK98}
{\sc Harada, K.} and {\sc Kawashima, N.}, 1998, {\em J. Phys. Soc. Jpn.\/}, 67,
  2768, cond-mat/9803090

\bibitem{KohnoT97}
{\sc Kohno, M.} and {\sc Takahashi, M.}, 1997, {\em Phys. Rev. B\/}, 56, 3212,
  cond-mat/9705148

\bibitem{EvertzM93}
{\sc Evertz, H.~G.} and {\sc Marcu, M.}, 1993, {\em Int. J. Mod. Phys. C\/}, 4,
  1147

\bibitem{SandvikSC97}
{\sc Sandvik, A.~W.}, {\sc Singh, R. R.~P.}, and {\sc Campbell, D.~K.}, 1997,
  {\em Phys. Rev. B\/}, 56, 14510

\bibitem{Otsuka01}
{\sc Otsuka, H.}, 2001, {\em Phys. Rev. B\/}, 6402, 0406

\bibitem{LuijtenB95}
{\sc Luijten, E.} and {\sc Bl\"ote, H. W.~J.}, 1995, {\em Int. J. Mod. Phys.
  C\/}, 6, 359

\bibitem{MiyaharaTJU98}
{\sc Miyahara, S.}, {\sc Troyer, M.}, {\sc Johnston, D.~C.}, and {\sc Ueda,
  K.}, 1998, {\em J. Phys. Soc. Jpn.\/}, 67, 3918, cond-mat/9807127

\bibitem{Nakamura97}
{\sc Nakamura, T.}, 1997, {\em Phys. Rev. B\/}, 57, R3197, cond-mat/9707019

\bibitem{NakamuraO97}
{\sc Nakamura, T.} and {\sc Okamoto, K.}, 1998, {\em Phys. Rev. B\/}, 58, 2411,
  cond-mat/9709295

\bibitem{ZhangKCG95}
{\sc Zhang, S.~W.}, {\sc Kawashima, N.}, {\sc Carlson, J.}, and {\sc
  Gubernatis, J.~E.}, 1995, {\em Phys. Rev. Lett.\/}, 74, 1500

\bibitem{KashurnikovPST99}
{\sc Kashurnikov, V.~A.}, {\sc Prokof'ev, N.~V.}, {\sc Svistunov, B.~V.}, and
  {\sc Troyer, M.}, 1999, {\em Phys. Rev. B\/}, 59, 1162, cond-mat/9802294

\bibitem{OnishiNKM99}
{\sc Onishi, H.}, {\sc Nishino, M.}, {\sc Kawashima, N.}, and {\sc Miyashita,
  S.}, 1999, {\em J. Phys. Soc. Jpn.\/}, 68, 2547, cond-mat/9903375

\bibitem{Syljuasen00}
{\sc Syljuasen, O.}, 2000, {\em Phys. Rev. B\/}, 61, R846, cond-mat/9907142

\bibitem{EvertzEV01}
{\sc Evertz, H.~G.}, {\sc Erkinger, H.~M.}, and {\sc von~der Linden, W.}, 2001,
  {\em Computer Simulations in Condensed Matter Physics XIV\/}, edited by D.~P.
  Landau et~al., Springer Proceedings in Physics

\bibitem{HaradaK02}
{\sc Harada, K.} and {\sc Kawashima, N.}, cond-mat/0205472, 2002,
  Coarse-grained loop algorithms for Monte Carlo simulation of quantum spin
  systems

\bibitem{HirschSSB82}
{\sc Hirsch, J.}, {\sc Sugar, R.}, {\sc Scalapino, D.}, and {\sc Blankenbecler,
  R.}, 1982, {\em Phys. Rev. B\/}, 26, 5033

\bibitem{AlvarezG00}
{\sc Alvarez, J.~V.} and {\sc Gros, C.}, 2000, {\em Eur. Phys. J. B\/}, 15,
  641, cond-mat/0002131

\bibitem{AlvarezG02}
{\sc Alvarez, J.} and {\sc Gros, C.}, cond-mat/0204320, 2002, Conductivity of
  quantum-spin chains: A Quantum Monte Carlo approach

\bibitem{Wiese93}
{\sc Wiese, U.~J.}, 1993, {\em Phys. Lett. B\/}, 311, 235

\bibitem{VebericPE00}
{\sc Veberic, D.}, {\sc Prelovsek, P.}, and {\sc Evertz, H.~G.}, 2000, {\em
  Phys. Rev. B\/}, 62, 6745, cond-mat/0002220

\bibitem{BrunnerAM00}
{\sc Brunner, M.}, {\sc Assaad, F.~F.}, and {\sc Muramatsu, A.}, 2000, {\em
  Phys. Rev. B\/}, 62, 15480, cond-mat/0002321

\bibitem{BrunnerCAM01}
{\sc Brunner, M.}, {\sc Capponi, S.}, {\sc Assaad, F.~F.}, and {\sc Muramatsu,
  A.}, 2001, {\em Phys. Rev. B\/}, 6318, 0511, cond-mat/0101462

\bibitem{Wiese98}
{\sc Wiese, U.~J.}, 1998, {\em Prog. Theor. Phys. Suppl.\/}, 483--494,
  hep-lat/9811025, and references therein

\bibitem{Lee02}
{\sc Lee, D.}, cond-mat/0202283, 2002, Permutation zones and the fermion sign
  problem

\bibitem{ProkofevS98b}
{\sc Prokof'ev, N.} and {\sc Svistunov, B.}, 1998, {\em Phys. Rev. Lett.\/},
  80, 4355, cond-mat/9706169

\bibitem{KashurnikovKS98}
{\sc Kashurnikov, V.}, {\sc Krasavin, A.}, and {\sc Svistunov, B.}, 1998, {\em
  Phys. Rev. B\/}, 58, 1826, cond-mat/9709035

\bibitem{MishchenkoPS01}
{\sc Mishchenko, A.}, {\sc Prokof'ev, N.}, and {\sc Svistunov, B.}, 2001, {\em
  Phys. Rev. B\/}, 64, 033101, cond-mat/0103234

\bibitem{KashurnikovPS02}
{\sc Kashurnikov, V.}, {\sc Prokof'ev, N.}, and {\sc Svistunov, B.},
  cond-mat/0202510, 2002, Revealing Superfluid--Mott-Insulator Transition in an
  Optical Lattice, to appear in PRL

\bibitem{GoeppertGPS98}
{\sc Goeppert, G.}, {\sc Grabert, H.}, {\sc Prokof'ev, N.~V.}, and {\sc
  Svistunov, B.~V.}, 1998, {\em Phys. Rev. Lett.\/}, 81, 2324, cond-mat/9802248

\bibitem{ProkofevS98a}
{\sc Prokof'ev, N.} and {\sc Svistunov, B.}, 1998, {\em Phys. Rev. Lett.\/},
  81, 2514, cond-mat/9804097

\bibitem{KaganKKPS98}
{\sc Yu.Kagan}, {\sc V.A.Kashurnikov}, {\sc A.V.Krasavin}, {\sc N.V.Prokof'ev},
  and {\sc Svistunov, B.}, cond-mat/9811090, 1998, Quasicondensation in 2D
  Interacting Bose Gas: Quantum Monte Carlo Study

\bibitem{ProkofevS99}
{\sc Prokof'ev, N.}, {\sc Svistunov, B.}, and {\sc Tupitsyn, I.}, 1999, {\em
  Phys. Rev. Lett.\/}, 82, 5092, cond-mat/9901083

\bibitem{KashurnikovMTK99}
{\sc Kashurnikov, V.~A.}, {\sc Mishchenko, A.~S.}, {\sc Tupitsyn, I.~S.}, and
  {\sc Kharchenko, Y.~G.}, 1999, {\em Phys. Low-Dimens. Struct.\/}, 5-6, 13

\bibitem{ProkofevS01}
{\sc Prokof'ev, N.} and {\sc Svistunov, B.}, 2001, {\em Phys. Rev. Lett.\/},
  87, 160601, cond-mat/0103146

\bibitem{HebertBSSTD01}
{\sc Hebert, F.}, {\sc Batrouni, G.~G.}, {\sc Scalettar, R.~T.}, {\sc Schmid,
  G.}, {\sc Troyer, M.}, and {\sc Dorneich, A.}, 2002, {\em Phys. Rev. B\/},
  65, 014513, cond-mat/0105450

\bibitem{SchmidTTD01}
{\sc Schmid, G.}, {\sc Todo, S.}, {\sc Troyer, M.}, and {\sc Dorneich, A.},
  2002, {\em Phys. Rev. Lett.\/}, 88, 167208, cond-mat/0110024

\bibitem{BernadetBMSTD01}
{\sc Bernardet, K.}, {\sc Batrouni, G.~G.}, {\sc Meunier, J.-L.}, {\sc Schmid,
  G.}, {\sc Troyer, M.}, and {\sc Dorneich, A.}, 2002, {\em Phys. Rev. B\/},
  65, 104519, cond-mat/0110314

\bibitem{BernadetBTD02}
{\sc Bernardet, K.}, {\sc Batrouni, G.~G.}, {\sc Troyer, M.}, and {\sc
  Dorneich, A.}, cond-mat/0204313, 2002, Destruction of diagonal and
  off-diagonal long range order by disorder in two-dimensional hard core boson
  systems

\bibitem{SandvikC99}
{\sc Sandvik, A.~W.} and {\sc Campbell, D.~K.}, 1999, {\em Phys. Rev. Lett.\/},
  83, 195, cond-mat/9902230

\bibitem{WesselH01}
{\sc Wessel, S.} and {\sc Haas, S.}, 2002, {\em Phys. Rev. B\/}, 65, 132402,
  cond-mat/0106385

\bibitem{WesselOH01}
{\sc Wessel, S.}, {\sc Olshanii, M.}, and {\sc Haas, S.}, 2001, {\em Phys. Rev.
  Lett.\/}, 87, 206407, cond-mat/0105464

\bibitem{Sandvik00}
{\sc Sandvik, A.~W.}, 2001, {\em Phys. Rev. Lett.\/}, 86, 3209,
  cond-mat/0010433

\bibitem{WesselNSH00}
{\sc Wessel, S.}, {\sc Normand, B.}, {\sc Sigrist, M.}, and {\sc Haas, S.},
  2001, {\em Phys. Rev. Lett.\/}, 86, 1086, cond-mat/0007228

\bibitem{Sandvik99a}
{\sc Sandvik, A.~W.}, cond-mat/9909230, 1999, The two-dimensional bond-diluted
  quantum Heisenberg model at the classical percolation threshold

\bibitem{Sandvik02}
{\sc Sandvik, A.~W.}, cond-mat/0206355, 2002, Multi-critical point in a diluted
  bilayer Heisenberg quantum antiferromagnet

\bibitem{Sandvik01a}
{\sc Sandvik, A.~W.}, cond-mat/0110510, 2001, Classical percolation transition
  in the diluted two-dimensional S=1/2 Heisenberg antiferromagnet, to appear in
  Phys. Rev. B

\bibitem{ShevchenkoSS00}
{\sc Shevchenko, P.~V.}, {\sc Sandvik, A.~W.}, and {\sc Sushkov, O.~P.}, 2000,
  {\em Phys. Rev. B\/}, 61, 3475, cond-mat/9905227

\bibitem{BergkvistHR02}
{\sc Bergkvist, S.}, {\sc Henelius, P.}, and {\sc Rosengren, A.},
  cond-mat/0206092, 2002, Ground state of the random-bond spin-1 Heisenberg
  chain

\bibitem{UedaTKSL96}
{\sc Ueda, K.}, {\sc Troyer, M.}, {\sc Kontani, H.}, {\sc Sigrist, M.}, and
  {\sc Lee, P.~A.}, 1996, {\em J. Korean Phys. Soc.\/}, 29, S128

\bibitem{KimT98}
{\sc Kim, J.~K.} and {\sc Troyer, M.}, 1998, {\em Phys. Rev. Lett.\/}, 80,
  2705, cond-mat/9709333

\bibitem{KatoTHKMT00}
{\sc Kato, K.}, {\sc Todo, S.}, {\sc Harada, K.}, {\sc Kawashima, N.}, {\sc
  Miyashita, S.}, and {\sc Takayama, H.}, 2000, {\em Phys. Rev. Lett.\/}, 84,
  4204, cond-mat/9905379

\bibitem{FrischmuthAT96}
{\sc Frischmuth, B.}, {\sc Ammon, B.}, and {\sc Troyer, M.}, 1996, {\em Phys.
  Rev. B\/}, 54, R3714

\bibitem{FrischmuthHSR97}
{\sc Frischmuth, B.}, {\sc Haas, S.}, {\sc Sierra, G.}, and {\sc Rice, T.~M.},
  1997, {\em Phys. Rev. B\/}, 55, R3340, cond-mat/9606183

\bibitem{SyljuasenCG97}
{\sc Syljuasen, O.~F.}, {\sc Chakravarty, S.}, and {\sc Greven, M.}, 1997, {\em
  Phys. Rev. Lett.\/}, 78, 4115, cond-mat/9701197

\bibitem{KimB00}
{\sc Kim, Y.~J.} and {\sc Birgeneau, R.~J.}, 2000, {\em Phys. Rev. B\/}, 62,
  6378, cond-mat/0004311

\bibitem{KimBKLESY99}
{\sc Y.~J.~Kim, R. J.~B.}, {\sc Kastner, M.~A.}, {\sc Lee, Y.}, {\sc Endoh,
  Y.}, {\sc Shirane, G.}, and {\sc Yamada, K.}, 1999, {\em Phys. Rev. B\/}, 60,
  3294, cond-mat/9902248

\bibitem{TworzydloDZ98}
{\sc Tworzydlo, J.}, {\sc van Duin, C. N.~A.}, and {\sc Zaanen, J.}, 1998,
  Spin-only approach to quantum magnetism in the ordered stripe state,
  cond-mat/9808034, proc. of 2nd Intern. Conf. on Stripes and High Tc
  Superconductivity, Rome '98, Journal of Superconductivity

\bibitem{TworzydloOvZ99}
{\sc Tworzydlo, J.}, {\sc Osman, O.~Y.}, {\sc van Duin, C. N.~A.}, and {\sc
  Zaanen, J.}, 1999, {\em Phys. Rev. B\/}, 59, 115, cond-mat/9804012

\bibitem{TroyerZU97}
{\sc Troyer, M.}, {\sc Zhitomirsky, M.~E.}, and {\sc Ueda, K.}, 1997, {\em
  Phys. Rev. B\/}, 55, R6117, cond-mat/9606089

\bibitem{JohnstonETal00b}
{\sc Johnston, D.} {\sc et~al.}, cond-mat/0001147, 2000, Magnetic
  Susceptibilities of Spin-1/2 Antiferromagnetic Heisenberg Ladders and
  Applications to Ladder Oxide Compounds

\bibitem{BeardBGW98}
{\sc Beard, B.~B.}, {\sc Birgeneau, R.~J.}, {\sc Greven, M.}, and {\sc Wiese,
  U.~J.}, 1998, {\em Phys. Rev. Lett.\/}, 80, 1742, cond-mat/9709110

\bibitem{KimLT97}
{\sc Kim, J.}, {\sc Landau, D.~P.}, and {\sc Troyer, M.}, 1997, {\em Phys. Rev.
  Lett.\/}, 79, 1583, cond-mat/9702138

\bibitem{Beard00}
{\sc Beard, B.~B.}, 2000, {\em Nucl. Phys. B-Proc. Suppl.\/}, 83-4, 682

\bibitem{YinTC98}
{\sc Yin, L.}, {\sc Troyer, M.}, and {\sc Chakravarty, S.}, 1998, {\em
  Europhys. Lett.\/}, 42, 559, cond-mat/9802196

\bibitem{KorotinEATK99}
{\sc Korotin, M.~A.}, {\sc Elfimov, I.~S.}, {\sc Anisimov, V.~I.}, {\sc Troyer,
  M.}, and {\sc Khomskii, D.~I.}, 1999, {\em Phys. Rev. Lett.\/}, 83, 1387,
  cond-mat/9901214

\bibitem{TroyerS98}
{\sc Troyer, M.} and {\sc Sachdev, S.}, 1998, {\em Phys. Rev. Lett.\/}, 81,
  5418, cond-mat/9807393

\bibitem{JohnstonETal00a}
{\sc Johnston, D.} {\sc et~al.}, 2000, {\em Phys. Rev. B\/}, 61, 9558,
  cond-mat/0003271

\bibitem{OnishiM00}
{\sc Onishi, H.} and {\sc Miyashita, S.}, 2000, {\em JPSJ\/}, 69, 2634,
  cond-mat/9912210

\bibitem{MatsumotoYTT01}
{\sc Matsumoto, M.}, {\sc Yasuda, C.}, {\sc Todo, S.}, and {\sc Takayama, H.},
  2001, {\em Phys. Rev. B\/}, 65, 014407, cond-mat/0107115

\bibitem{NakamuraT01}
{\sc Nakamura, M.} and {\sc Todo, S.}, cond-mat/0112377, 2001, Order Parameter
  to Characterize Valence-Bond-Solid States in Quantum Spin Chains, to appear
  in Phys.\ Rev.\ Lett.

\bibitem{NakamuraT02}
{\sc Nakamura, M.} and {\sc Todo, S.}, cond-mat/0201204, 2002, Novel Order
  Parameter to Characterize Valence-Bond-Solid States

\bibitem{FrischmuthS97}
{\sc Frischmuth, B.} and {\sc Sigrist, M.}, 1997, {\em Phys. Rev. Lett.\/}, 79,
  147, cond-mat/9702215

\bibitem{FrischmuthSAT99}
{\sc Frischmuth, B.}, {\sc Sigrist, M.}, {\sc Ammon, B.}, and {\sc Troyer, M.},
  1999, {\em Phys. Rev. B\/}, 60, 3388, cond-mat/9808027

\bibitem{AmmonS99}
{\sc Ammon, B.} and {\sc Sigrist, M.}, 1999, {\em J. Phys. Soc. Jpn.\/}, 68,
  1018, cond-mat/9902252

\bibitem{NishinoOYM00}
{\sc Nishino, M.}, {\sc Onishi, H.}, {\sc Yamaguchi, K.}, and {\sc Miyashita,
  S.}, 2000, {\em Phys. Rev. B\/}, 62, 9463, cond-mat/0002082

\bibitem{YasudaTMT01a}
{\sc Yasuda, C.}, {\sc Todo, S.}, {\sc Matsumoto, M.}, and {\sc Takayama, H.},
  2002, {\em J. Phys. Chem. of Solids\/}, 63, 1607, cond-mat/0109137

\bibitem{YasudaTMT02}
{\sc Yasuda, C.}, {\sc Todo, S.}, {\sc Matsumoto, M.}, and {\sc Takayama, H.},
  2002, {\em Prog. Theor. Phys. Suppl.\/}, 145, 339, cond-mat/0204397

\bibitem{IinoI96}
{\sc Iino, Y.} and {\sc Imada, M.}, 1996, {\em J. Phys. Soc. Jpn.\/}, 65, 3728,
  cond-mat/9609038

\bibitem{MiyazakiTOUY97}
{\sc Miyazaki, T.}, {\sc Troyer, M.}, {\sc Ogata, M.}, {\sc Ueda, K.}, and {\sc
  Yoshioka, D.}, 1997, {\em J. Phys. Soc. Jpn.\/}, 66, 2580, cond-mat/9706123

\bibitem{GrevenB98}
{\sc Greven, M.} and {\sc Birgeneau, R.~J.}, 1998, {\em Phys. Rev. Lett.\/},
  81, 1945, cond-mat/9803064

\bibitem{ImadaI97}
{\sc Imada, M.} and {\sc Iino, Y.}, 1997, {\em J. Phys. Soc. Jpn.\/}, 66, 568,
  cond-mat/9702158

\bibitem{TodoTK01}
{\sc Todo, S.}, {\sc Takayama, H.}, and {\sc Kawashima, N.}, 2001, {\em Phys.
  Rev. Lett.\/}, 86, 3210, cond-mat/0101226

\bibitem{YasudaTHKMT01}
{\sc Yasuda, C.}, {\sc Todo, S.}, {\sc Harada, K.}, {\sc Kawashima, N.}, {\sc
  Miyashita, S.}, and {\sc Takayama, H.}, 2001, {\em Phys. Rev. B\/}, 6314,
  0415, cond-mat/0010397

\bibitem{YasudaTMT01b}
{\sc Yasuda, C.}, {\sc Todo, S.}, {\sc Matsumoto, M.}, and {\sc Takayama, H.},
  2001, {\em Phys. Rev. B\/}, 6409, 2405, cond-mat/0101337

\bibitem{OnishiM02}
{\sc Onishi, H.} and {\sc Miyashita, S.}, cond-mat/0207025, 2002, Quantum
  Narrowing Effect in a Spin-Peierls System with Quantum Lattice Fluctuation

\bibitem{KuhneL99}
{\sc Kuhne, R.~W.} and {\sc Low, U.}, 1999, {\em Phys. Rev. B\/}, 60, 12125,
  cond-mat/9905337

\bibitem{RaabLUK01}
{\sc Raas, C.}, {\sc Löw, U.}, {\sc Uhrig, G.~S.}, and {\sc Kühne, R.~W.},
  2002, {\em Phys. Rev. B\/}, 65, 144438, cond-mat/0110298

\bibitem{AlvarezG01}
{\sc Alvarez, J.} and {\sc (Saarbruecken), C.~G.}, 2002, {\em Phys. Rev.
  Lett.\/}, 88, 077203, cond-mat/0105585

\bibitem{IkegamiMR98}
{\sc Ikegami, T.}, {\sc Miyashita, S.}, and {\sc Rieger, H.}, 1998, {\em J.
  Phys. Soc. Jpn.\/}, 67, 2671, cond-mat/9803270

\bibitem{PichYRK98}
{\sc Pich, C.}, {\sc Young, A.~P.}, {\sc Rieger, H.}, and {\sc Kawashima, N.},
  1998, {\em Phys. Rev. Lett.\/}, 81, 5916, cond-mat/9812414

\bibitem{KohnoTH98}
{\sc Kohno, M.}, {\sc Takahashi, M.}, and {\sc Hagiwara, M.}, 1998, {\em Phys.
  Rev. B\/}, 57, 1046, cond-mat/9709173

\bibitem{HagiwaraNKKNST98}
{\sc Hagiwara, M.}, {\sc Narumi, Y.}, {\sc Kindo, K.}, {\sc Kohno, M.}, {\sc
  Nakano, H.}, {\sc Sato, R.}, and {\sc Takahashi, M.}, 1998, {\em Phys. Rev.
  Lett.\/}, 80, 1312, cond-mat/9708127

\bibitem{OnishiM01}
{\sc Onishi, H.} and {\sc Miyashita, S.}, 2001, {\em Phys. Rev. B\/}, 64,
  014405, cond-mat/0012377

\bibitem{Nishiyama98}
{\sc Nishiyama, Y.}, 1998, {\em Eur. Phys. J. B\/}, 6, 335

\bibitem{TodoKT00}
{\sc Todo, S.}, {\sc Kato, K.}, and {\sc Takayama, H.}, 2000, {\em J. Phys.
  Soc. Jpn. Suppl.\/}, 69A, 355

\bibitem{VebericPE02}
{\sc Veberic, D.}, {\sc Prelovsek, P.}, and {\sc Evertz, H.}, 2001, {\em Open
  Problems in Strongly Correlated Electron Systems (Bled 2000)\/}, edited by
  J.~Bonca, P.~Prelovsek, A.~Ramsak, and S.~Sarkar (Kluwer), cond-mat/0203519

\bibitem{FrischmuthMT99}
{\sc Frischmuth, B.}, {\sc Mila, F.}, and {\sc Troyer, E.}, 1999, {\em Phys.
  Rev. Lett.\/}, 82, 835, cond-mat/9807179

\bibitem{MilaFDT99}
{\sc Mila, F.}, {\sc Frischmuth, B.}, {\sc Deppeler, A.}, and {\sc Troyer, M.},
  1999, {\em Phys. Rev. Lett.\/}, 82, 3697, cond-mat/9809094

\bibitem{ZaanenOKNT01}
{\sc Zaanen, J.}, {\sc Osman, O.~Y.}, {\sc Kruis, H.~V.}, {\sc Nussinov, Z.},
  and {\sc Tworzydlo, J.}, 2001, {\em Philos. Mag. B-Phys. Condens. Matter
  Stat. Mech. Electron. Opt. Magn. Prop.\/}, 81, 1485, cond-mat/0102103

\bibitem{TroyerWA02}
{\sc Troyer, M.}, {\sc Wessel, S.}, and {\sc Alet, F.}, cond-mat/0207138, 2002,
  A quantum Monte Carlo algorithm to calculate the free energy of a quantum
  system

\bibitem{Binder}
See e.g. {\em Monte Carlo Methods in Statistical Physics}, 2nd ed., ed. by K.
  Binder (Springer Berlin, New York 1986); {\em Applications of the Monte Carlo
  Method in Statistical Physics} 2nd ed., ed. by K. Binder (Springer Berlin,
  New York 1987); K. Binder and D.W. Heermann, {\em Monte Carlo Simulation in
  Statistical Physics} (Springer Berlin, New York 1986).

\bibitem{LandauBinder-book00}
{\sc Landau, D.~P.} and {\sc Binder, K.}, 2000, {\em A Guide to Monte Carlo
  Simulations in Statistical Physics\/} (Cambridge University Press)

\bibitem{BinderL01}
{\sc Binder, K.} and {\sc Luijten, E.}, 2001, {\em Physics Reports\/}, 344, 179

\bibitem{MetropolisRRTT53}
{\sc Metropolis, N.}, {\sc Rosenbluth, A.}, {\sc Rosenbluth, M.}, {\sc Teller,
  A.}, and {\sc Teller, E.}, 1953, {\em J. Chem. Phys.\/}, 21, 1087

\bibitem{AllenTildesley-book87}
{\sc Allen, M.} and {\sc Tildesley, D.}, 1987, {\em Computer Simulations of
  Liquids\/}, chap.~6 (Oxford University Press)

\bibitem{HohenbergH77}
{\sc Hohenberg, P.~C.} and {\sc Halperin, B.~I.}, 1977, {\em Rev. Mod.
  Phys.\/}, 49, 435

\bibitem{ProppW96}
{\sc Propp, J.~G.} and {\sc Wilson, D.~B.}, 1996, {\em Random Structures and
  Algorithms\/}, 9(1\&2), 223, (Paper and current annotated bibliography are
  available at \mbox{http://dbwilson.com/exact})

\bibitem{jackknife}
See e.g.\ {\sc Yang, C.K.} and {\sc Robinson, D.H.}, {\em Understanding and
  Learning Statistics by Computer}, World Scientific Series in Computer
  Sciences, Vol.\ 4, 1986

\end{thebibliography}

\end{document}